\documentclass[letterpaper,11pt]{article}
\pdfoutput=1 
\usepackage{jcappub} 
\usepackage[utf8]{inputenc}
\usepackage[T1]{fontenc} 
\usepackage[dvipsnames]{xcolor}
\usepackage{multirow}
\usepackage{slashed}
\usepackage[normalem]{ulem}
\usepackage{sidecap}
\usepackage{mathtools}
\usepackage{caption}
\usepackage{extarrows}
\usepackage{subcaption}
\usepackage{multicol}
\usepackage[toc]{appendix}
\usepackage{amssymb}
\usepackage{makecell}
\newcommand\underrel[3][]{\mathrel{\mathop{#3}\limits_{%
      \ifx c#1\relax\mathclap{#2}\else#2\fi}}}

\newcommand{\CL}{{\tt ${\mathcal C}$osmo${\mathcal L}$attice}}

\def\figureautorefname~#1\null{Fig.\,#1\null}
\def\tableautorefname~#1\null{Tab.\,#1\null}

\def\equationautorefname~#1\null{Eq.\,(#1)\null}

\title{Geometric reheating of the Universe}

\author{Daniel G. Figueroa}
\author{and Nicol\'{a}s Loayza}
\affiliation{Instituto de F\'{i}sica Corpuscular (IFIC), Universitat de Val\`{e}ncia-CSIC, Parc Cient\'{i}fic UV, C/ Catedr\'{a}tico Jos\'{e} Beltr\'{a}n 2,
E-46980 Paterna, Spain}

\emailAdd{daniel.figueroa@ific.uv.es}
\emailAdd{nicolas.loayza@ific.uv.es}

\date{\today}

\abstract{We study the post-inflationary energy transfer from the inflaton ($\phi$) into a scalar field ($\chi$) non-minimally coupled to gravity through $\xi R|\chi|^2$, considering models with inflaton potential $V_{\rm inf} \propto |\phi|^{\,p}$ around $\phi = 0$. This corresponds to the paradigm of {\it geometric preheating}, which we extend to its non-linear regime via lattice simulations. Considering $\alpha$-attractor T-model 
potentials as a proxy, we study the viability of proper {\it reheating} for $p=2, 4, 6$, determining whether radiation domination (RD) due to energetic dominance of $\chi$ over $\phi$, can be achieved. 
For large inflationary scales $\Lambda$, reheating is frustrated for $p = 2$, it can be partially achieved for $p = 4$, and it becomes very efficient for $p = 6$. 
Efficient reheating can be however blocked if $\chi$ sustains self-interactions (unless these are extremely feeble), or if $\Lambda$ is low enough, so that inflaton fragmentation brings the universe rapidly into RD. 
Whenever RD is achieved, either due to reheating or to inflaton fragmentation, we characterize the energy and time scales of the problem, as a function 
of $\Lambda$ and $\xi$. 
} 

\begin{document}
\maketitle
\flushbottom

\section{Introduction}

Viable inflationary scenarios must be compatible with observations of the cosmic microwave background (CMB)~\cite{Martin:2013tda,Planck:2018jri,Planck:2018vyg}, and in particular with the latest constraint on the B-mode polarization~\cite{BICEP:2021xfz,Tristram:2021tvh}, which sets an upper bound on the inflationary Hubble rate as $H_* \lesssim 4.4\times10^{13}$ GeV. Whilst CMB observations support strongly the idea of an early inflationary period, there are no stringent constraints on the stages following afterwards. The only clear requisite is that 
a period of \textit{reheating} must ensue after inflation, during which the Universe must reach a radiation dominated (RD) thermal state, prior to Big Bang Nucleosynthesis (BBN)~\cite{Kawasaki:1999na,Kawasaki:2000en,Hannestad:2004px,Hasegawa:2019jsa}. 

The possibilities for reheating depend largely upon the manner in which the inflaton interacts with other particle species, which we refer to as {\it daughter} fields. At one extreme lie direct couplings, allowing (non-)perturbative 
decays of the inflaton into the daughter species~\cite{Linde:1981mu,Albrecht:1982mp,Dolgov:1982th,Abbott:1982hn,Traschen:1990sw,Kofman:1994rk,Shtanov:1994ce,Kaiser:1995fb,Kofman:1997yn,Greene:1997fu,Kaiser:1997mp,Kaiser:1997hg,Greene:1998nh,Greene:2000ew,Peloso:2000hy,Felder:2000hj,Felder:2001kt,Copeland:2002ku,GarciaBellido:2002aj,Rajantie:2000nj,Copeland:2001qw,Smit:2002yg,GarciaBellido:2003wd,Tranberg:2003gi,Skullerud:2003ki,vanderMeulen:2005sp,DiazGil:2007dy,DiazGil:2008tf,Dufaux:2010cf,Berges:2010zv,Tranberg:2017lrx,Deskins:2013lfx,Adshead:2015pva,Adshead:2016iae,Lozanov:2016hid,Lozanov:2016pac,Lozanov:2017hjm,Figueroa:2017qmv,Adshead:2017xll,Adshead:2018doq,Cuissa:2018oiw,Adshead:2019lbr,Adshead:2019igv,Figueroa:2019jsi,Antusch:2020iyq,Antusch:2021aiw}. At the other extreme lie scenarios where only gravity mediates between the inflationary and the daughter sectors, see e.g.~\cite{Ford:1986sy,Spokoiny:1993kt,Bassett:1997az,Tsujikawa:1999jh,Tsujikawa:1999iv,Tsujikawa:1999me,DeCross:2015uza,DeCross:2016fdz,DeCross:2016cbs,Figueroa:2016dsc,Opferkuch:2019zbd,Dimopoulos:2018wfg,Bettoni:2021zhq,Figueroa:2021iwm,Figueroa:2024asq,Ema:2016dny,Nguyen:2019kbm,vandeVis:2020qcp}. In either type of scenario, non-perturbative {\it preheating} 
effects can be present, leading to very efficient exponential transfer of energy into the daughter fields, see Refs.~\cite{Allahverdi:2010xz,Amin:2014eta,Lozanov:2019jxc,Allahverdi:2020bys} for reviews. 

In this work we focus on gravity-mediated (p)reheating scenarios, which we 
refer to as {\it gravitational (p)reheating}. We consider a daughter scalar field $\chi$ non-minimally coupled to gravity through an interaction of the form $\xi |\chi|^2 R$, with $R$ the Ricci scalar and $\xi$ a dimensionless coupling constant. The presence of such term is actually required by the renormalizability of the theory in curved spacetime~\cite{Parker2009,Birrell1984}, with $\xi$ a running parameter that cannot be set to zero at all energy scales.  We will refer to $\chi$ as a non-minimally coupled (NMC) field. 

An early formulation of gravitational reheating~\cite{Ford:1986sy} relied on two ingredients, the excitation of a NMC scalar field towards the end of inflation~\cite{Ford:1986sy,Spokoiny:1993kt,Peebles:1998qn}, and the occurrence of a period of {\it kination} domination (KD) immediately after inflation. Under those circumstances, the initially subdominant energy of the NMC field, assumed in the form of radiation, eventually becomes the dominant energy component of the Universe. This idea, first proposed in Ref.~\cite{Ford:1986sy} and later extended in~\cite{Spokoiny:1993kt,Damour:1995pd}, plays a key role in {\it Quintessential inflation} scenarios~\cite{Peebles:1998qn,Peloso:1999dm,Huey:2001ae,Majumdar:2001mm,Dimopoulos:2001ix,Wetterich:2013jsa,Wetterich:2014gaa,Hossain:2014xha,Rubio:2017gty}. The original proposal has been shown~\cite{Figueroa:2018twl}, however, to be inconsistent with CMB/BBN constraints~\cite{Caprini:2018mtu,Clarke:2020bil}, due to an excess of gravitational waves (GWs) accumulated during KD~\cite{Giovannini:2009kg, Boyle:2007zx,Figueroa:2019paj,Bernal:2020ywq}. Furthermore, Ref.~\cite{Figueroa:2016dsc} has also shown that the NMC field does not behave as radiation during KD (as assumed in~\cite{Ford:1986sy,Spokoiny:1993kt,Peebles:1998qn}), but rather experiences a tachyonic instability, due to the flip of sign of $R \propto (1-3w)$ for a stiff equation of state $w > 1/3$. This key effect is precisely the underlying idea of {\it Ricci reheating}~\cite{Figueroa:2016dsc,Opferkuch:2019zbd,Dimopoulos:2018wfg,Bettoni:2021zhq,Figueroa:2024asq}, where a NMC scalar is exponentially excited during KD, leading to successful reheating, and to a suppression of the aforementioned excess of GWs.

Interestingly, while gravitational reheating can be realized by the excitation of NMC scalars during a sustained period of KD, just as described above, NMC scalars can be also naturally excited after inflation, in a broader set of scenarios. In particular, if the part of the inflaton potential that dictates the dynamics after inflation is characterized by a monomial shape, inflaton oscillations ensue as soon as inflation ends. This induces an oscillatory behavior of $R$, leading to a tachyonic excitation of NMC scalars, whenever $R$ becomes negative during the oscillations. As a consequence of the tachyonic regime, there is an exponential growth of the energy of the NMC field. This mechanism, first studied and coined as {\it geometric preheating} in Ref.~\cite{Bassett:1997az}, see also Ref.~\cite{Tsujikawa:1999jh}, has been later on considered in higher order curvature inflationary models~\cite{Tsujikawa:1999iv,Fu:2019qqe}, and generalized to multi-field inflationary scenarios~\cite{DeCross:2015uza,DeCross:2016fdz,DeCross:2016cbs,Nguyen:2019kbm,vandeVis:2020qcp}. 

In this paper we focus on geometric preheating effects in single field inflationary scenarios, choosing an inflaton potential that has a monomial shape during the post-inflationary dynamics. In this regard, $\alpha$-{\it attractor} T-models~\cite{Kallosh:2013hoa} are inflationary scenarios particularly useful, as they have potentials that approach a plateau at large field values, but follow a monomial form as $V_{\rm inf} \propto |\phi|^{\,p}$ 
for small field values. Particularly interesting is the fact that these models offer the possibility to tune the energy scale of inflation to arbitrary values. In this work, we exploit this fact and investigate geometric preheating-like effects at different energy scales, including scales well below the current upper bound imposed by CMB observations. We study the dynamics when they become non-linear, determining whether proper {\it reheating}, i.e.~onset of radiation domination (RD) due to energetic dominance of $\chi$ over $\phi$, can be achieved. We promote, in this way, the idea of geometric preheating into {\it geometric reheating}, studying the circumstances under which reheating is successfully achieved. 

Using $\alpha$-attractor T-model 
potentials as a proxy, we have studied the non-linear dynamics of the system for $p=2, 4, 6$, as a function of the energy scale $\Lambda$. We find that in high energy scale scenarios, reheating is never achieved for $p = 2$, it can be partially achieved for $p = 4$, and it becomes very efficient for $p = 6$. If the NMC field sustains self-interactions of the form $V_{\rm NMC} \propto \lambda\chi^4$, reheating will be frustrated even for $p = 6$, unless the interactions are extremely feeble, with $\lambda \lll 1$. If the energy scale of inflation $\Lambda$ is smaller than a certain threshold, the Universe enters very rapidly into RD, due to efficient inflaton fragmentation after inflation. As a result, there is no efficient transfer of energy into the NMC field in this case, and reheating (into $\chi$) never completes. In this paper, we have characterized the energy and time scales of the problem whenever RD is achieved, either due to proper reheating or due to inflaton fragmentation, as a function of $\Lambda$ and $\xi$.

We highlight that while the initial tachyonic growth of $\chi$ corresponds to a linear regime of the dynamics of the system, the realization of proper reheating or of any of its frustration mechanisms (either due to self-interactions or due to inflaton fragmentation), are events characterized by the development of non-linearities in the dynamics. Namely, a regularisation effect of the tachyonic mass eventually becomes manifest, stopping the exponential energy transfer into $\chi$. This can be due to:\vspace{0.2cm}

\noindent\hspace{0.5cm}$i)$ {\it Gravitational backreaction}: when the NMC field becomes the dominant energy component in the universe and proper reheating is achieved.\vspace{0.2cm}

\noindent~~~$ii)$ {\it Self-interaction backreaction}: when self-interactions of the NMC field are present and they block eventually the energy transfer into $\chi$.\vspace{0.2cm}

\noindent\hspace{0.3cm}$iii)$ {\it Inflaton fragmentation}: when inflaton backreacts onto itself and fragmentation becomes efficient, preventing further energy transfer into $\chi$.\vspace{0.2cm} 

In the our work we capture all three regimes in our simulations, in the cases of interest. This manuscript is divided as follows. In section~\ref{sec:ScaleofInflation}, we introduce the $\alpha$-{\it attractor} T-model family of inflationary potentials, and analyze their CMB constraints. We also discuss the inflaton dynamics after inflation, including both oscillatory and fragmentation regimes, depending on the inflationary energy scale $\Lambda$. In section~\ref{sec:LinearRH} we present the linear dynamics of a NMC daughter field after inflation, i.e.~we describe geometric preheating during inflaton oscillations. In section~\ref{sec:LatticeRH} we present our lattice results for the full non-linear dynamics of the system, in the case when the inflaton condensate oscillates many times after inflation. This includes gravitational backreaction (when geometric reheating is achieved) and NMC self-interaction backreaction (when reheating is frustrated). We also discuss how inflaton fragmentation affects either case, when the energy scale of inflation is sufficiently low. In section~\ref{sec:discussion} we summarize our findings and discuss potential implications. 

{\it Conventions --.} We assume summation over repeated indices. We denote the reduced Planck mass by $m_p = 2.435\times 10^{18}$ GeV .

\section{Inflationary model}\label{sec:ScaleofInflation}
We consider a family of inflationary potentials based on $\alpha$-attractor T-models~\cite{Kallosh:2013hoa}
\begin{equation}\label{eq:potential}
V_{\rm inf}(\phi ) = \frac{\Lambda^4}{p} \tanh^p\left(\frac{|\phi|}{M}\right) \simeq 
\left\lbrace\begin{array}{cl}
\frac{\Lambda^4}{p} & ,~|\phi| \gg M\vspace{2mm}\\
\frac{\Lambda^4}{p}\left(\frac{|\phi|}{M}\right)^p & ,~|\phi| \ll M\\
\end{array}\right.
\end{equation}
with $p$ a dimensionless parameter, and $\Lambda$ and $M$ dimension one parameters which control the height and the width of the potential,   respectively.  We shall work with $p \geq 2$, so that $\partial V_{\rm inf}/\partial\phi$ and $\partial^2 V_{\rm inf}/\partial\phi^2$ are well defined when $\phi = 0$. Potentials~(\ref{eq:potential}) exhibit the characteristic property that for large field values, $|\phi|\gg M$, they approach a plateau as $V_{\rm inf}(\phi) \rightarrow \Lambda^4/p$, whereas for small field values, $|\phi|\ll M$, they adopt a monomial shape as $V_{\rm inf}(\phi) \propto |\phi|^{\,p}$.  

As a first approach to understand the inflationary dynamics, we obtain the {\it potential slow-roll} parameters, which read
\begin{eqnarray}
    \epsilon_V \equiv \frac{m_p^2}{2}\left( \frac{V_{\rm inf}'(\phi)}{V_{\rm inf}(\phi)} \right)^2 = \frac{2 p^2 \text{csch}^2(\frac{2\phi}{M})}{(M/m_p)^2}\,, ~~~\eta_V \equiv m_p^2\frac{V_{\rm inf}''(\phi)}{V_{\rm inf}(\phi)} = \frac{4p(p-\cosh(\frac{2\phi}{M})) \text{csch}^2(\frac{2\phi}{M})}{(M/m_p)^2}\,.\nonumber\\
\end{eqnarray}
From the condition $\epsilon_{V} = 1$, we immediately obtain the inflaton amplitude ${\phi}_{\rm end}$
at the end of inflation\footnote{Such value is only approximated, as the end of inflation should be rather identified with the moment when $\epsilon_{H} \equiv -d\log H/d\log N = 1$, with $H$ the Hubble rate and $N$ the number of efoldings. The condition $\epsilon_{V} = 1$ is simply an approximation to this.}. Indicating with a sub-index $_k$ the time when a scale $k$ crossed the Hubble radius $- N_k$ efolds before the end of inflation, from the expression $N_k \simeq -\int_{{\phi}_{\rm end}}^{{\phi}_{k}}{d{\phi}\over\sqrt{2\epsilon_V}}$ we can obtain the inflaton amplitude ${\phi}_{k}$ as
\begin{equation}\label{eq:phik60efoldsappr}
{\phi_{k}(M,p)\over m_p} \simeq \frac{\mathcal{M}}{2}\text{arccosh}\left\lbrace\frac{\mathcal{M}\sqrt{2p^2 + \mathcal{M}^2}-4pN_{k}}{\mathcal{M}^2}\right\rbrace\,,
\end{equation}
where, without loss of generality, we have chosen $N_k < 0$ and the positive branch ${\phi} > 0$, during inflation. For later convenience we have also introduced the dimensionless mass-parameter
\begin{eqnarray}
\mathcal{M} \equiv {M\over m_p} \,.
\end{eqnarray}
 The slow roll framework relates the amplitude $A_s$ and spectral index $n_s$ of the scalar perturbations produced during inflation, as well as the amplitude of inflationary tensor perturbations, parameterized through the tensor-to-scalar ratio $r \equiv A_t/A_s$, 
with the inflaton potential and its derivatives, 
via the expressions~\cite{Baumann:2009ds}
\begin{eqnarray}\label{eq:SRapprox}
A_s \simeq \frac{1}{24\pi^2\epsilon_{V_k}}\frac{V_{\rm inf}({\phi}_{k})}{m_p^4}\,,~~~~ n_s - 1 \simeq 2\eta_{V_k} - 6\epsilon_{Vk}\,,~~~~ r \simeq 16\epsilon_{V_k}\,.
\end{eqnarray}  
Introducing the value ${\phi}_{k}$ from (\ref{eq:phik60efoldsappr}) into expressions~(\ref{eq:SRapprox}), leads to
\begingroup
\allowdisplaybreaks
\begin{eqnarray}\label{eq:SRpredictionA}
\hspace{-1cm}A_s(\Lambda,M,p) &=& 
{\Lambda^4\over m_p^4}\frac{\Big[8pN_k^2+\mathcal{M}^2\Big(p -4 N_k \sqrt{1+\frac{2p^2}{{\mathcal M}^2}}\,\Big)\Big]\Big[{\mathcal M}^2\Big(\sqrt{1+\frac{2p^2}{{\mathcal M}^2}}-1\Big) - 4pN_k \Big]^p}{24\pi^2p^2\mathcal{M}^2\Big[{\mathcal M}^2\Big(\sqrt{1+\frac{2p^2}{{\mathcal M}^2}}+1\Big) - 4pN_k\Big]^p},\\
\label{eq:SRpredictionNs}
n_s(M,p) &=& \frac{8p(N_k +2) N_k  - {\mathcal M}^2\Big(4(N_k + 1 )\sqrt{1+\frac{2p^2}{{\mathcal M}^2}} + p\Big)}{8pN_k^2 + {\mathcal M}^2 \Big(p - 4 N_k\sqrt{1+\frac{2p^2}{{\mathcal M}^2}} \Big)}\,,
\\
\label{eq:SRpredictionR}
r(M,p) &=& \frac{16p{\mathcal M}^2}{8pN_k^2 + {\mathcal M}^2 \Big(p - 4 N_k \sqrt{1+\frac{2p^2}{{\mathcal M}^2}}\Big)}\,.
\end{eqnarray}
\endgroup
Though not very illuminating by themselves, expressions (\ref{eq:SRpredictionA})-(\ref{eq:SRpredictionR}) can be used readily to constrain the parameters $\lbrace {\Lambda}, {M}, p \rbrace$ that characterize the potential~(\ref{eq:potential}), by confronting them 
against CMB observations. In particular, current CMB data constrain scalar and tensor perturbations at the pivot scale $k_{\rm CMB}=0.05$ Mpc$^{-1}$ as~\cite{Planck:2018jri,BICEP:2021xfz,Tristram:2021tvh}
\begin{eqnarray}\label{eq:CMBconstraintsA}
    A_s &=& 2.099^{+0.296}_{-0.292}\cdot 10^{-9}~(68\% ~{\rm CL})\,,~~~
    \\
    \label{eq:CMBconstraintsNs}
    n_s &=& 0.9649 \pm 0.0042~(68\% ~{\rm CL})\,,\\
    \label{eq:BmodeConstraints}
    r &<& 0.032 ~(95\% ~{\rm CL})\,. 
\end{eqnarray}
As $r(M,p)$ in~Eq.~(\ref{eq:SRpredictionR}) is an increasing function of $M$ for fixed $p$, the upper bound  on the tensor-to-scalar ratio (\ref{eq:BmodeConstraints}) sets an upper bound as $M \leq M_{\rm max}^{(p)}$, for a given $p$, independently of $\Lambda$.  Saturating the bound ~(\ref{eq:BmodeConstraints}) and setting $-N_k = -N_{\rm CMB} = 60$, leads to $M_{\rm max}^{(2)} \simeq 8.79m_p$, $M_{\rm max}^{(4)} \simeq 8.22m_p$, $M_{\rm max}^{(6)} \simeq 8.08m_p$, for $p = 2, 4, 6$, respectively. Furthermore, we observe that $n_s(M,p)$, c.f.~Eq.~(\ref{eq:SRpredictionNs}), is always well within the boundaries indicated in Eq.~(\ref{eq:CMBconstraintsNs}), for the allowed range $M \leq M_{\rm max}^{(p)}$, for any $p$, and independently of $\Lambda$. In other words, our inflationary modelling is compatible with CMB constraints on the spectral index by construction. Finally, the value of $\Lambda$, which controls the energy scale of inflation, can be solved from the expression of the scalar amplitude in Eq.~(\ref{eq:SRpredictionA}), allowing us to obtain $\Lambda$ as a function of $\lbrace M,p\rbrace$, for fixed value of $A_s$,
\begin{eqnarray}\label{eq:LambdaMp}
\hspace{-3mm}
{{\Lambda^4}(M,p;A_s)\over m_p^4}  = \frac{24\pi^2 A_s \mathcal{M}^2p^2\,\Big[{\mathcal M}^2\Big(\sqrt{1+\frac{2p^2}{{\mathcal M}^2}}+1\Big) - 4pN_k\Big]^p}{\Big[8pN_k^2+\mathcal{M}^2\Big(p -4 N_k \sqrt{1+\frac{2p^2}{{\mathcal M}^2}}\,\Big)\Big]\Big[{\mathcal M}^2\Big(\sqrt{1+\frac{2p^2}{{\mathcal M}^2}}-1\Big) - 4pN_k \Big]^p} \,.
\end{eqnarray}
By inspecting Eq.~(\ref{eq:LambdaMp}) we observe that $\Lambda(M,p;A_s)$ is a growing function of $M$, for  given values of $p$ and $A_s$. Therefore, fixing $A_s$, say to the central value 
in Eq.~(\ref{eq:CMBconstraintsA}), and taking $M = M_{\rm max}^{(p)}$, leads to an upper bound on the scale of inflation as
$\Lambda \leq \Lambda_{\rm max}^{(p)}$, with $\Lambda_{\rm max}^{(2)} \simeq 7.16\cdot10^{-3}\,m_p$, $\Lambda_{\rm max}^{(4)} \simeq 8.47\cdot10^{-3}\,m_p$, $\Lambda_{\rm max}^{(6)} \simeq 9.37\cdot10^{-3}\,m_p$, for $p = 2, 4, 6$, respectively. Equivalently, we can also obtain an approximate expression for the Hubble rate $H$ at the Hubble crossing time of CMB scales. Evaluating (\ref{eq:phik60efoldsappr}) at $N_k = N_{\rm CMB} = -60$, using~(\ref{eq:LambdaMp}), and the fact that deep inside inflation the potential energy dominates over the kinetic, i.e.~$H_{\rm inf}^2 \simeq \frac{1}{3 m_p^2} V_{\rm inf}(\phi_{\rm CMB})$, leads to
\begin{eqnarray}\label{eq:Hubble}
    \frac{H^2_{\rm inf}(M,p;A_s)}{m_p^2} & &\simeq \frac{8p \; \mathcal{M}^2 A_s \; \pi^2}{28800\;p + \mathcal{M}^2\Big(p + 240 \sqrt{1+ \frac{2p^2}{\mathcal{M}^2}}\,\Big) } \,,
\end{eqnarray}
in correspondence with the idea that decreasing $M$ reduces the energy scale of inflation. As we will see next, reducing the value of $M$ does not only lower the inflationary energy scale, but also affects noticeably the way the inflaton oscillates around its minimum after inflation.

In summary, even though the exact details depend on the choice of $p$, CMB constraints imply a maximum mass scale roughly as $M \leq M_{\rm max} \sim 8 m_p$, and a maximum energy scale as $\Lambda \leq \Lambda_{\rm max} \sim 0.01 m_p$, or alternatively as $H_{\rm inf} \leq H_{\rm max} \sim 5 \cdot 10^{13}$ GeV. Interestingly, nothing prevents us from taking the limits ${M} \ll {M}_{\rm max}$ (equivalently  ${\Lambda} \ll {\Lambda}_{\rm max}$). One of the advantages of the chosen model is precisely that we can make the energy scale of inflation to be well below its current upper bound, while respecting all CMB constraints. If we want to lower the energy scale
of inflation we just need to decrease ${M}$. The limit of {\it small scale} ${M} \ll m_p$, can be easily obtained from expressions~(\ref{eq:SRpredictionA})-(\ref{eq:SRpredictionR}), and (\ref{eq:Hubble}), as
\begin{eqnarray}\label{eq:AsympNs}
n_s(M,p) && \xlongrightarrow[~{M\over m_p} \ll 1~]{}{29\over30} \simeq 0.967\,, 
\\
\label{eq:AsympR}
r(M,p) &&  \xlongrightarrow[~{M\over m_p} \ll 1~]{}{({M/m_p})^2\over1800} = 5.56\cdot 10^{-4}\left({M\over m_p}\right)^2\,, 
\\
\label{eq:AsympLambda}
{\Lambda(M,p)\over m_p} &&  \xlongrightarrow[~{M\over m_p} \ll 1~]{} \left({ 24\pi^2 p A_s\over 28800} \right)^{1/4}\sqrt{M\over m_p} \simeq 2.04\cdot 10^{-3}\,p^{1/4}\sqrt{M\over m_p}\,,
\\
\label{eq:AsympHubble}
H_{\rm inf}(M,p) && \xlongrightarrow[~{M\over m_p} \ll 1~]{} 5.8 \times 10^{12} \left({M\over m_p}\right) \; \text{GeV}  
\,,
\end{eqnarray}
We note that for $M = m_p$, the exact expressions of $n_s(M,p)$, $r(M,p)$ and 
$\Lambda(M,p;A_s)$,  [c.f.~Eqs.~(\ref{eq:SRpredictionNs}),(\ref{eq:SRpredictionR}) and (\ref{eq:LambdaMp}), respectively], approach the asymptotic expressions in the $rhs$ of Eqs.~(\ref{eq:AsympNs})-(\ref{eq:AsympLambda}) to better than $0.1\%$, $0.2\%$ and $2\%$, while for $M = 0.1 m_p$, they approach the asymptotic behavior to better than $0.002\%$, $0.03\%$ and $0.1\%$, respectively.

We note that expressions (\ref{eq:phik60efoldsappr}), (\ref{eq:SRpredictionA})-(\ref{eq:SRpredictionR}) and (\ref{eq:LambdaMp})-(\ref{eq:Hubble}), as well as their asymptotic expressions (\ref{eq:AsympNs})-(\ref{eq:AsympHubble}), are all based on the SR description. In reality, for given $p$, once we determine $\Lambda$ as a function of $M$ with the SR approximation, i.e.~via Eq.~(\ref{eq:LambdaMp}), we can compute the dynamics of the homogeneous inflaton by solving its Klein-Gordon equation (for each $M$), and determine when inflation really ends according to the condition $\epsilon_{H} \equiv -d\log H/d\log N = 1$, with $H$ the Hubble rate and $N$ the number of efoldings. This allow us to determine the correct value $\phi_{\rm CMB}$ of the inflaton $60$ e-folds before the end of inflation [instead of using approximation~\eqref{eq:phik60efoldsappr}], for given $\lbrace p, M, \Lambda \rbrace$. Using Eq.~(\ref{eq:SRapprox}) and the observed value~(\ref{eq:CMBconstraintsA}) of the scalar perturbation amplitude $A_s$, leads to correct the value of $\Lambda$. We repeat this process iteratively, till there is no appreciable change in $\phi_{\rm CMB}$ and $\Lambda$. As a result of this procedure, the value of $\Lambda$ is corrected with respect to the SR prediction~(\ref{eq:LambdaMp}) by $ \sim 1 \% $ in a first iteration, and by $ \sim 0.001 \% $ in a second iteration, for all values of $M$ used. We conclude therefore that two iterations are more than enough.

\subsection{Inflaton and Ricci curvature oscillations}
\label{subsec:InflatonOscillations}

\begin{figure*}[tbp]
\begin{center}
\includegraphics[width=0.45\textwidth,height=5cm]{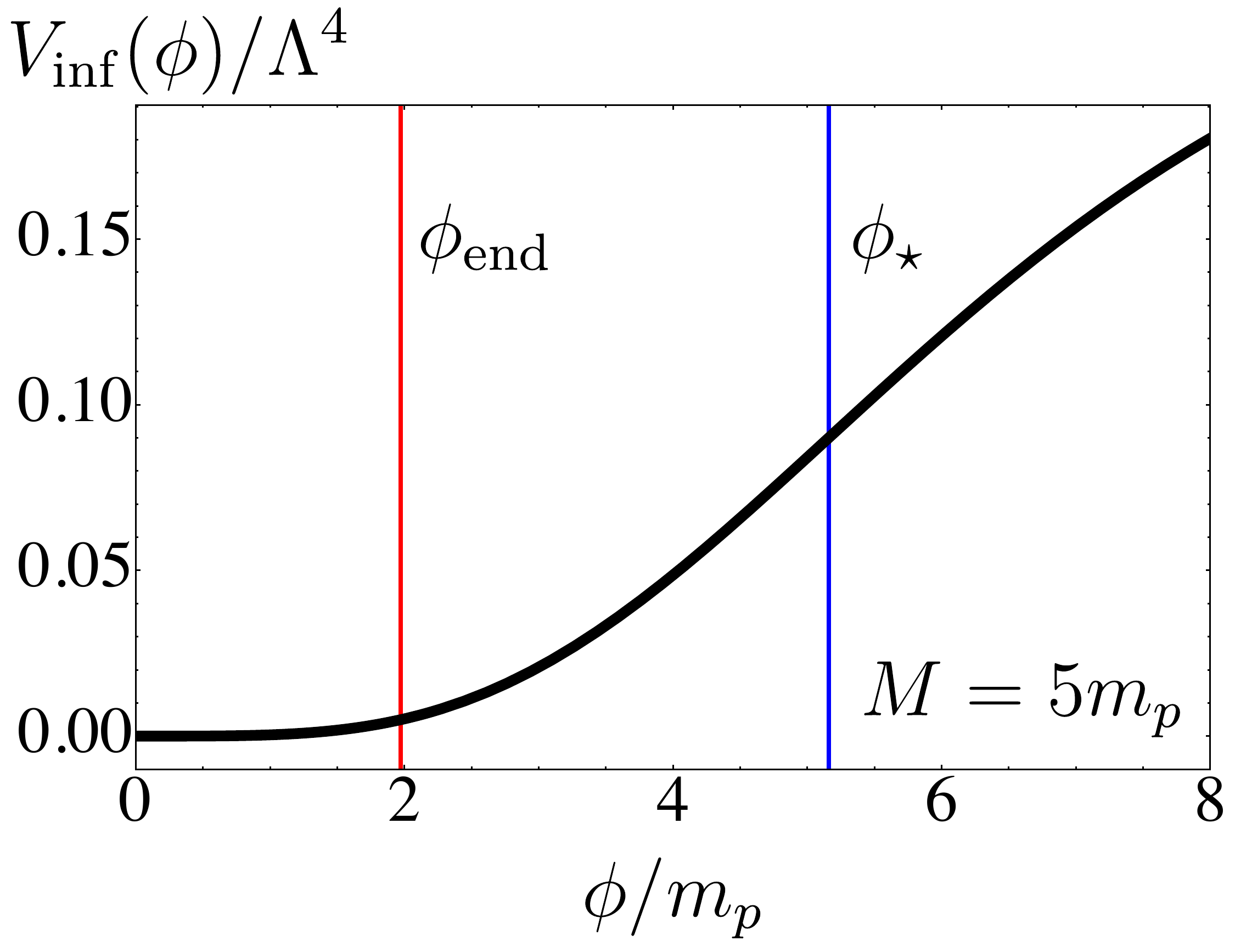}
\includegraphics[width=0.45\textwidth,height=5cm]{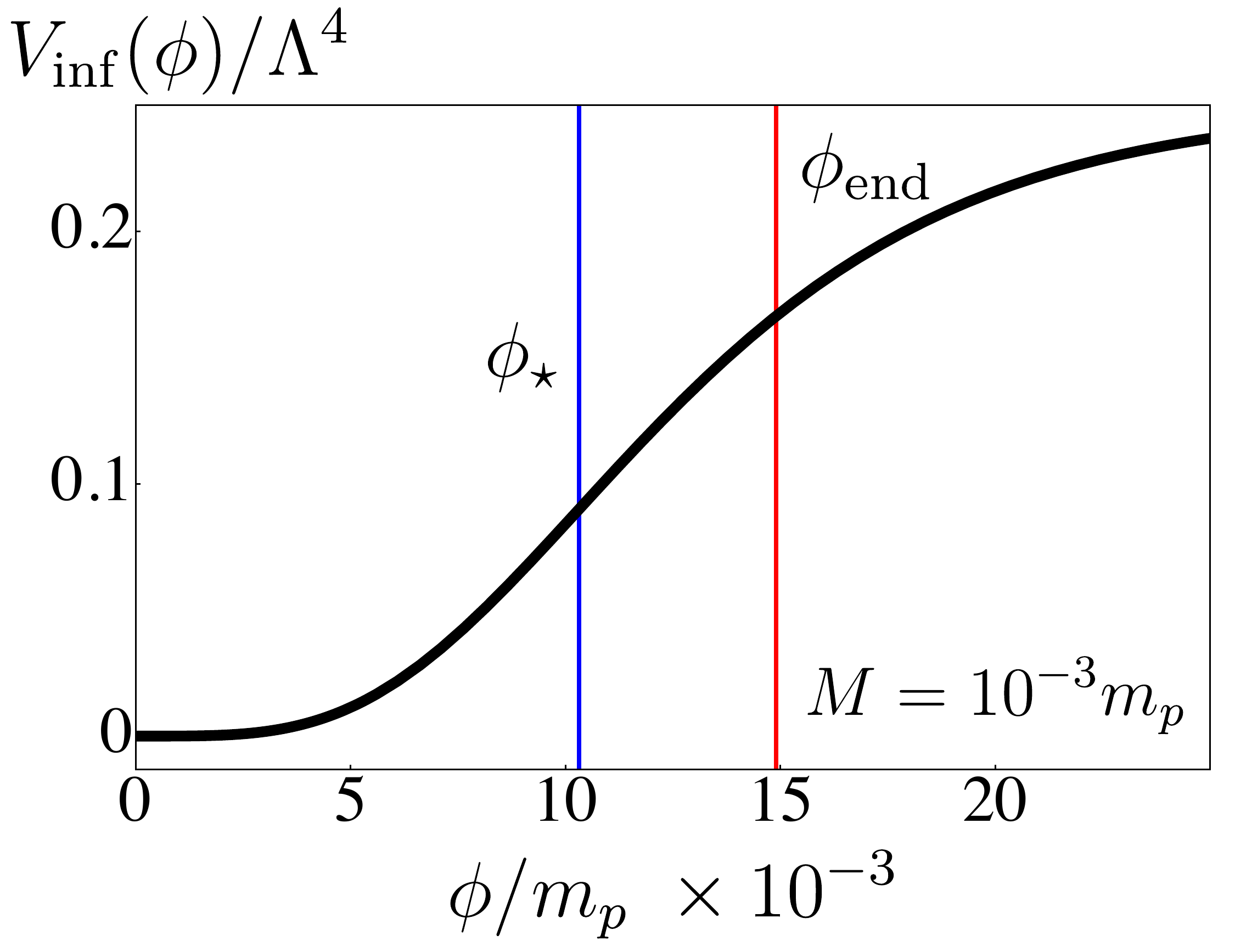}  
\end{center}
\vspace*{-0.5cm}
\caption{Inflaton potential profile for $p=4$, {\it Left: }${M}=5m_p$ and {\it Right: }${M}=10^{-3}m_p$. The choice of ${M}=5m_p$ shows that the field value at the end of inflation (${\phi}_{\rm end}$) is below its inflection point (${\phi}_{\star}$), this means that the inflaton will only probe the monomial potential of the form $\propto {\phi}^p$ when it oscillates. Alternatively, for ${M}=10^{-3}m_p$ the inflaton ends inflation at a value higher than the inflection point meaning that the first oscillations of the inflaton will probe the change of inflection in the potential. This explains why the first oscillations are potential dominated.} \label{fig:PotentialM5andM0p1} 
\vspace*{-0.3cm}
\end{figure*}

Towards the end of inflation the energy budget of the universe is initially dominated by the inflaton, which can be regarded as homogeneous\footnote{We will assume the inflaton to remain homogeneous all throughout subsection~\ref{subsec:InflatonOscillations}. While this is a valid approximation for sufficiently large energy scales, in subsection~\ref{subsec:InflatonFragmentation} we will learn that self-interactions of the inflaton will eventually force it to become inhomogeneous (the sooner the smaller the mass scale $M$). For sufficiently low scales, the analysis of subsection~\ref{subsec:InflatonOscillations} will become eventually invalidated. We will return to this in subsection~\ref{subsec:InflatonFragmentation} and at the end of Section~\ref{sec:LatticeRH}.}.  
Following the end of inflation, the homogeneous inflaton amplitude starts oscillating around the minimum of potential~(\ref{eq:potential}), located at $\phi = 0$. The nature of the oscillations depend crucially on the choice of $M$, as this scale determines
whether the inflaton amplitude at the end of inflation, ${\phi}_{\rm end}$, is above or below the {\it inflection point} of the potential, 
for which $V_{\rm inf}''({\phi}_{\star}) = 0$. In light of Eq.~(\ref{eq:potential}), we obtain
\begin{eqnarray}
    {{\phi}_{\star}\over {M}} = \text{arcsinh}\left\lbrace\sqrt{\frac{p-1}{2}}\right\rbrace\,,
\end{eqnarray}
so that the potential is divided in two field regions, ${\phi} < {\phi}_{\star}$ with positive curvature, $V_{\rm inf}''> 0$, and ${\phi} > {\phi}_{\star}$ with negative curvature, $V_{\rm inf}''< 0$. For {\it small scale} scenarios with ${M} \lesssim m_p$, it holds that ${\phi}_{\rm end} > {\phi}_{\star}$, whereas for {\it large scale} scenarios with $m_p \lesssim {M} \leq {M}_{\rm max}$ (i.e.~those close to saturate the $r$ bound), it holds the opposite situation, ${\phi}_{\rm end} < {\phi}_{\star}$. The boundary between small and large scales is always of the order of ${M} \sim m_p$, but the exact number depends on the choice of $p$, so we simply differentiate loosely large from small scale scenarios, by indicating ${M} \gtrsim m_p$ and ${M} \lesssim m_p$, respectively. In Fig.~\ref{fig:PotentialM5andM0p1} we plot, as an example, the inflaton potential for $p = 4$, indicating ${\phi}_{\rm end}$ and ${\phi}_{\star}$ with vertical lines. In the left panel we see that for a large scale model, with ${M}=5m_p$, inflation ends below the inflection point, i.e.~${\phi}_{\rm end} < {\phi}_{\star}$. In the right panel we observe that for a small scale model, with ${M} = 10^{-3}m_p$, inflation ends above the inflection point, ${\phi}_{\rm end} > {\phi}_{\star}$.

In large scale scenarios, the inflaton amplitude oscillates, consequently, probing only the positive curvature region of the potential, $|{\phi}| < {\phi}_{\star}$. Taking the small field limit ${\phi} < {M}$ of~(\ref{eq:potential}), and making use of the fact that the amplitude of the inflaton at the onset of oscillations is $\sim m_p$, we infer that the initial oscillation frequency is of the order $\Omega_{\rm osc}(\phi_{\rm end}) \sim (\Lambda/m_p)^2 (M/m_p)^{-p/2}m_p$. Due to the expansion of the universe, the homogeneous field amplitude decays $\propto a^{-6/(p+2)}$ and, as a consequence, the oscillatory frequency is damped as $\Omega_{\rm osc}(\phi) \propto a^{-3(p-2)/(p+2)}$~\cite{Antusch:2020iyq,Antusch:2021aiw}. However, as the Hubble rate also decays as $H \propto a^{-3p/(2+p)}$, their ratio grows as $\Omega_{\rm osc}/H \propto a^{6/(2+p)}$, 
which indicates that the inflaton performs more and more oscillations within a Hubble time, as time goes by.

In small scale scenarios, the inflaton amplitude probes, during oscillations, both positive ($|{\phi}| < {\phi}_{\star}$) and negative (${\phi}_{\star} < |{\phi}| < {\phi}_{\rm end}$) curvature regions of the potential. The smaller the scale ${M}$, the more time the field spends in the negative region, as compared to the positive region. The fraction of time spent within one oscillation in the negative curvature decreases however in time, as the inflaton amplitude decays due to the expansion of the universe. In addition, the effective frequency of oscillation could be approximated as $\Omega_{\rm eff}\sim V'(\phi)/\phi$, which for $\phi \sim M$ it approaches a value independent of $M$. This means that the smaller the value of $M$, the faster the oscillations are compared to the expansion of the universe, as it holds that $\Omega_{\rm eff}/H_{\rm inf} \propto 1/M$. 
Independently of whether we are in large or small scale oscillatory regimes, we can always define an {\it oscillation-averaged} version of the (homogeneous) pressure and energy densities of the inflaton, $\overline{p}_{\phi}, \overline{\rho}_{\phi}$, with $\overline{~^{{\,}^{\,}}}$ denoting oscillation averaging. This leads to define an {\it effective equation of state} (EoS) as ${w}_{\rm osc} = {\overline{ p}_{\phi}}/{\overline{\rho}_{\phi}}$. Even though $\overline p_{\phi}$ and $ \overline \rho_{\phi}$ are in general functions of time, in large scale scenarios the EoS takes a simple constant value as $w_{\rm osc} \equiv (p-2)/(p+2)$~\cite{Turner:1983he}, reflecting the fact that only the positive curvature region of the potential, with $V_{\rm inf} \propto |{\phi}|^p$, is sampled during oscillations. In the case of small scale scenarios, as the fraction of time (per oscillation) spent  in the negative curvature of the potential decreases in time, ${w}_{\rm osc}$ becomes a increasing function in time, as the oscillations carry on. However, after a number of oscillations the inflaton amplitude becomes smaller than ${\phi}_{\star}$, and from then onward the inflaton only oscillates within the positive curvature region. As a consequence, $w_{\rm osc}$ tends to the same constant $p$-dependent value characteristic of large scale scenarios. In summary, we can write
\begin{eqnarray}\label{eq:HomEoS}
    {w}_{\rm osc} \equiv \frac{\overline{ p}_{\phi}}{\overline{\rho}_{\phi}} \simeq 
    \left\lbrace
    \begin{array}{cll}
     \text{\Large$ \frac{p-2}{p+2} $}  & \hspace*{-2mm}({\rm as}~|{\phi}| < {\phi}_{\star}~{\rm always}) & \,,~ {M} \gtrsim m_p 
     \vspace{0.4cm}\\
     \Big(
     \begin{array}{c}
     {\text{\scriptsize increasing}}\vspace*{-1mm}\\
     {\text{\scriptsize function}}
     \end{array}
     \Big) &
      \xlongrightarrow
     [~({\text{\scriptsize Once}}~|{\phi}| < {\phi}_{\star})~]{}~ \text{\Large$ \frac{p-2}{p+2} $} & \,,~ {M} \lesssim m_p 
    \end{array}\right.
\end{eqnarray}

An interesting consequence of the oscillations of the inflaton is that the Ricci scalar $R$ must also oscillate. This will have important consequences for the scenarios we want to study, later on, in Sects.~\ref{sec:LinearRH} \&~\ref{sec:LatticeRH}. In order to see this, we note that we can write the Ricci scalar in a (flat) Friedmann–Lemaître–Robertson–Walker (FRLW) background metric $g_{\mu\nu} \equiv {\rm diag}(-a(\eta)^{2\alpha},a(\eta)^2,a(\eta)^2,a(\eta)^2)$, 
as~\cite{Figueroa:2021iwm}
\begin{equation}\label{eq:RicciforFLRW}
    R = \frac{6}{a^{2\alpha}}\left[\frac{a''}{a}-(\alpha-1)\left(\frac{a'}{a}\right)^2\right]\:,
\end{equation}
where $a(\eta)$ is the scale factor, $\eta$ is an $\alpha$-time variable (which becomes e.g.~cosmic time for $\alpha = 0$, or conformal time for $\alpha = 1$), and $'$ denotes derivatives respect to $\eta$. At the same time, we note that the Ricci scalar is determined initially only by the inflaton, so one should be able to write an expression for $R$ depending solely on the inflaton energy contributions. To see this explicitly, we need to write first the Friedman equations sourced by the inflaton, 
\begin{eqnarray}\label{eq:FriedmannInflaton}
{a''\over a} &=& \frac{a^{2 \alpha}}{3 m_p^2}\Big\langle (\alpha-2)K(\phi) + \alpha G(\phi) + (\alpha + 1)V_{\rm inf}(\phi) \Big\rangle\,,\\
\label{eq:FriedmannInflatonII}
\left({a'\over a}\right)^2 &=&  \frac{a^{2 \alpha}}{3 m_p^2}
\Big\langle K(\phi) + G(\phi) + V_{\rm inf}(\phi) \Big\rangle
\end{eqnarray}
where $\langle ... \rangle$ represents volume averaging, and $K(\phi) \equiv \frac{1}{2 a^{2\alpha} } {\phi'}^2$, $G(\phi) \equiv \frac{1}{2 a^2}  (\vec\nabla \phi)^2$ and $V_{\rm inf}(\phi)$ are the inflaton's kinetic, gradient and potential energy densities, respectively. Plugging Eqs.~(\ref{eq:FriedmannInflaton})-(\ref{eq:FriedmannInflatonII}) into Eq.~(\ref{eq:RicciforFLRW}), 
leads to
\begin{eqnarray}\label{eq:RicciAsKinPot}
     R = \frac{2}{m_p^2} \Big\langle 2V_{\rm inf}(\phi) + G(\phi) - K(\phi) \Big\rangle  
     \quad \xlongrightarrow
     [~\vec\nabla\phi \rightarrow 0~]{}\quad {2\over m_p^2}\Big(2V_{\rm inf}(\phi)-K(\phi)\Big)\,,
\end{eqnarray}
where in the expression in the $rhs$, we have exploited the fact that, as long as the inflaton remains homogeneous, $\langle V_{\rm inf}(\phi) \rangle = V_{\rm inf}(\phi)$, $\langle K(\phi) \rangle = K(\phi)$ and $\langle G(\phi) \rangle = 0$. From~(\ref{eq:RicciAsKinPot}) we see that the oscillations of the homogeneous inflaton are imprinted in $R$ via a competition between twice the potential energy ($2V_{\rm inf}(\phi)$) 
and the kinetic energy ($K(\phi)$). 

In order to understand the behavior of the curvature as a function of ${M}$, we note that we can write an exact relation between $R$, the Hubble rate $H^2$ ($= (\dot a/a)^2$ in cosmic time), and the background EoS $w$, using Eq.~\eqref{eq:RicciforFLRW} and the Friedmann equations, as
\begin{eqnarray}\label{eq:EoSfromRonH2} 
    {R\over H^2} = 3(1-3w)\,.
\end{eqnarray}
While expression~(\ref{eq:EoSfromRonH2}) is valid for arbitrary matter content in a FLRW expanding background, i.e.~it does not require inflaton domination, neither homogeneity of the fields, we can specialize it to our initial case of a homogeneous inflaton, and furthermore take an oscillation-average on both sides. This leads to
\begin{eqnarray}\label{eq:RvsW}
    {\overline{R}\over \overline{H^2}} = 3(1-3 w_{\rm osc}) \simeq 
    \left\lbrace
    \begin{array}{cll}
     \text{6\Large$ \frac{(4-p)}{(2+p)} $}  & \hspace*{2mm}({\rm as}~|{\phi}| < {\phi}_{\star}~{\rm always}) & \,,~ {M} \gtrsim m_p 
     \vspace{0.4cm}\\
     \Big(
     \begin{array}{c}
     {\text{\scriptsize decreasing}}\vspace*{-1mm}\\
     {\text{\scriptsize function}}
     \end{array}
     \Big) &
      \xlongrightarrow
      [~({\text{\scriptsize Once}}~|{\phi}| < {\phi}_{\star})~]{}~ \text{6\Large$ \frac{(4-p)}{(2+p)} $} & \,,~ {M} \lesssim m_p
    \end{array}\right.
\end{eqnarray}
with $\overline{H^2} \equiv 3m_p^2\,\overline{\rho}_\phi$ and $w_{\rm osc}$ given by Eq.~(\ref{eq:HomEoS}). Eq~(\ref{eq:RvsW}) is very informative, as it indicates that the condition to obtain an oscillation-averaged negative curvature, $\overline{R} < 0$, relies on having a {\it stiff} effective EoS as $w_{\rm osc} > 1/3$. For the oscillatory regime that samples only the positive curvature region $V''> 0$ of the potential, the EoS is given by $w_{\rm osc} = (p-2)/(p+2)$, c.f.~Eq.~(\ref{eq:HomEoS}), which implies that $\overline{R} < 0$ requires $p > 4$. 

In summary, potentials of the form $\propto {\rm tanh}^p(\phi/{M})$ with $p > 4$, irrespective of the choice of ${M}$, eventually yield an oscillation-averaged negative curvature $\overline{R} < 0$: while for high scale models the condition $\overline{R} < 0$ will be established immediately, upon onset of oscillations, for low scale models there is first a period of oscillations with positive and time decaying $\overline{R}$, and eventually, once the inflaton amplitude falls below the inflection point, the (Hubble normalized) oscillation-averaged curvature settles to a negative constant value as $\overline{R}/\overline{H^2} = 6(4-p)/(2+p) < 0$, c.f.~Eq.~(\ref{eq:RvsW}). As we will see in Sects.~\ref{sec:LinearRH} \& \ref{sec:LatticeRH}, this interesting {\it delay} on the onset of $\overline{R} < 0$, characteristic of low scale scenarios with $p > 4$, will have important consequences for enabling reheating in our scenario. 
\begin{figure*}[tbp]
\begin{center}
\includegraphics[width=0.45\textwidth,height=4cm]{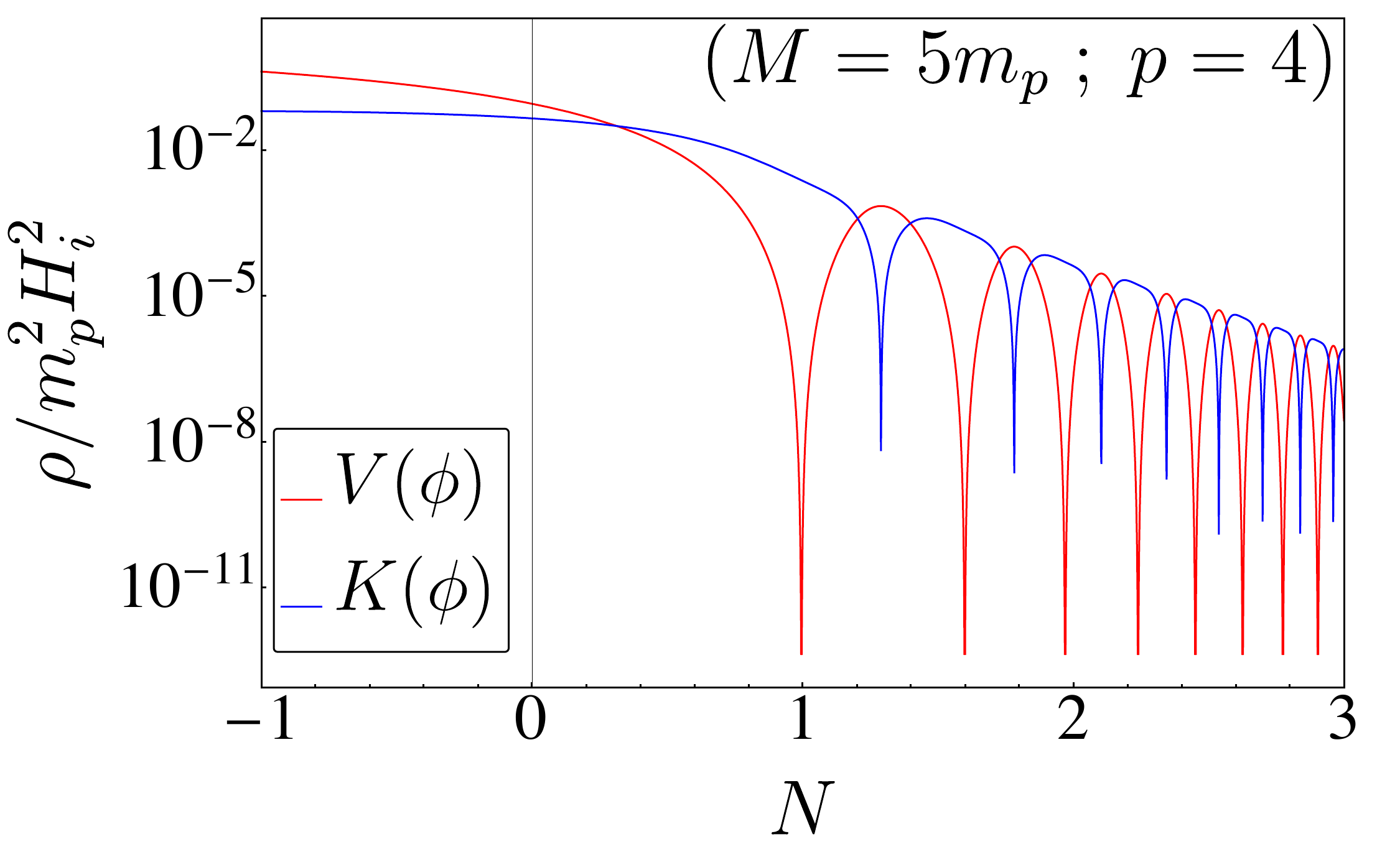} 
\includegraphics[width=0.45\textwidth,height=4cm]{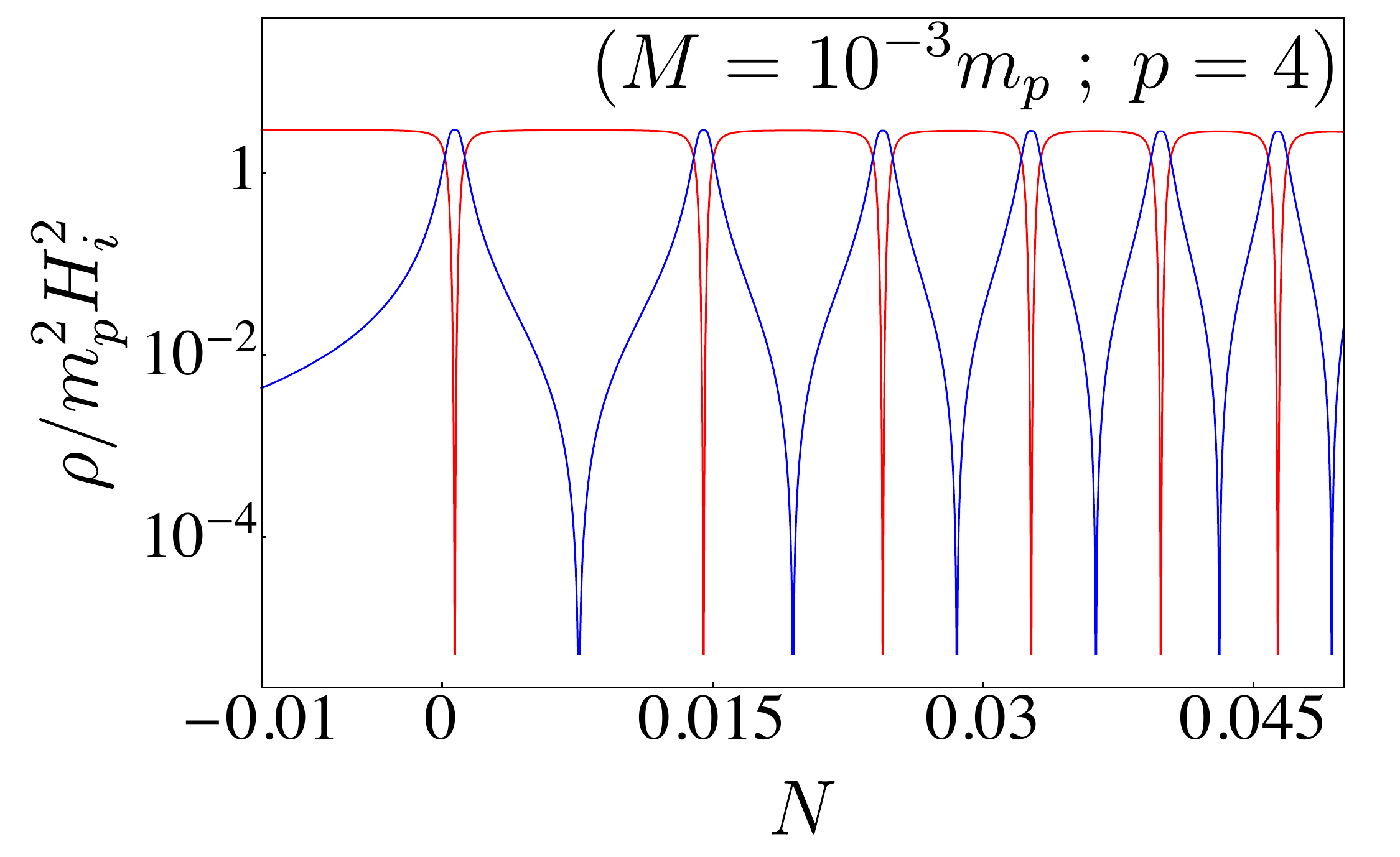} 
\includegraphics[width=0.45\textwidth,height=4cm]{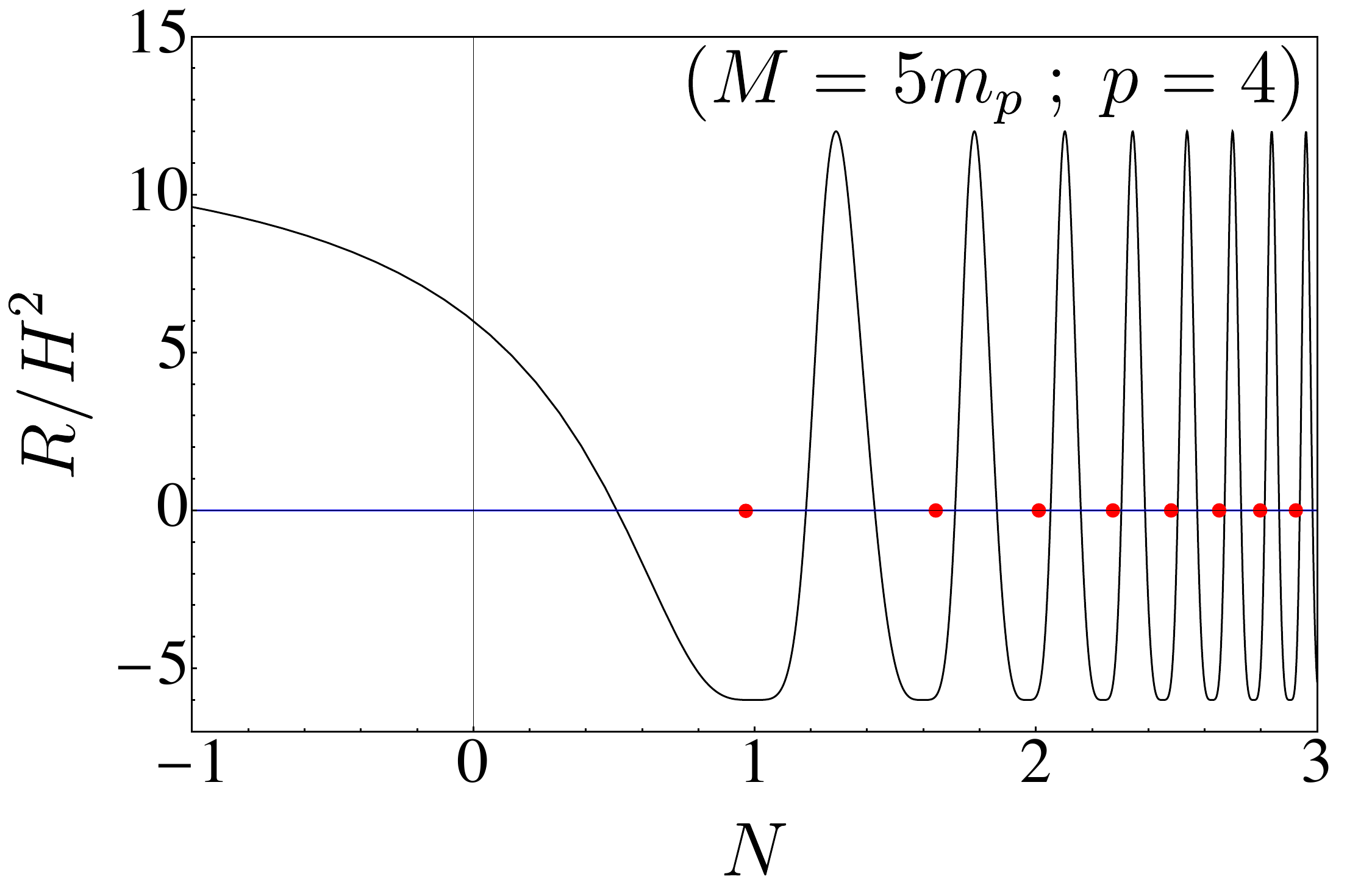}
\includegraphics[width=0.45\textwidth,height=4cm]{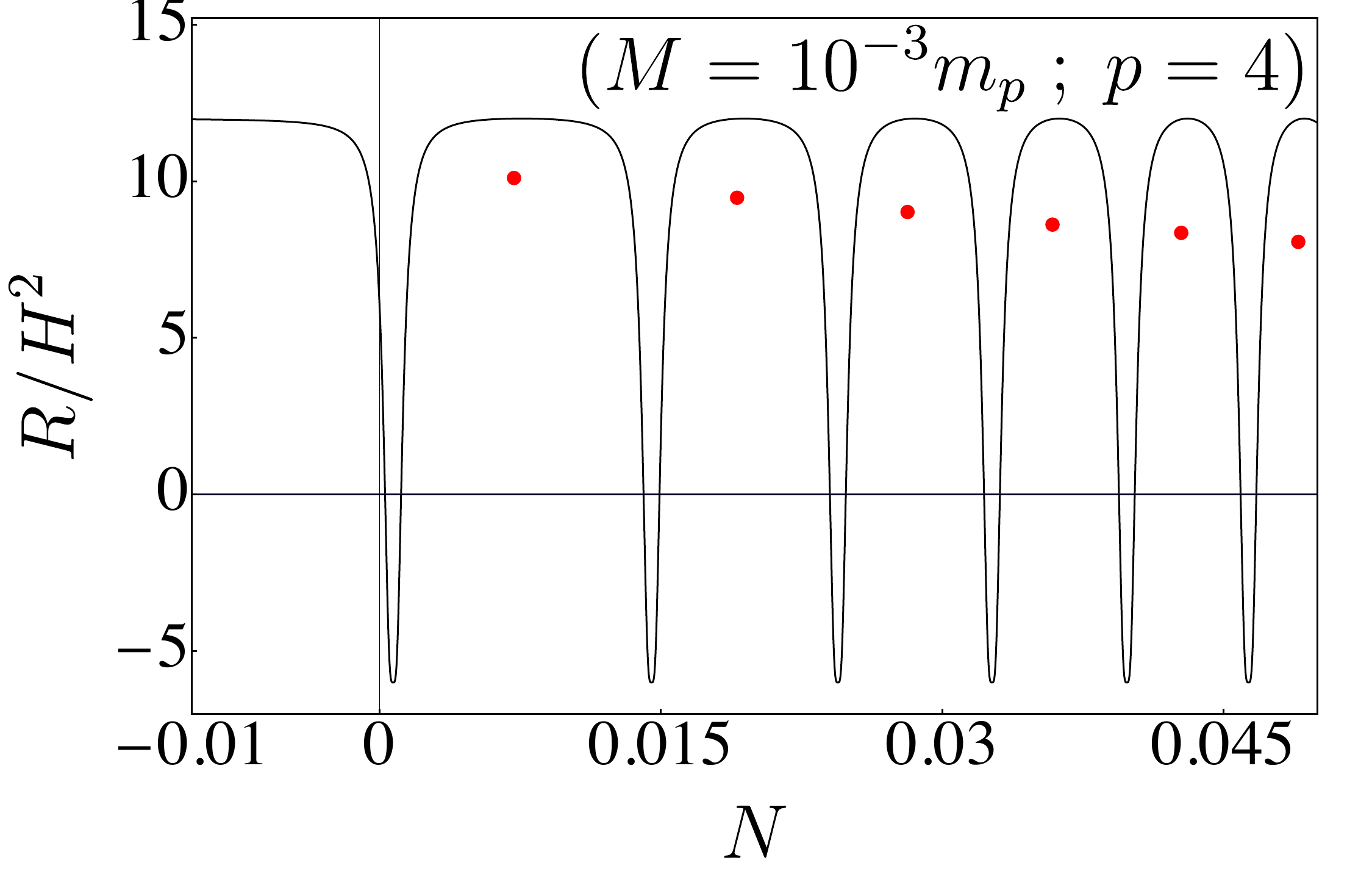}
\end{center}
\vspace*{-0.5cm}
\caption{Time evolution of the kinetic and potential energy of the inflaton and $R/H^2$ for $p=4$, for ${M}=5m_p$ ({\it Left Panel}) and ${M}=10^{-3}m_p$ ({\it Right Panel}). We can see that whenever the kinetic energy dominates over the potential, the Ricci scalar becomes negative. Red dots indicate the ratio of the oscillation averaged quantities, $\overline{R}/\overline{H^2}$, {\it c.f.~}Eq.~(\ref{eq:RvsW}).}\label{fig:RicciandKinPotvsM} \vspace*{-0.3cm}
\end{figure*}
\begin{figure*}[tbp]
\begin{center}
    \includegraphics[width=1\textwidth,height=4cm]{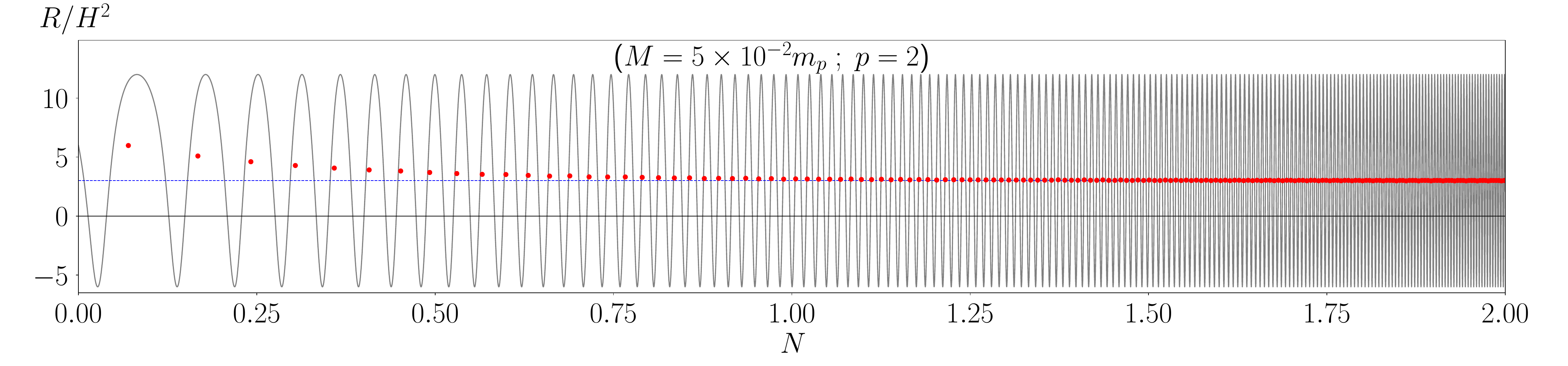}
    \includegraphics[width=1\textwidth,height=4cm]{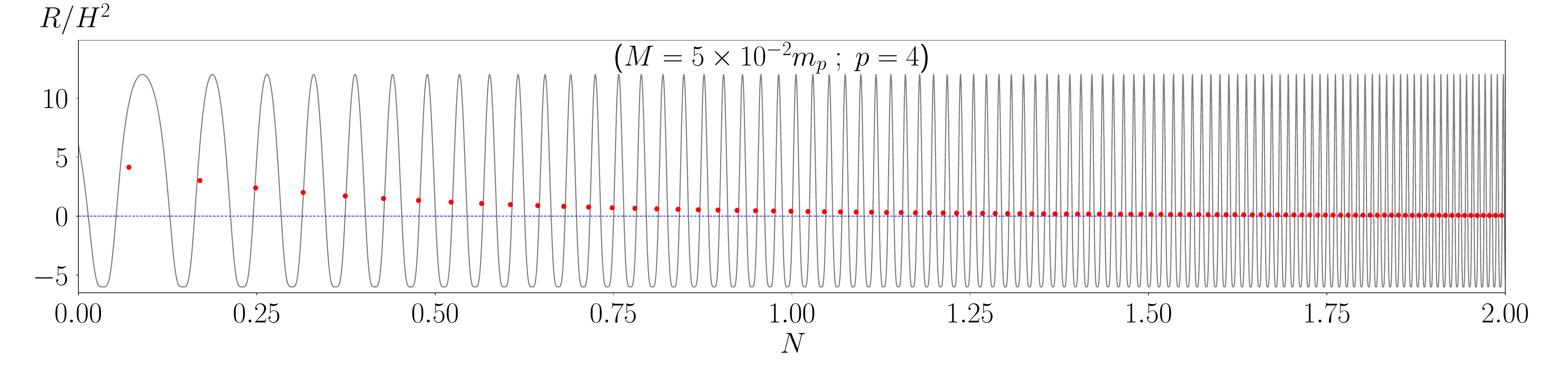}  
    \includegraphics[width=1\textwidth,height=4cm]{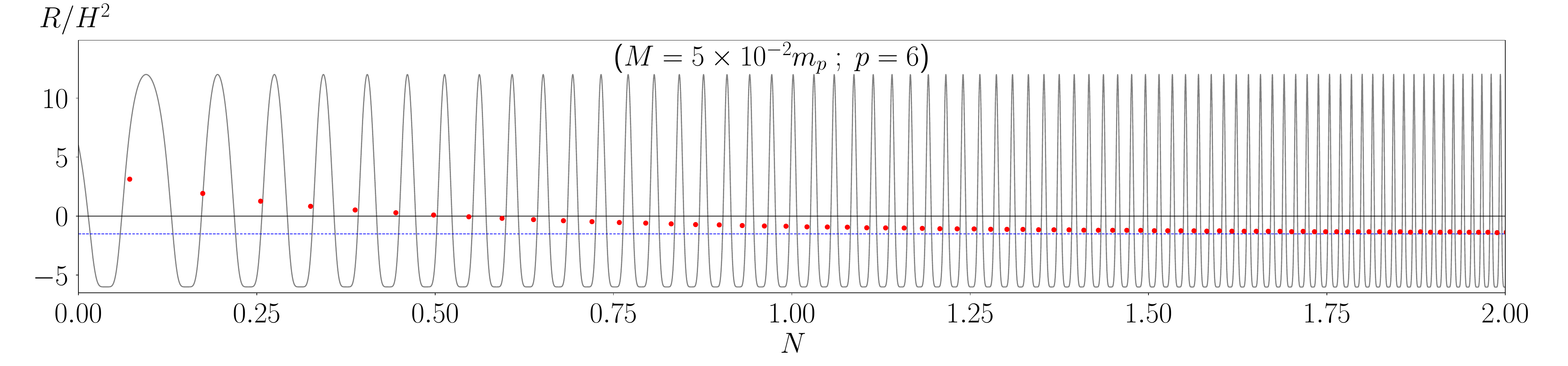}
     
    \end{center}
    \vspace*{-0.5cm}
    \caption{Time evolution of $R/H^2$ for low scale scenarios with $M=5\cdot 10^{-2}m_p$, for $p=2, 4, 6$ ({\it top to bottom}). Red dots indicate the ratio of the oscillation averaged quantities, $\overline{R}/\overline{H^2}$,  whereas blue dashed lines represent the asymptotic value~\eqref{eq:RvsW}.} \label{fig:lowScaleRicci2} \vspace*{-0.3cm}
\end{figure*}
Although less interesting for our later purposes of reheating, we also note that potentials of the form $\propto {\rm tanh}^p({\phi}/{M})$ with $p \leq 4$, always yield, for large scale models, an oscillation-averaged curvature as $\overline{R}/\overline{H^2} = 6(4-p)/(2+p)$, which is positive ($\overline{R} > 0$) for $p < 4$ and vanishing\footnote{The fact that $\overline{R} = 0$ for $p = 4$ in large scale models can be anticipated on the basis that $V \propto {\phi}^4$ is a scale-free potential with oscillations leading to a radiation dominated background, for which one expects $\overline{R} = 0$.} ($\overline{R} = 0$) for $p = 4$. In the case of low scale models with $p \leq 4$, we expect that initially $\overline{R}$ is positive and decreases in time, and eventually, once the condition $|{\phi}| < {{\phi}}_{\star}$ is reached, the averaged curvature settles to the same asymptotically expression of large scale models, $\overline{R}/\overline{H^2} = 6(4-p)/(2+p)$, which again is positive for $p < 4$, and vanishing for $p = 4$.

As an example, in Fig.~\ref{fig:RicciandKinPotvsM} we show the potential and kinetic energy of the homogeneous inflaton during the first of oscillations after inflation, for scenarios with $p = 4$. We choose representative cases of large scale and low scale scenarios, with ${M} = 5 m_p$ (left panels) and ${M} = 10^{-3} m_p$ (right panels). In the large scale scenario, we see that the kinetic energy is balanced by the potential energy within each oscillation, in such a way that $\overline{R} = 0$. In the low scale scenario, we see instead that the potential dominates over the kinetic energy within each oscillation, and hence $\overline{R} > 0$. However, as remarked before, the inflaton amplitude in this case eventually decays below the inflection point, leading to an asymptotic vanishing curvature $\overline{R} = 0$. While such feature cannot be appreciated in the bottom-right panel of Fig.~\ref{fig:RicciandKinPotvsM}, it can be clearly observed in middle panel of Fig.~\ref{fig:lowScaleRicci2}, where the behavior of the system for a low scale scenario with $p=4$ is depicted up to much later times. Furthermore, the corresponding behavior of $\overline{R}$ for low scale scenarios with $p = 2$ and $p = 6$, can also be observed in the top and bottom panels of Fig.~\ref{fig:lowScaleRicci2},  respectively.

\subsection{Inflaton fragmentation}
\label{subsec:InflatonFragmentation}

In the previous section~\ref{subsec:InflatonOscillations} we have assumed that the inflaton remained homogeneous at all times. In this section we will see that such circumstance can not always be kept, as the inflaton condensate may be fragmented into inhomogeneous fluctuations due to its own self-interactions, see e.g.~\cite{Lozanov:2016hid,Lozanov:2017hjm}. In order to understand inflaton fragmentation, we write first the EoM for the inflaton as
\begin{eqnarray}
    \phi'' + (3-\alpha)\left(\frac{a'}{a}\right)\phi' - a^{-2(1-\alpha)}\nabla^2\phi &=& - a^{2\alpha}V_{,\phi}\,.
\end{eqnarray}
Setting $\alpha=3(p-2)/(p+2)$, we introduce the following variables
\begin{eqnarray}
    \varphi = \frac{1}{m_p} a^{\frac{6}{p+2}}\phi \quad d\tilde{\eta} = H_*  a^{\frac{3(2-p)}{p+2}} d{\eta} \quad d\tilde{x}_i =  H_* d{x}_i\;.
\end{eqnarray}
with $H_*$ the Hubble rate deep inside the slow-roll regime of inflation. This leads to rewrite the the inflaton's EoM as
\begin{eqnarray}
    \varphi'' - a^{-\frac{16-4p}{2+p}}\tilde{\nabla}^{2} \varphi + (\tilde{V}_{,\varphi} + F(a)) = 0 \; ,
\end{eqnarray}
where
\begin{eqnarray}
    F(a) = \frac{6}{p+2}\left(\frac{p-4}{p+2}\left(\frac{a'}{a}\right)^2 - \frac{a''}{a}\right) \; ,
\end{eqnarray}
and $\tilde{\nabla}_i \equiv \partial/\partial\tilde x_i$ and $\tilde V \equiv V_{\rm inf}(\phi(\varphi))/(m_p^2H_*^2)$. We split now the field into homogeneous and inhomogeneous parts, $\bar{\varphi}(\tilde{\eta})$ and $\delta \varphi(\tilde{\eta},\tilde{x}_i)$, respectively, 
\begin{eqnarray}
\varphi(\tilde{\eta},\tilde{x}_i) \equiv \bar{\varphi}(\tilde{\eta})  + \delta \varphi(\tilde{\eta},\tilde{x}_i)\;,
\end{eqnarray} 
and think of the inhomogeneous part as a perturbation over the background, i.e.~$|\delta \varphi(\tilde{\eta},\tilde{x}_i)| \ll |\bar{\varphi}(\tilde{\eta})|$, at least initially. We can distinguish two regimes. For $M \gtrsim m_p$,  inflaton oscillations take place sampling only the positive curvature part of the potential, which can be written in dimensionless form as $\tilde{V}(\varphi) = {1\over p} \mu^2 \varphi^p$, with $\mu \equiv {(\Lambda^2 m_p^{p/2-1})}/{(M^{p/2} H_*)}$  acting as an effective dimensionless ``mass'' parameter of the inflaton fluctuations. The EoM in momentum space for the inflaton fluctuations then reads
\begin{eqnarray}\label{eqn:InflatonPertEoM}
    \delta {\varphi}_k '' + \omega_k^2 \delta {\varphi}_k = 0 \;,~~~\omega_k^2 = {\kappa}(a)^2 + (p-1)\mu^2 \bar{\varphi}^{p-2} + F(a)\;,
\end{eqnarray}
where
\begin{eqnarray}
    {\kappa}(a) = \frac{k}{H_*} a^{\frac{2p-8}{2+p}}\;,
\end{eqnarray}
with $k$ the comoving momenta. In this case, the oscillations of the inflaton will induce parametric resonance of the inflaton fluctuations, given that $F(a)$ becomes irrelevant rapidly, soon after the onset of oscillations. Eq~\eqref{eqn:InflatonPertEoM} admits solutions of the form $\varphi_k \sim e^{\mu_k \tilde{\eta}}$, and by means of a Floquet chart analysis we conclude that the exponent peaks at $\mu_k \sim 0.036$ for $p=3.6$~\cite{Antusch:2021aiw}. This mechanism is, however, not particularly efficient in fragmenting the inflaton condensate, as the usual time scale for inflaton fragmentation, i.e.~for the fluctuations to backreact on the inflaton background, is typically $N_{\rm frag} > 10$ e-folds, see e.g.~\cite{Lozanov:2016hid,Lozanov:2017hjm,Antusch:2020iyq,Antusch:2021aiw}.    

In the case of low scale scenarios $M \lesssim m_p$, the initial oscillations are dominated initially by the negative curvature branch of the potential, as the inflaton ends inflation above the inflection point. In this situation, we cannot approximate the potential as a monomial, but we can still gain insight by analyzing a more generic EOM of the fluctuation in this regime,  
\begin{eqnarray}\label{eqn:InflatonPertEoM2}
    \delta {\varphi}_k '' + \omega_k^2 \delta {\varphi}_k = 0\,,~~~ \omega_k^2 = {\kappa}(a)^2 +  m_{\delta \varphi}^2 + F(a)\;,~~~  m_{\delta \varphi}^2 \equiv \frac{\partial^2 \tilde{V}}{\partial {\varphi}^2}
\end{eqnarray}
It is useful to express quantities in the small scale limit $M \ll m_p$. To begin with, we consider $H_*$ to be dominated by the potential energy, which for $\phi > M$ approximates to the asymptotic value $V_{\rm inf} \approx \Lambda^4/p$, so that $H_*^2 \approx {\Lambda^4}/{(3 p m_p^2)}$. The dimensionless potential then reads $\tilde{V}({\varphi}) \sim 3 \tanh^p({|{\varphi}|}/{\mathcal{M}})$, where we have set the initial scale factor to $a=1$ and introduced $\mathcal{M} = M/m_p$. The aim now is to understand how the initial effective mass of the inflaton perturbations behaves. We derive an approximated mass scale as
\begin{eqnarray}
    m_{\delta \varphi}^2 = \frac{\partial^2 \tilde{V}}{\partial {\varphi}^2} = \frac{12 p}{\mathcal{M}^2}
    \frac{\left(p-\cosh\left(\frac{2{\varphi}}{\mathcal{M}}\right)\right)\tanh^p\left(\frac{{\varphi}}{\mathcal{M}}\right)}{ \sinh^2\left(\frac{2{\varphi}}{\mathcal{M}}\right)} ~~\xrightarrow{\mathcal{M} \ll 1}~~ -\frac{\sqrt{72}}{\mathcal{M}}  
\end{eqnarray}
where the asymptotic value given, corresponds to the limit $\mathcal{M} \ll 1$ when the hyperbolic functions are evaluated at the background field amplitude satisfying $\epsilon_V({\varphi}_{\rm end}) = 1$. We remark the negative sign, which was expected as the oscillations start above the inflection point  in this case. This implies that, initially, modes with ${\kappa}^2 \lesssim \sqrt{72}\mathcal{M}^{-1}$, will have $\omega^2_{k} <0 $, and hence will experience a tachyonic growth. In light of this linearized analysis, we see that inflaton modes will start growing as soon as inflation ends, and their tachyonic growth will be stronger the smaller the value of $\mathcal{M}$. This might lead eventually to a fragmentation of the inflaton condensate, if the fluctuations grow enough so that they become comparable to the background amplitude. In order to obtain a proper characterization of the time scale of inflaton fragmentation, we need therefore to go beyond the previous linear analysis, and run lattice simulations that fully capture the non-linear dynamics of the system. 
\begin{figure*}[tbp]
\begin{center}
    \includegraphics[width=.6\textwidth,height=5cm]{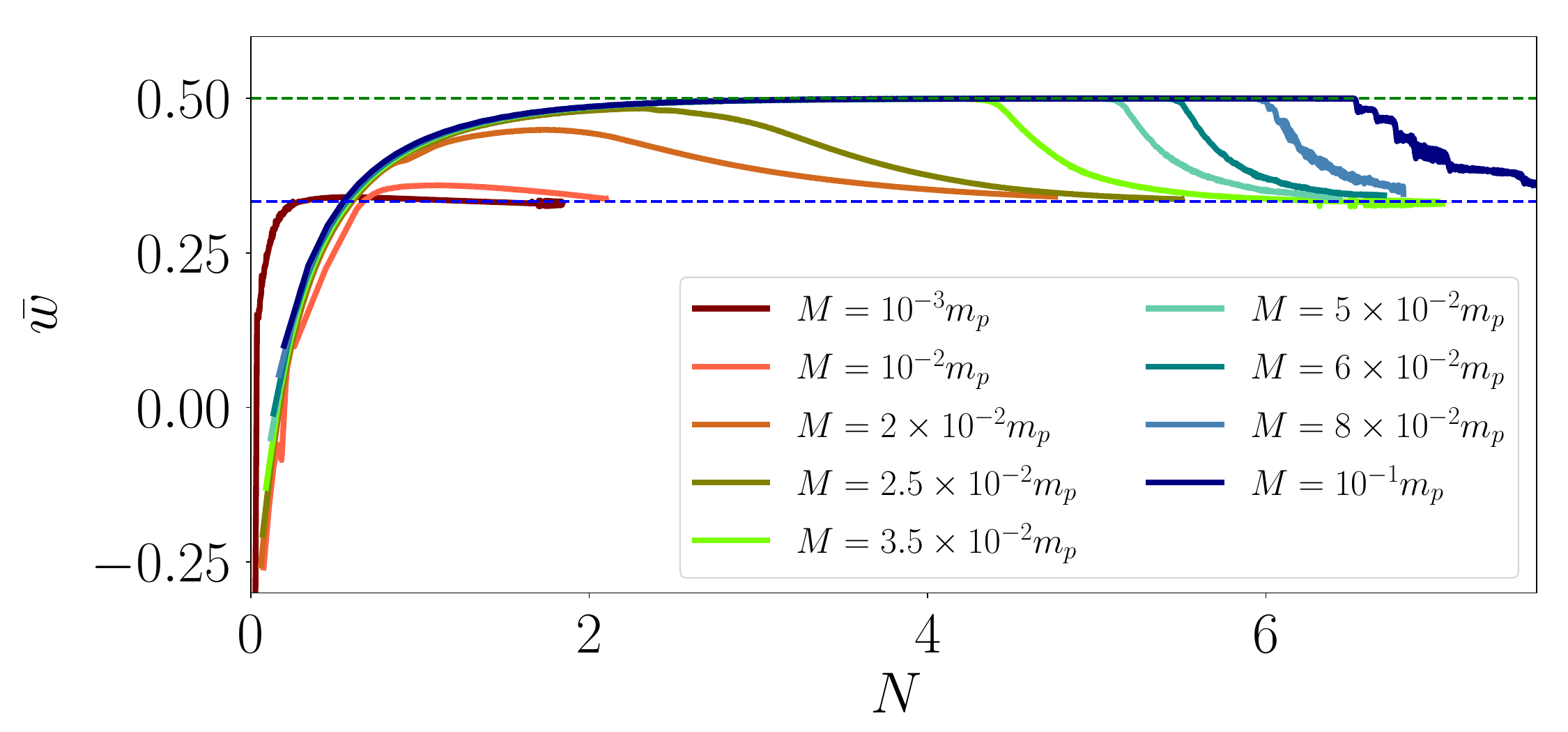}
    \end{center}
    \vspace*{-0.5cm}
    \caption{Time evolution of oscillation averaged equation of state $\bar{w}$ for low scale scenarios with $M \in [10^{-3},10^{-1}]m_p$, for $p= 6$. Green dashed lines represent the asymptotic value~\eqref{eq:RvsW}, Blue dashed line corresponds to radiation-domination EoS.} \label{fig:lowScaleEoSfrag} \vspace*{-0.3cm}
\end{figure*}

Using the public code \CL~\cite{Figueroa:2021yhd,Figueroa:2020rrl,Figueroa:2023xmq}, we have simulated and measured the fragmentation time scale of the inflaton\footnote{We have characterized the inflaton fragmentation time only for $p=6$, as this is the most promising case to lead to proper reheating at low energy scales, see Section~\ref{sec:LatticeRH}.}. We have run simulations in lattices with $N=385$ sites/dimension, for different values of the mass scale ranging as $\mathcal{M} \in [10^{-1},10^{-3}]$. Starting each simulation at the end of inflation (defined properly by $\epsilon_H = 1$), we have run them till the system reaches radiation domination (RD) due to full inflaton fragmentation. To characterize the time scale of inflaton fragmentation we followed the evolution of the inflaton background equation-of-state
\begin{eqnarray}
    w =\frac{\langle p\rangle}{\langle \rho \rangle} = \frac{\langle K(\phi)-\frac{1}{3}G(\phi) -V(\phi)\rangle}{\langle  K(\phi)+G(\phi)+V(\phi) \rangle}\,,
\end{eqnarray}
with $\langle ... \rangle$ denoting volume averaging. We define the fragmentation time scale $N_{\rm frag}$ following~\cite{Lozanov:2016hid}, as the number of efoldings since the end of inflation, when the envelope of the evolution of $w$ (which is highly oscillatory), falls below 95~\% of its maximum value. As shown in Fig.~\ref{fig:lowScaleEoSfrag}, the fragmentation time decreases drastically for the smallest $M$ values, following the logic commented in our linear analysis, where we showed that the tachyonic growth (that eventually leads to fragmentation) is stronger for the lower mass scales $\mathcal{M}$. The cases depicted in the figure, where fragmentation takes longer to be achieved, are due to the low efficiency of the inflaton self-resonance: in such cases, the initial tachyonic growth is not enough to fragment the inflaton condensate before the background inflaton starts oscillating only below the inflection point. From then onward, only parametric resonance acts, and hence the inflaton condensate takes longer to really fragment. 

In the coming Sections~\ref{sec:LinearRH} and~\ref{sec:LatticeRH}, we will study the efficiency of geometric preheating of a NMC scalar field, based on the assumption that the inflaton remains homogeneous during its oscillations. At the end of Section~\ref{sec:LatticeRH}, we will re-introduce the notion of inflaton fragmentation and we will study how this affects the achievement of proper reheating, depending on the mass scale $M$.

\section{Geometric preheating: linear regime}\label{sec:LinearRH}
Our starting point is to consider the action
\begin{equation}\label{eq:Action}
    S =  \int d^4x \sqrt{-g} \left(\frac{1}{2}m_p^2 R - \frac{1}{2}\xi R \chi^2 - \frac{1}{2}g^{\mu\nu} \partial_{\mu}\chi\partial_{\nu}\chi - \frac{1}{2}g^{\mu\nu} \partial_{\mu}\phi\partial_{\nu}\phi - V(\phi,\chi) \right)\:,
\end{equation}
which describes the dynamics of the inflaton $\phi$ and a scalar spectator field $\chi$ with non-minimal interactions with gravity as $\xi \chi^2 R$, with $R$ the {\it Ricci} scalar and $\xi$ the interaction strength. From now on we will refer to $\chi$ as the non-minimally coupled (NMC) scalar field. The potential $V = V_{\rm inf}(\phi) + V_{\rm NMC}(\chi,...)$ includes the inflaton potential $V_{\rm inf}(\phi)$ given by~(\ref{eq:potential}) and interactions of $\chi$ represented by $V_{\rm NMC}(\chi,...)$. The equation of motion of the NMC field in $\alpha$-time $\eta$, can be obtained straight forward by varying the action~(\ref{eq:Action}) with respect to $\chi$,
\begin{eqnarray}\label{eq:NonminimalfieldEoM}
     \chi'' + (3-\alpha)\mathcal{H}\chi' - a^{-2(1-\alpha)}\nabla^2\chi + a^{2\alpha}\left(\xi R \chi + V_{,\chi} \right) = 0\:,
\end{eqnarray}
with $\mathcal{H} \equiv a'/a$ and $'$ representing derivatives with respect to $\eta$. 

In order to understand the dynamics of the NMC field during the oscillations of the inflaton $\phi$ following the end of inflation, we first canonically quantize $\chi$ in Eq.~\eqref{eq:NonminimalfieldEoM}
\begin{equation}\label{eq:FourierTransformNMC}
    \chi = \int \frac{d^3 k}{2\pi^3} \Big[\chi_k \hat{a}_{\bf k} e^{i {\bf k} \cdot {\bf x}} + \chi_k^* \hat{a}^{\dagger}_{\bf k} e^{-i {\bf k} \cdot {\bf x}}\Big]\:,
\end{equation}
with $\hat{a}_{\bf k},\hat{a}^{\dagger}_{\bf k}$ standard creation/annihilation operators satisfying
$[\hat{a}_{\bf k},\hat{a}^{\dagger}_{\bf k'}] = (2\pi)^3 \delta({\bf k}-{\bf k'})$, and with the field modes initially normalized such that $\chi_k\chi_k'^* - \chi_k'\chi_k^* = i$. In order to understand the initial evolution of the modes $\chi_k(\eta)$, we choose to work in conformal time $\tau$ (i.e.~$\alpha = 1$) and consider a conformal field redefinition as $\widetilde{\chi}(\tau) = a \chi$. The modes $\widetilde{\chi}_k$ satisfy then approximately the equation
\begin{equation}\label{eq:modeseqnconfchi}
   \widetilde{\chi}_k'' + \omega_k^2\,\widetilde{\chi}_k = 0\,,~~~~~~ {\rm with}~~\omega_k^2 \equiv k^2 + a^2 \Big(\xi-\frac{1}{6}\Big)R\,,
\end{equation}
where we have neglected interactions of $\chi$ (i.e.~$\partial_\chi V = 0$), and we have used the fact that in conformal time $a^3R = 6a''$. Initially, any backreaction of $\chi$ into the Ricci scalar can be ignored, so we shall think of $R$ during the early stages after inflation, as a time dependent function sourced only by the homogeneous inflaton. This turns (\ref{eq:modeseqnconfchi}) into a linear equation for $\widetilde\chi_k$, and upon inspection, we see that for $\xi > 1/6$,  whenever $R$ is negative, $\widetilde\chi_k$ enjoys a {\it tachyonic} instability for modes up to a cutoff $k \lesssim k_*$, with
\begin{equation}\label{eq:UVcutoff}
 k_* = \sqrt{(\xi-1/6) |R| a^2}
\end{equation}
As a result, infrared modes $k < k_*$ experience an exponential growth as $\tilde\chi_k \sim e^{|\omega_k|\tau}$, due to the negativeness of $\omega_k^2 < 0$. 
As Eq.~(\ref{eq:EoSfromRonH2}) indicates that the most negative value of the Ricci scalar over the Hubble rate squared is bounded\footnote{The fact that $w \leq 1$ emerges as a natural upper bound on the equation of state of the Universe, from the requirement that the sound speed $c_{\text{s}}^2=\partial p/\partial \rho \equiv w \leq 1$ of a fluid does not exceed the speed of light. } from below as $R/H^2>-6$, we can take Eq.~\eqref{eq:UVcutoff} and normalize it by $aH$, to realize that
\begin{equation}\label{eq:limit}
    \frac{k_*}{aH} 
    < \sqrt{6\xi - 1}\:.
\end{equation}
Using the expression $\overline{R}/\overline{H^2} = 6(4-p)/(2+p)$ from Eq.~(\ref{eq:EoSfromRonH2}), valid when the inflaton oscillates only wandering around the positive curvature region of the potential, we can even estimate directly the scale, in this case, as
\begin{equation}\label{eq:kStarApprox}
    \frac{k_*}{aH} \simeq \sqrt{(6\xi-1)}\sqrt{|4-p|\over 2+p}\,,
\end{equation}
which of course respects Eq.~(\ref{eq:limit}) $\forall\,p$. In other words, we obtain that as long as $\xi \gg 0.1$, it roughly holds that $k_* \lesssim \sqrt{6\xi}aH$. While for $1/6 < \xi \lesssim \mathcal{O}(1)$ sub-horizon modes are not excited, sub-horizon modes will be greatly excited if $\xi \gg 1$, up to a scale $\sim 1/\sqrt{6\xi}$ times shorter than the Hubble radius. 

In order to find the evolution of the modes $\chi_k \equiv \widetilde \chi_k/a$, we follow the prescription of Ref.~\cite{Figueroa:2021iwm}. This amounts to considering as initial condition the {\it Bunch-Davies} (BD) vacuum for modes deep inside the Hubble radius during inflation. This can be found by solving Eq.~\eqref{eq:modeseqnconfchi} in the limit $k^2 \gg a^2 R$, so that
\begin{eqnarray}\label{eq:BD}
    \widetilde{\chi}_k ( - k\tau \gg 1) \approx \frac{1}{\sqrt{2k}} e^{-ik\tau} \approx \frac{1}{\sqrt{2k}} e^{\frac{ik}{aH}} \:,
\end{eqnarray}
where in the last equality we have used that deep inside the slow-roll regime of inflation it holds that $\tau \approx -1/(aH)$. In order to ensure that the initial condition of the modes is well represented by the BD condition~(\ref{eq:BD}), we start evolving each mode during inflation when they are initially (at least) a factor $\beta = 10^3$ times inside the Hubble radius\footnote{In practice $\beta \gtrsim 100$ also works fine, but to be on the safest side, we take $\beta = 10^3$.}. 
We consider a range of modes spanning 3 decades and adjust the initial time so that the most ultraviolet mode of such range (with $\beta = 10^6$ initially) will leave the Hubble radius just at the very end of inflation. We need the scale factor therefore to grow a factor $\sim 10^6$, which in terms of efolds is obtained from $-N_{\rm init} \approx \ln(10^6) \approx 14$ efoldings before the end of inflation. In practice, we evolve the modes during approximately $\Delta N_{\rm tot} = -N_{\rm init} + N_{\rm final}$ efoldings in total, from $N_{\rm init} \simeq -14$ deep inside inflation till $N = 0$ at the end of inflation, and typically during another $N_{\rm final} = 2-4$ e-folds after inflation, depending on the case, in order to capture the impact of the evolution of $R$ into the post-inflationary dynamics of $\chi$. 

\begin{figure*}[tbp] 
\begin{center}
    \includegraphics[width=0.45\textwidth,height=4.5cm]{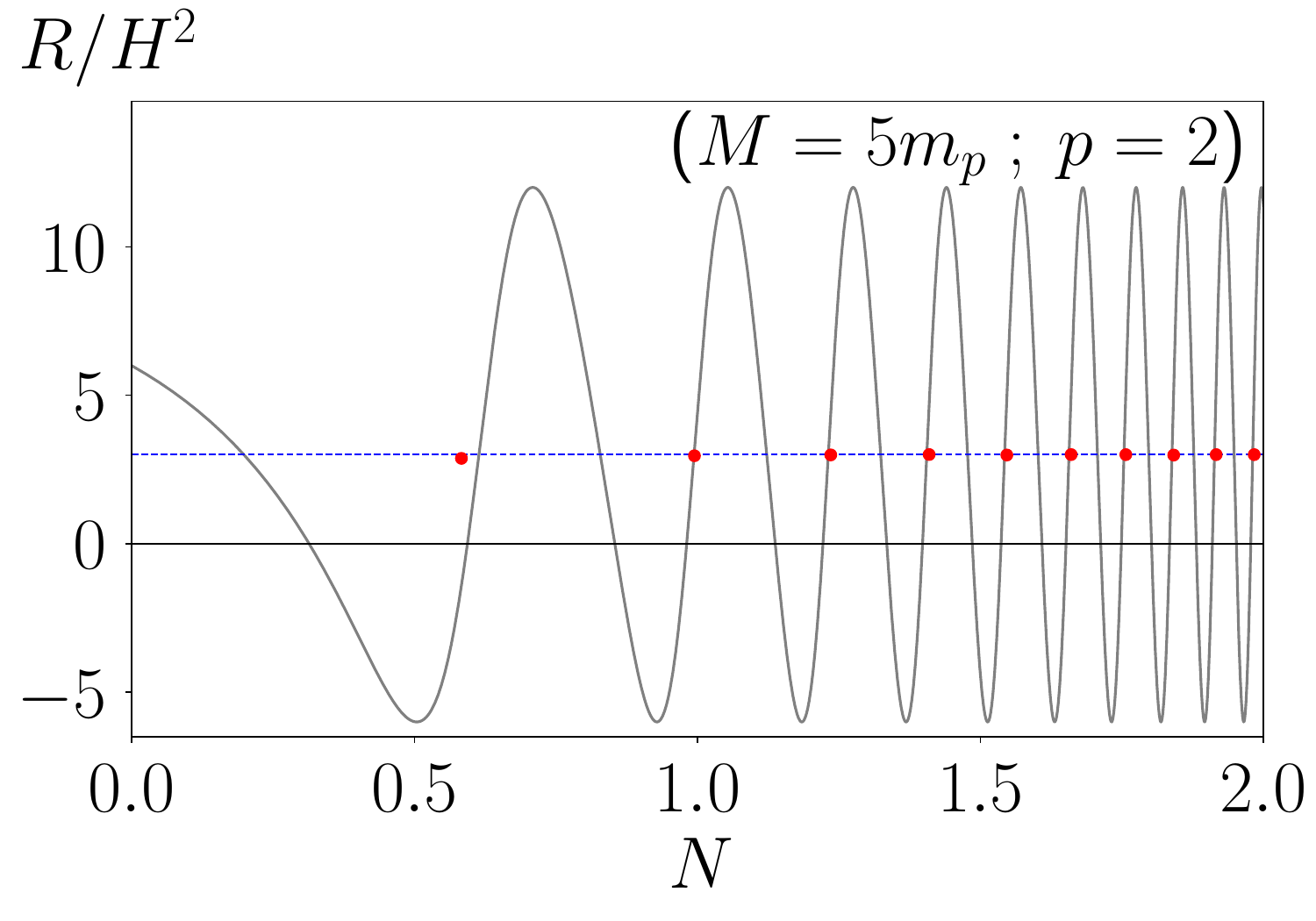}  
    \includegraphics[width=0.5\textwidth,height=4.5cm]{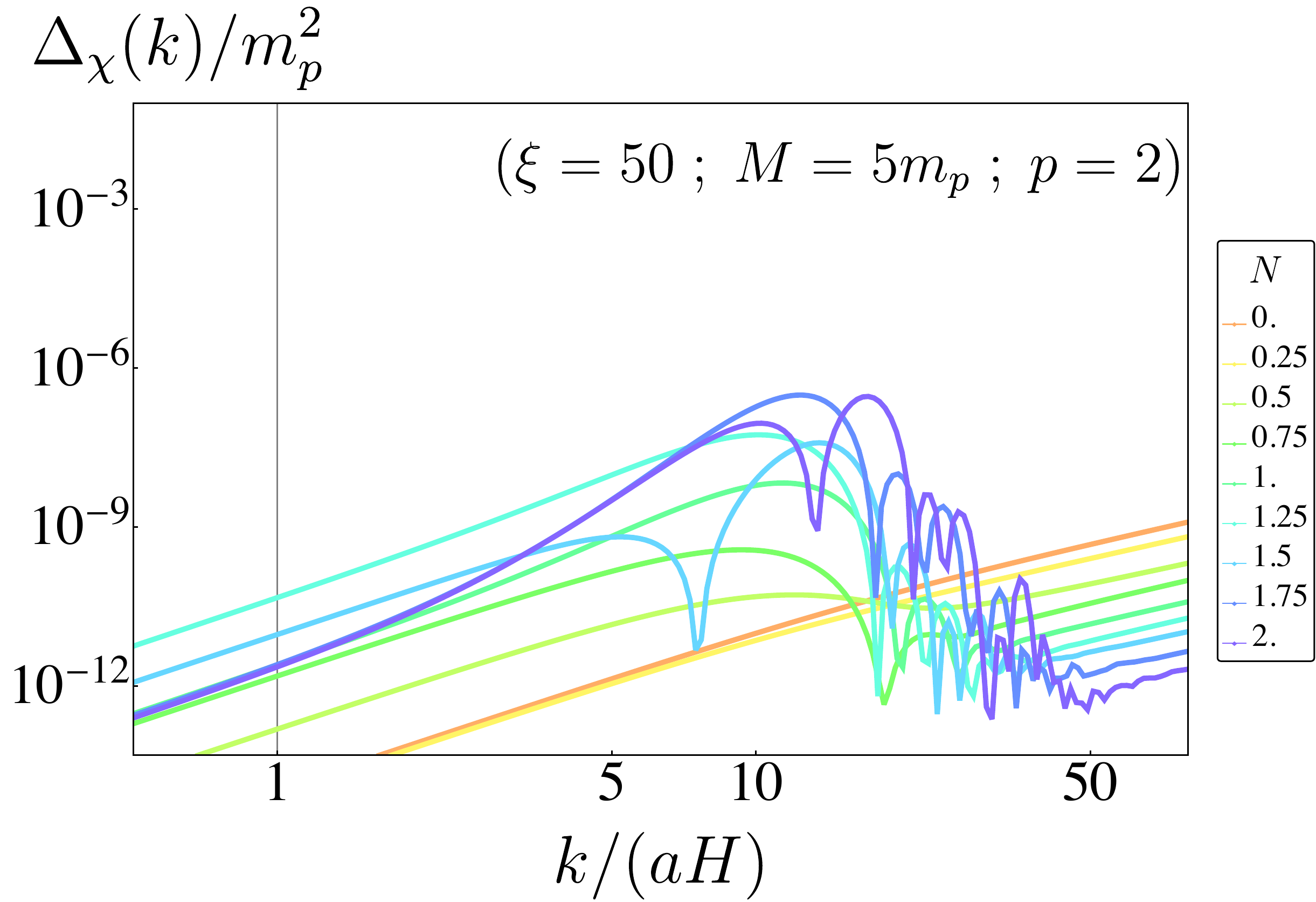}
    \includegraphics[width=0.45\textwidth,height=4.5cm]{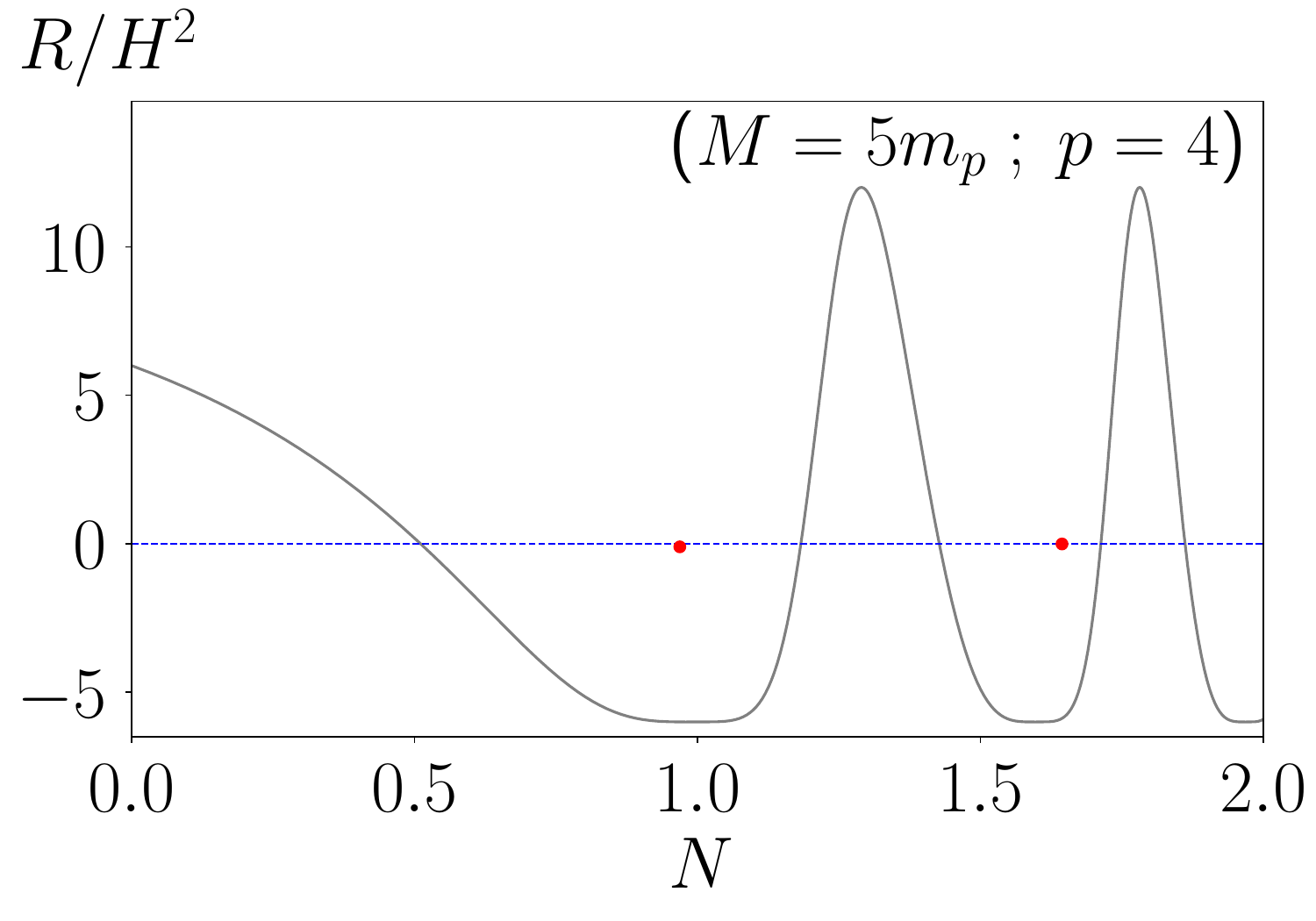}  
    \includegraphics[width=0.5\textwidth,height=4.5cm]{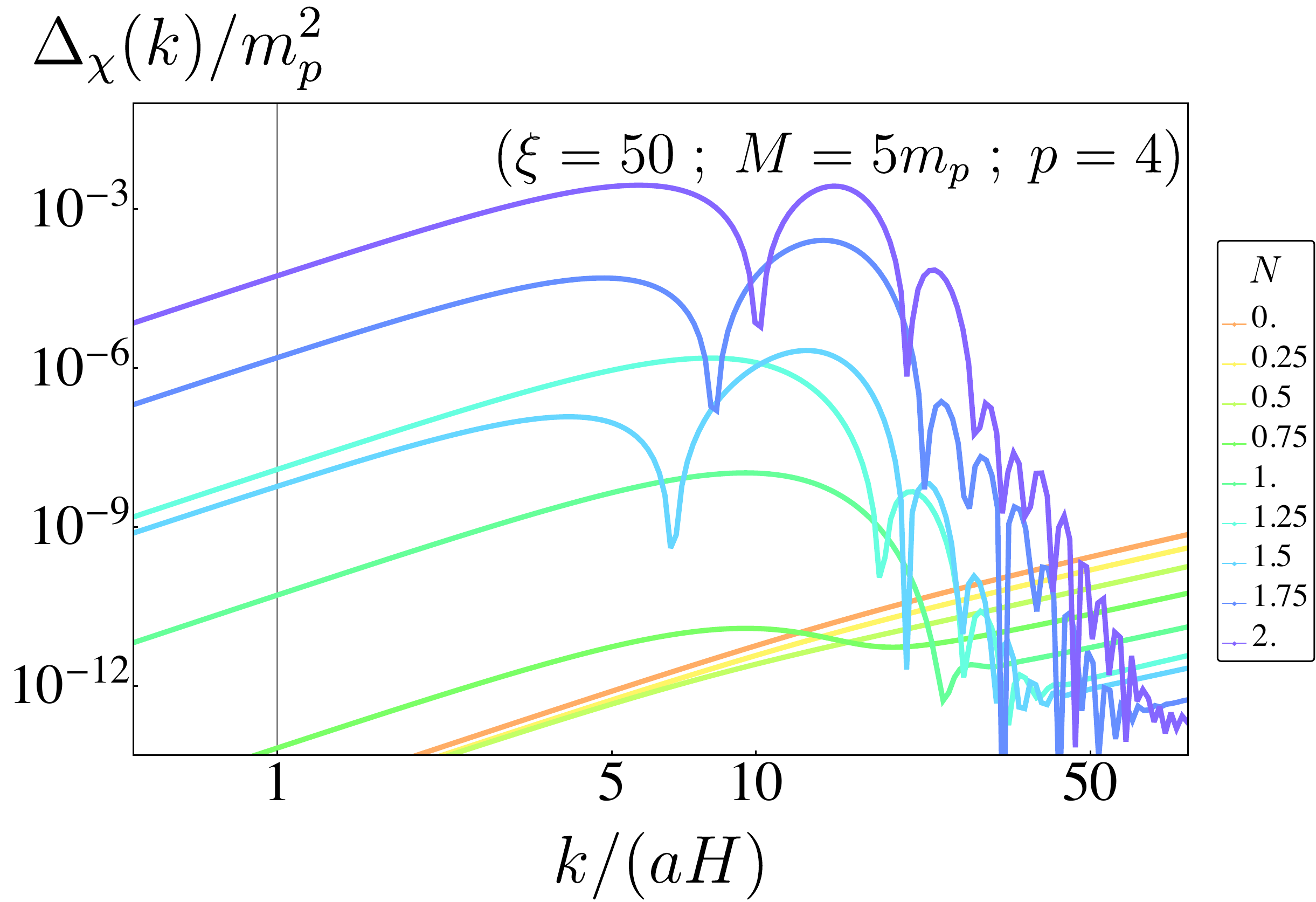}
    \includegraphics[width=0.45\textwidth,height=4.5cm]{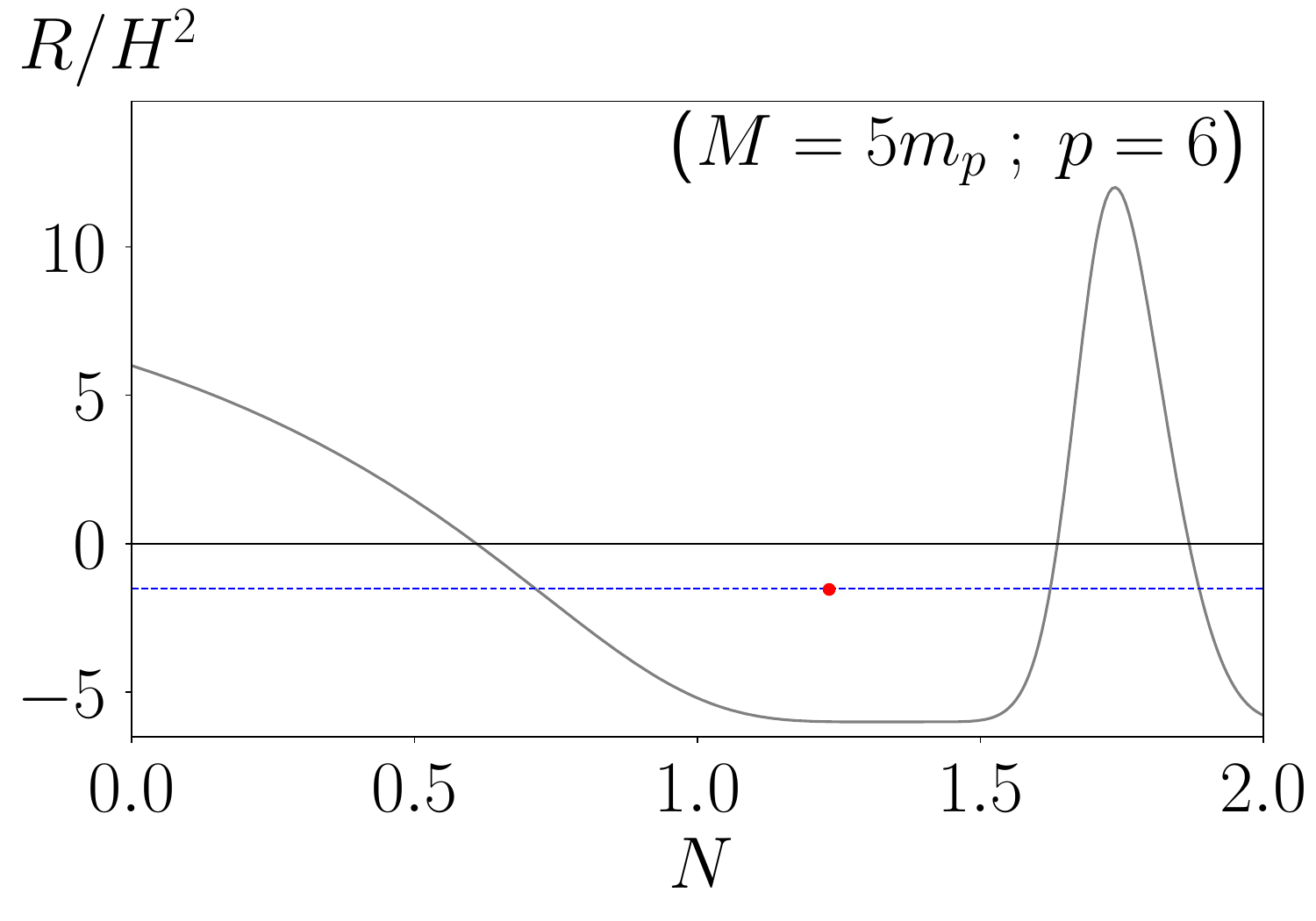} 
    \includegraphics[width=0.5\textwidth,height=4.5cm]{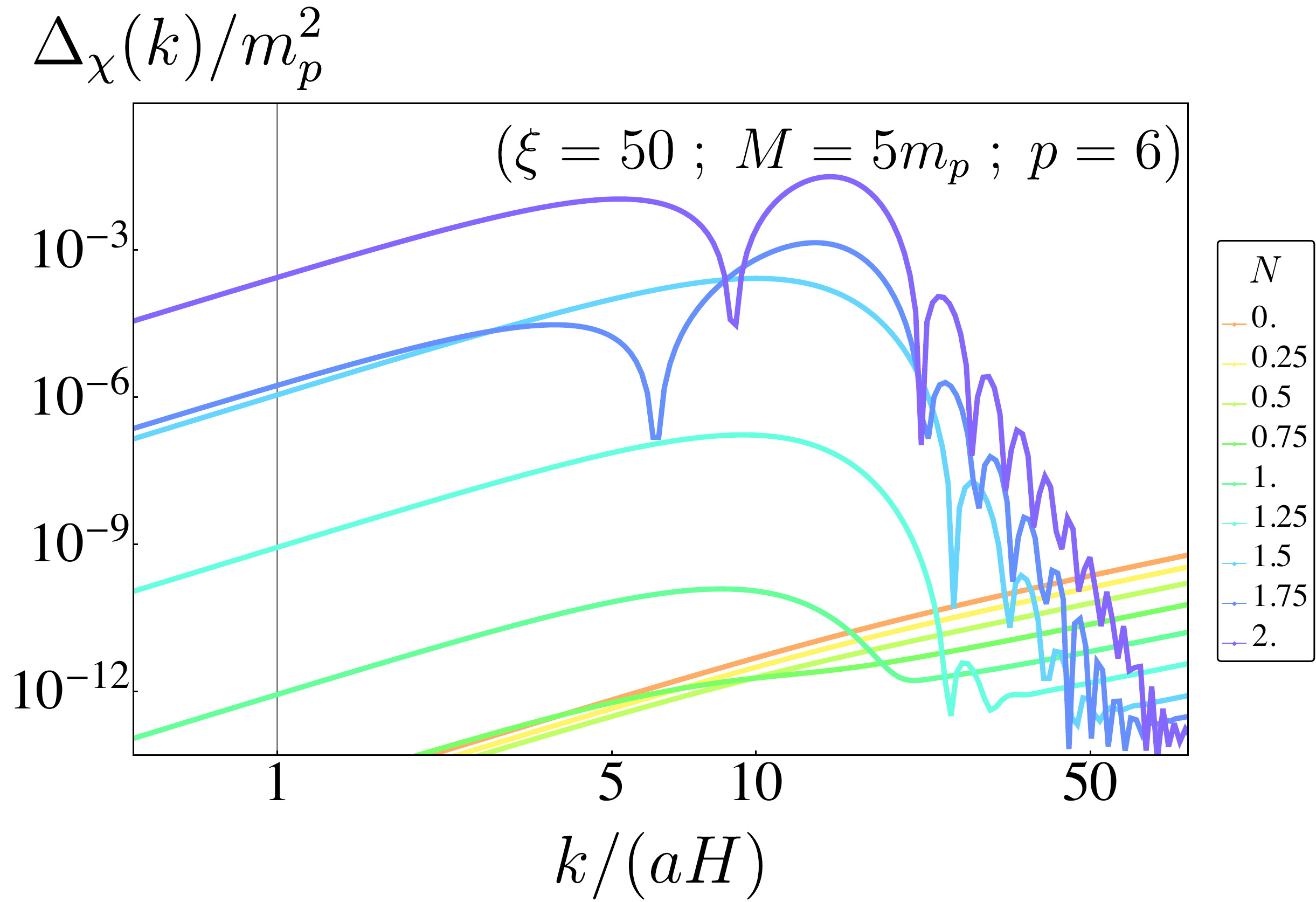} 
    \end{center}
    \caption{{\it Left-Panels: } Time Evolution of $R/H^2$ for $M=5\,m_p$ and $p=\{2,4,6\}$. Red dots indicate $R/H^2$ average per oscillation and blue dashed line is given by the asymptotic value of Eq.~\eqref{eq:RvsW}. {\it Right-Panels:} Time evolution of the matter power spectrum of the NMC field $\Delta_{\chi}(k,N)/m_p^2$ for each of the cases with $\xi=50$.} \label{fig:ScalarPSandRicci} \vspace*{-0.3cm}
\end{figure*}

In the left panels of Fig.~\ref{fig:ScalarPSandRicci}, we show the oscillations of the Ricci scalar, normalized to the Hubble rate, $R/H^2$, against the number of e-folds after inflation. In the right panels of the same figure, in order to illustrate the growth of the tachyonic modes of $\chi$, we show the power spectrum $\Delta_{\chi}(k,\tau)$, defined in relation to the variance of the $\chi$ field as
\begin{eqnarray}
    \langle \chi^2 \rangle= \langle 0|\hat{\chi}(\tau,x)\hat{\chi}(\tau,x)|0\rangle = \int \frac{dk}{k} \Delta_{\chi}(k,\tau)\:.
\end{eqnarray}
The first feature worth noticing is how different the oscillatory regimes of the Ricci scalar are, depending on $p$, see left panels of Fig.~\ref{fig:ScalarPSandRicci}. For the chosen large scale example $M=5m_p$, the duration of a semi-oscillation with $R < 0$ is longer for $p=6$ than for $p=4$, and much longer than for $p=2$. This means that the tachyonic growth of the modes is more effective the higher the value of $p$. This is reflected in how much faster the power spectrum grows for $p=6$ in comparison to $p=4$ and $p=2$, see right panels of Fig.~\ref{fig:ScalarPSandRicci}.

\begin{figure*}[tbp]
\begin{center}
    \includegraphics[width=0.45\textwidth,height=4.65cm]{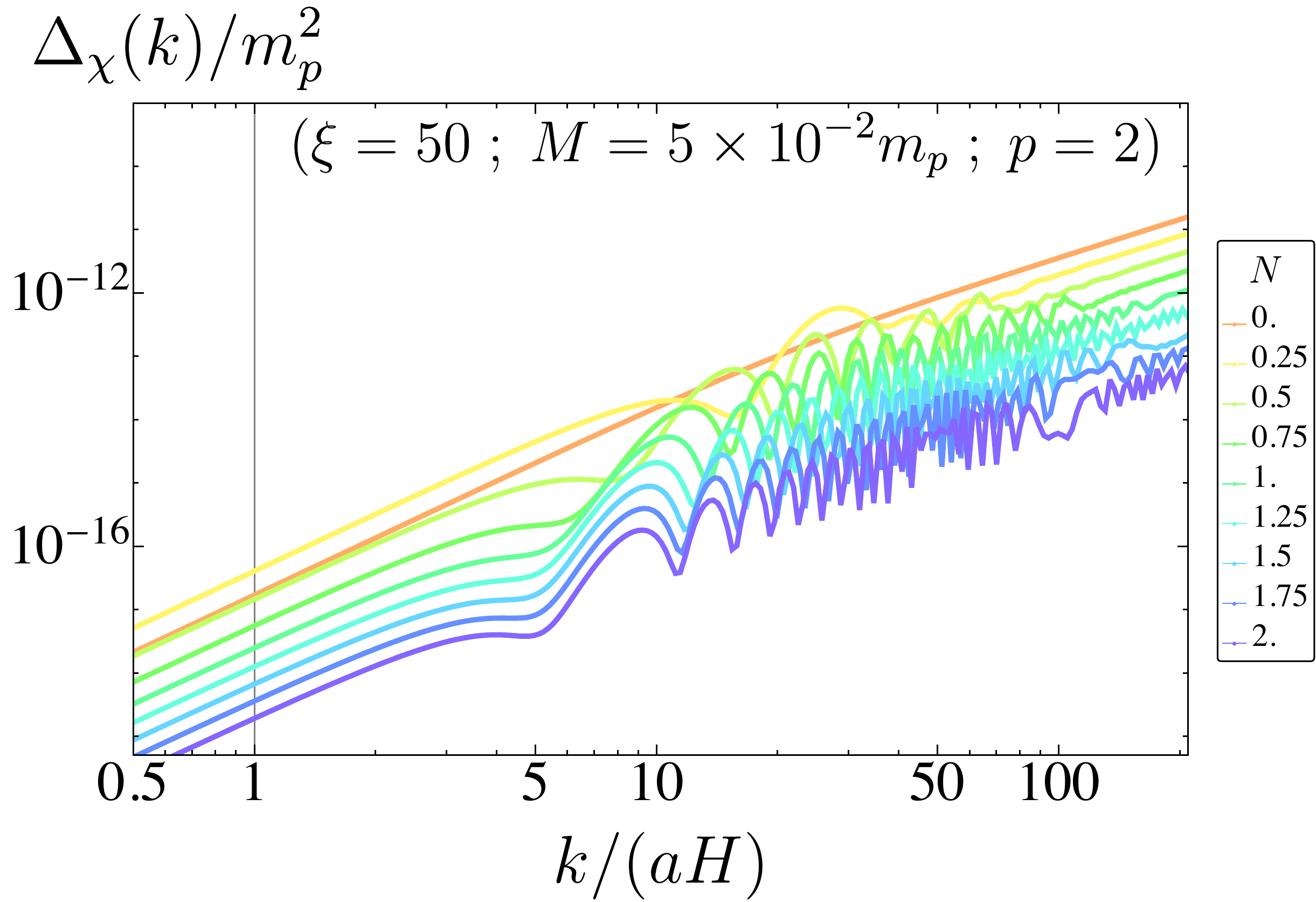}
    \includegraphics[width=0.45\textwidth,height=4.65cm]{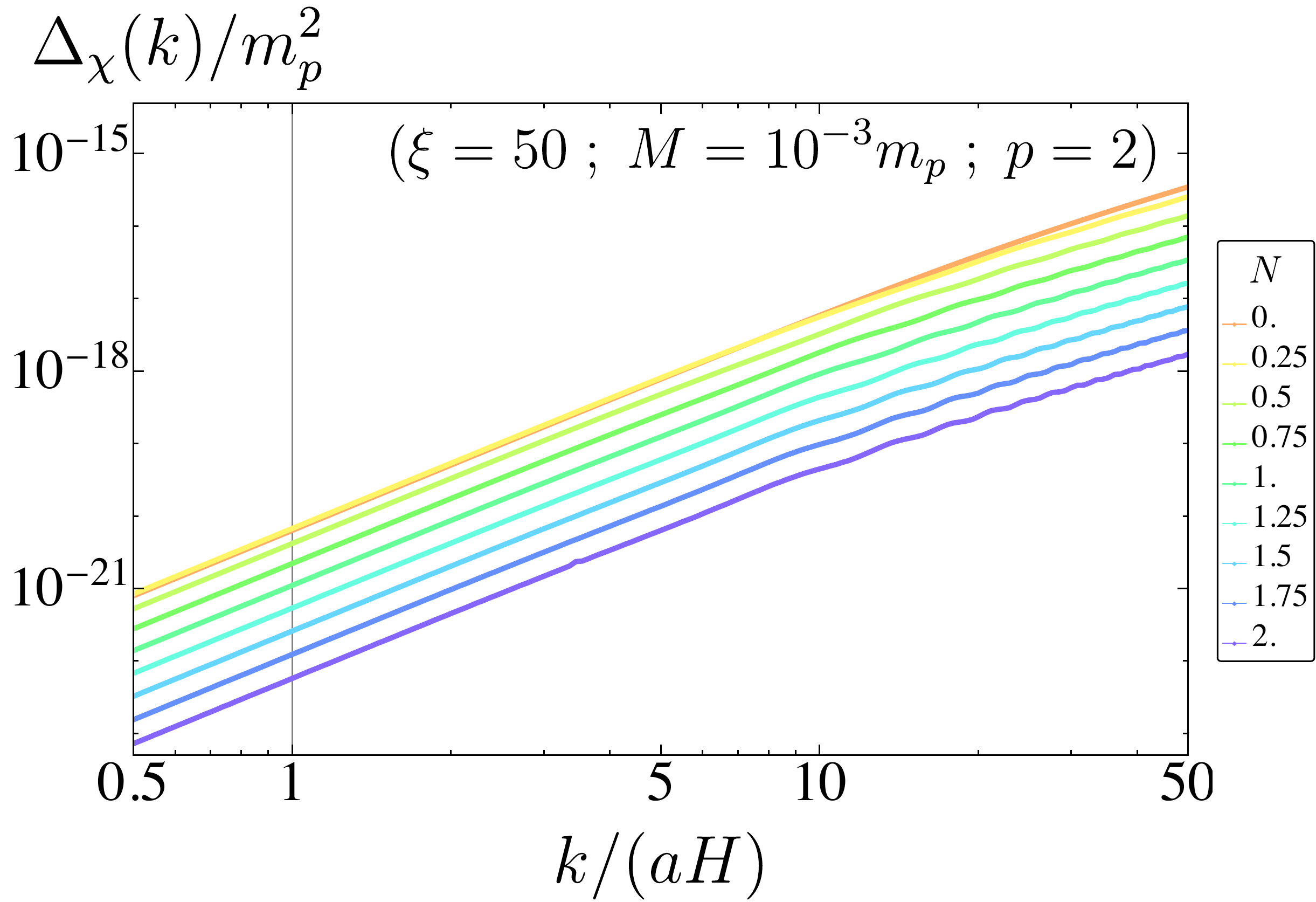}
    \includegraphics[width=0.45\textwidth,height=4.65cm]{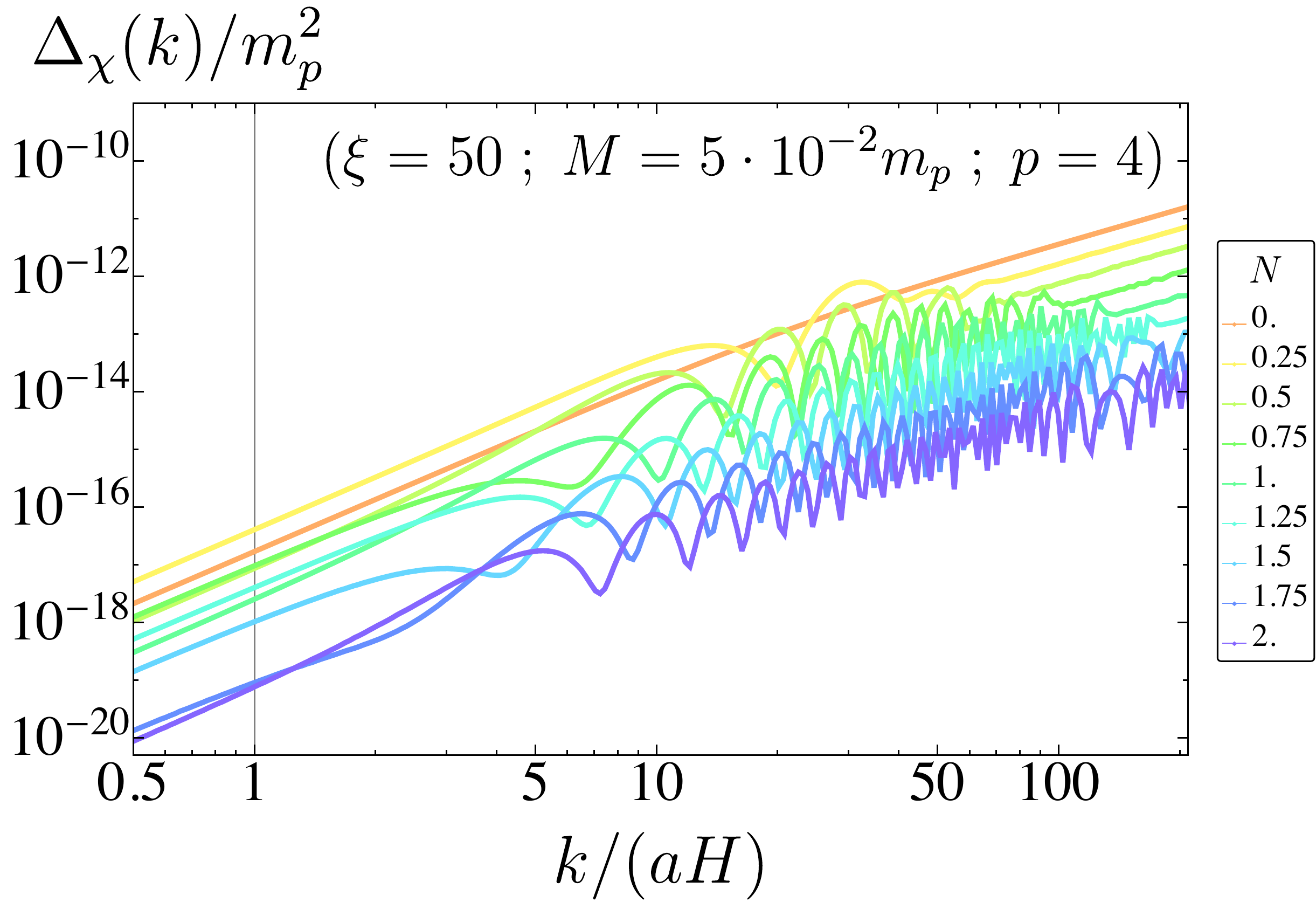}
    \includegraphics[width=0.45\textwidth,height=4.65cm]{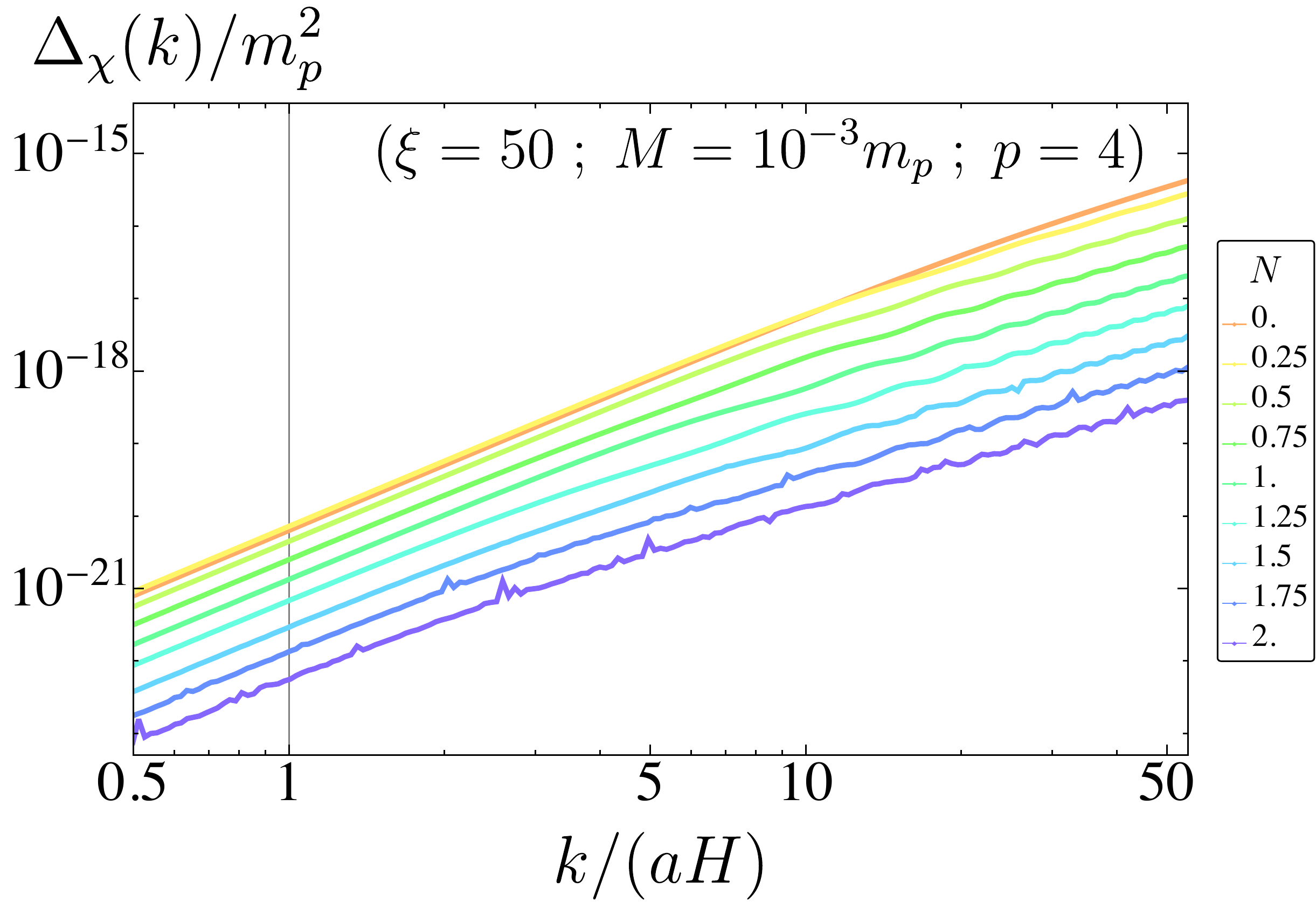}
    \includegraphics[width=0.45\textwidth,height=4.65cm]{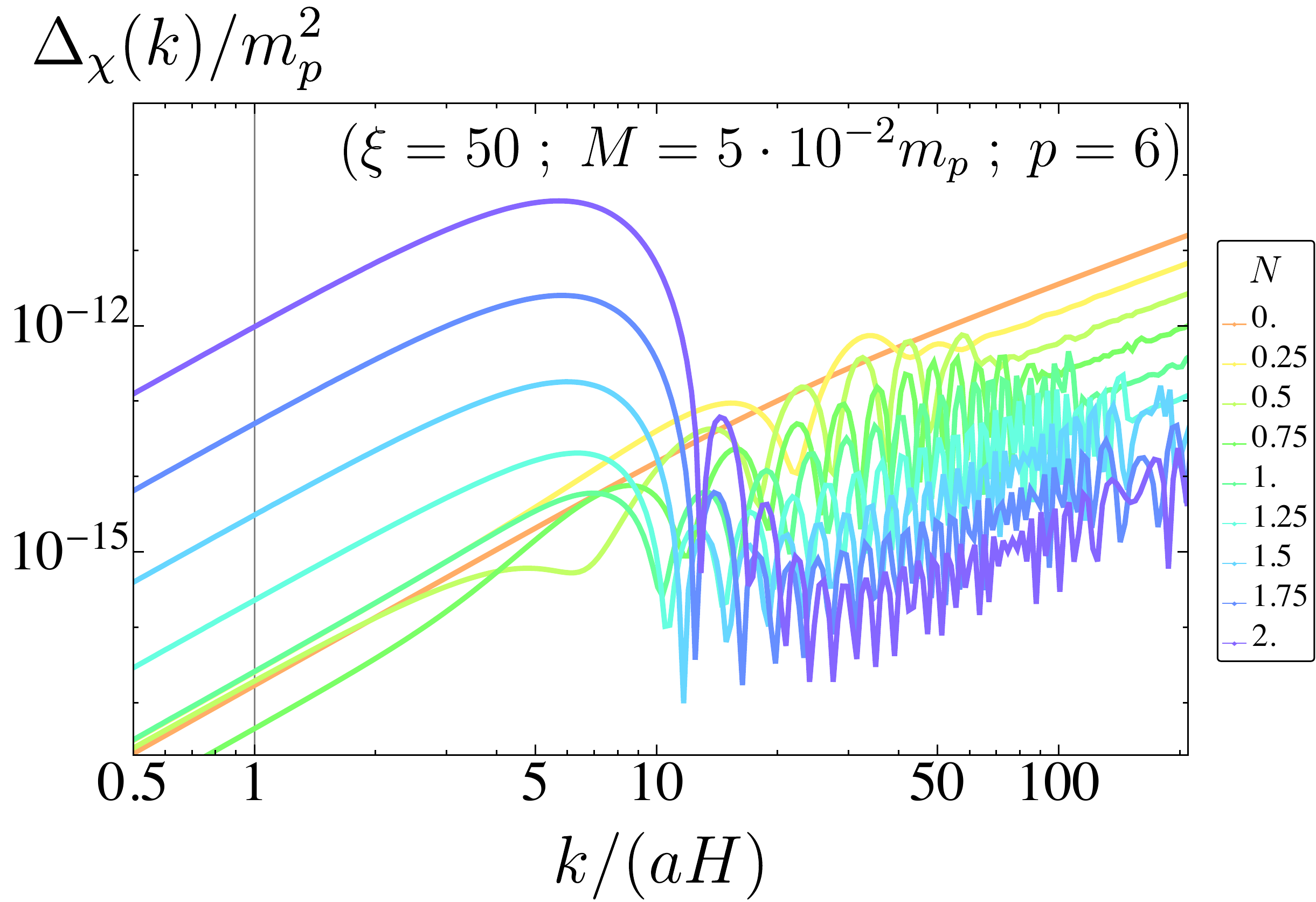}
    \includegraphics[width=0.45\textwidth,height=4.65cm]{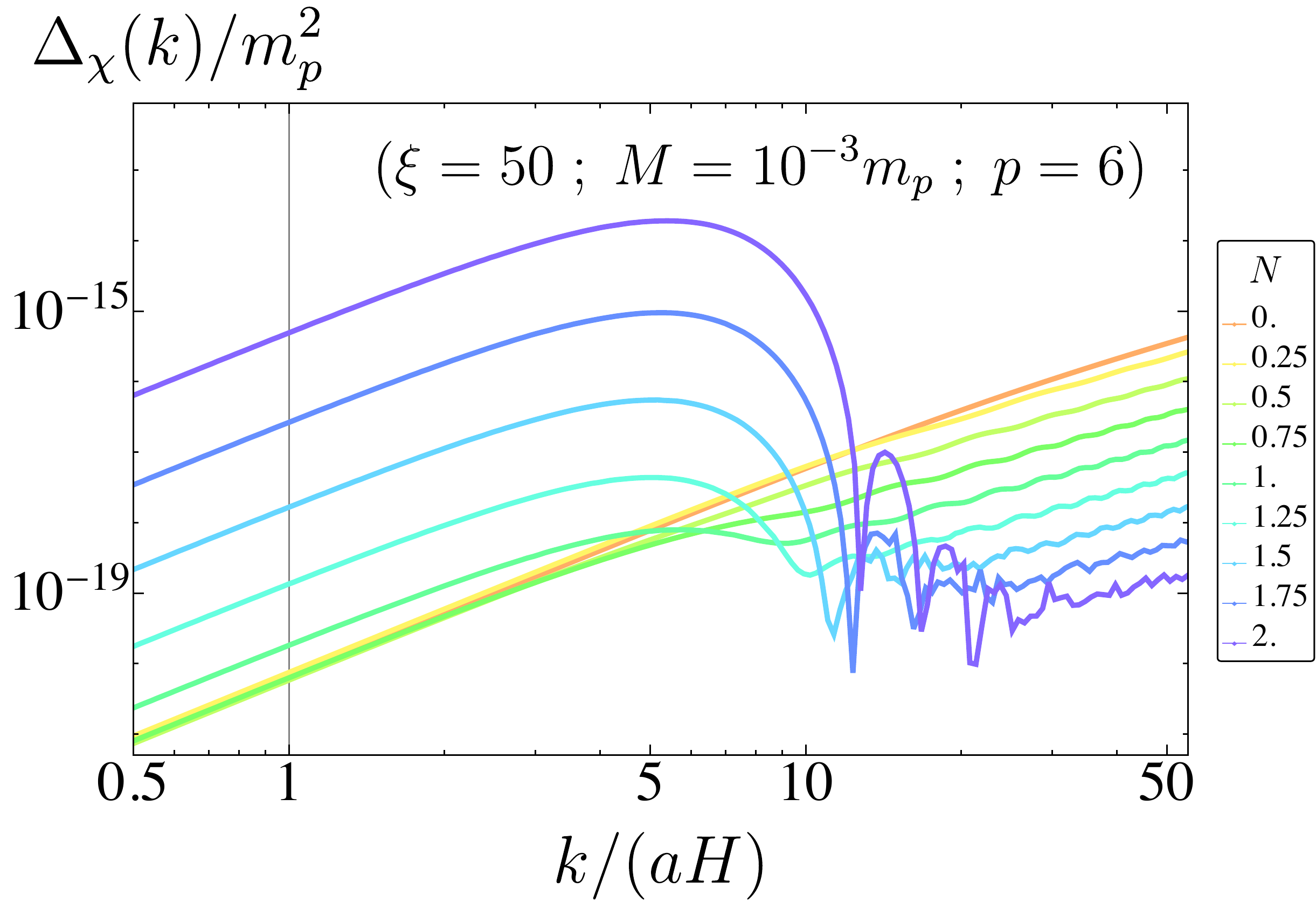}
    \end{center}
    \vspace*{-0.5cm}
    \caption{Time evolution of the power spectrum of the NMC field $\Delta_{\chi}(k,N)/m_p^2$.} \label{fig:lowScalePS} \vspace*{-0.3cm}
\end{figure*}

In Fig.~\ref{fig:lowScalePS} we plot the growth of $\chi$ modes for low scale scenarios with $M=0.05m_p$ (left panels) and $M=0.001m_p$ (right panels), for $p = 2, 4$ and $6$ (top to bottom panels). The case $M=0.05m_p$ corresponds in fact to the scenarios considered in Fig.~\ref{fig:lowScaleRicci2} for which we already showed the ratio $\overline R/\overline H^2$.
As explained in Sect.~\ref{sec:ScaleofInflation} and noted in Fig.~\ref{fig:lowScaleRicci2}, the oscillation-averaged Ricci $\overline R$ in these scenarios remains a positive decaying function while $\phi$ samples the potential above $\phi_{\star}$ (indicated by red squares), but the ratio $\overline R/\overline H^2$ decays eventually down to the asymptotic value~\eqref{eq:RvsW}. In Fig.~\ref{fig:lowScalePS} we observe that $\chi$ modes are not much amplified when $M = 0.05m_p$ for $p=2$ and $p=4$, as $\overline R$ remains essentially non-negative (see top and middle panels of Fig.~\ref{fig:lowScaleRicci2}). However, for $p=6$, while modes do not grow initially while $R$ remains positive, as soon as the averaged Ricci becomes negative (see bottom panel of Fig.~\ref{fig:lowScaleRicci2}), the tachyonic instability leads to an exponential growth of the $\chi$ modes. By comparing left and right panels in Fig.~\ref{fig:lowScalePS}, we also observe how different it is the efficiency on the mode's growth, depending on the choice of $M$. Decreasing $M$ makes less efficient the tachyonic instability, to the point that for $p=2$ and $p=4$, modes are not excited at all. For $p = 6$, the tachyonic instability is triggered later the smaller the value of $M$ is, and hence for a given time elapse after the end of inflation, the $\chi$ modes will have grown less for smaller values of $M$.

We have analyzed so far the presence of a tachyonic instability in the dynamics of the NMC field, due to the oscillations of $R$. A relevant aspect to address is the duration of the instability. Clearly, the tachyonic growth cannot be sustained indefinitely, because either: $i)$ the instability mechanism gradually switches off during the linear regime due to the expansion of the universe, or $ii)$ the energy transferred into the NMC field becomes sizeable, and the overall dynamics become non-linear, which also switches off the instability. While the case when the tachyonic instability is switched off during the linear regime is of less interest than the transition to non-linearities, it is however easy to understand, and we discuss it first. As long as the Ricci scalar is sourced solely by the homogeneous inflaton, the dynamics of $\chi$ are in a linear regime. In particular, looking at Eq.~(\ref{eq:modeseqnconfchi}) and assuming $\xi \gg 1/6$, we see that the instability will cease for a given mode $k$, whenever $a^2 |R| \lesssim k^2/\xi$. Considering the oscillation-averaged Ricci curvature~(\ref{eq:RvsW}), we see that $a^2 |\overline{R}| \sim \mathcal{H}^2 \sim {1/\tau^2}$, where $\mathcal{H} \equiv a'/a \sim 1/\tau$ is the Hubble rate in conformal time $\tau$. This means that within a (conformal) period of time since the end of inflation, of the order of $\tau_{k} \sim \sqrt{\xi}/k$, a mode $k$ typically ceases to be tachyonic. As a consequence, within a time $\tau_{\rm max} \sim \sqrt{\xi}/\mathcal{H}$, all sub-horizon modes $k \gtrsim \mathcal{H}$ have ceased to be tachyonic, and the instability is effectively switched off. Depending on the choice of $p$ and the strength of $\xi$, this might happen before the energy transferred into $\chi$ becomes sizeable as compared to the energy in $\phi$. This can be appreciated, for example, in the case of $p = 2$ in the left-middle panels of Fig.~\ref{fig:Plotsxi500p2}. There, even for couplings as large as $\xi = 500$, the energy transfer into the NMC field ceases around $N_{\rm max} \lesssim 1$ efoldings after the end of inflation, either for large ($M = 5 m_p$, left panel) or small ($M = 0.1m_p$, right panel) scale scenarios. In the case of $p = 4$, we also appreciate in the top panels of Fig.~\ref{fig:Plotsxi300p4}, that even for $\xi = 300$, the NMC energy density ceases to grow around $N_{\rm max} \lesssim 1$ efoldings after the end of inflation. The exact time of the instability switch-off and the fraction of energy transferred on that moment, depend on $M$, and we see that the instability switches off well inside the linear regime for small ($M = 0.1m_p$, right panel) and intermediate ($M = m_p$, middle panel) scale scenarios, but marginally at the onset of the non-linear regime for large scale scenarios ($M = 5m_p$, left panel).

The case when the tachyonic instability is switched off due to the onset of non-linear dynamics is more complex to describe, so we rather discuss it in section~\ref{sec:LatticeRH}. We anticipate here, in any case, that in certain cases the transfer of energy into the NMC field is so effective, that $\chi$ will backreact on the the background dynamics, driving the inflaton-NMC coupled dynamics into a non-linear regime. This is actually the most interesting regime, as it will allow us to determine the circumstances for achieving proper {\it reheating} of the universe, i.e.~whether the transfer of energy may lead to an energetic dominance of NMC scalar $\chi$ over the inflaton $\phi$. In the next section we will then switch to study the 2-field system with the aid of lattice simulations, as this technique will allow us to fully explore the non-linear regime, and hence to determine the circumstances under which reheating can be achieved. Before that, we can still gain some insight of the up-coming non-linear dynamics, by noting that the total energy momentum tensor of the system can be split into separate parts contributed by the inflaton and NMC field, $T_{\mu\nu} \equiv T_{\mu\nu}^{\chi} + T_{\mu\nu}^{\phi}$. The part of the NMC field reads~\cite{Figueroa:2021iwm}
\begin{equation}\label{eq:TmunuNMCfield}
    T_{\mu\nu}^{\chi} = \partial_{\mu}\chi\partial_{\nu}\chi - g^{\mu\nu}\Big(\frac{1}{2}g^{\rho\sigma}\partial_{\rho}\chi\partial_{\sigma}\chi+V_{\rm NMC}\Big)+\xi\Big(G_{\mu\nu}+g_{\mu\nu}g^{\alpha\beta}\nabla_{\alpha}\nabla_{\beta} -\nabla_{\mu}\nabla_{\nu}\Big)\chi^2\:,
\end{equation}
which, if {\it traced} (with respect to the background metric), gives
\begin{equation}\label{eq:TraceNMC}
    T_{\chi} \equiv g^{\mu\nu}T_{\mu\nu}^{\chi} = (6\xi-1)(\partial^{\mu}\chi\partial_{\mu}\chi+\xi R \chi^2) + 6\xi\chi \partial_{\chi}V_{\rm NMC}-4V_{\rm NMC}(\chi) \:.
\end{equation}
Following~\cite{Figueroa:2021iwm}, we can write the Ricci scalar from the trace part of Einstein's equations, $m_p^2 R = -(T_{\chi} + T_\phi)$. Using $T_\phi = \partial^{\mu}\phi\partial_{\mu}\phi-4V_{\rm inf}(\phi)$ and expression (\ref{eq:TraceNMC}), this leads to 
\begin{eqnarray}\label{eq:FullR}
    R =\frac{(1-6\xi)\langle\partial^{\mu}\chi\partial_{\mu}\chi\rangle +4\langle V \rangle-6\xi\langle \chi V_{,\chi}\rangle + \langle\partial^{\mu}\phi\partial_{\mu}\phi\rangle}{m_p^2+(6\xi-1)\xi\langle \chi^2 \rangle}\:,
\end{eqnarray}
where $\langle ... \rangle $ stands for volume averaging, and $V = V_{\rm inf}(\phi) + V_{\rm NMC}(\chi)$. Eq.~\eqref{eq:FullR} indicates that as long as all the NMC terms $\lbrace (1-6\xi)\langle\partial^{\mu}\chi\partial_{\mu}\chi\rangle,6\xi\langle \chi V_{,\chi}\rangle, (6\xi-1)\xi\langle \chi^2 \rangle \rbrace$ remain small, the Ricci scalar will be determined by the inflaton-dominated expression given in (\ref{eq:RicciAsKinPot}). However, for sufficiently large $\xi$, it is possible that either (or a combination) of the NMC terms  becomes sizeable, signaling the onset of the non-linear regime. We describe this circumstance in the following section.

\section{Geometric reheating: non-linear regime}\label{sec:LatticeRH}

In order to study the non-linear evolution of the system including the backreaction of the NMC field on the Ricci scalar, and hence on the expanding background and on the inflaton dynamics, it is necessary to use lattice simulations. We use \CL~\cite{Figueroa:2021yhd,Figueroa:2020rrl,Figueroa:2023xmq} and adopt the lattice formulation for NMC dynamics introduced in~\cite{Figueroa:2021iwm}. We study the efficiency of preheating and whether proper reheating is achieved, for $p=2$, $p=4$ and $p=6$, as a function of the non-minimal coupling $\xi$ and of the mass scale $M$. In order to initialize the $\chi$ field in the lattice at a time $\eta_{\rm sim}$ after the end of inflation, we set vanishing zero-mode and initial fluctuations with power spectrum $\Delta_{\chi}(k,\eta_{\rm sim})$ computed from the linear analysis. We choose $\eta_{\rm sim}$ as a moment when, due to the tachyonic excitation during the linear regime, a peak has clearly emerged in the $\chi$ spectrum, but the NMC field is still energetically very subdominant compared to the inflaton. At the same time we initialize the inflaton in the lattice by providing a homogeneous part (amplitude and velocity) and a set of fluctuations on top. While the homogeneous part is read from the solution to its homogeneous Klein-Gordon equation solved during the linear regime,  the fluctuations are added on top of the homogeneous component using a spectrum mimicking quantum vacuum fluctuations (see e.g.~section 7 of~\cite{Figueroa:2020rrl} for details on this procedure). We evolve the lattice version given in Ref.~\cite{Figueroa:2021iwm} of the following set of coupled differential equations
\begin{eqnarray}    \label{eq:LatticeEoMchi}
     \chi'' + (3-\alpha)\left(\frac{a'}{a}\right)\chi' - a^{-2(1-\alpha)}\nabla^2\chi &=& - a^{2\alpha}\left(\xi R \chi + V_{,\chi}\right) \; ,\\
     \label{eq:LatticeEoMphi}
     \phi'' + (3-\alpha)\left(\frac{a'}{a}\right)\phi' - a^{-2(1-\alpha)}\nabla^2\phi &=& - a^{2\alpha}V_{,\phi} \; ,\\
     \label{eq:FriedmannSimulation}
     \frac{a''}{a} + (1-\alpha)\left(\frac{a'}{a}\right)^2 &=& \frac{a^{2\alpha}}{6} R \; ,  
\end{eqnarray}
with $R$ computed from volume averages $\langle ... \rangle$ as
\begin{eqnarray}\label{eq:RiccifromMatter}
R = {\Big[(6\xi-1)\Big(\frac{\left\langle\chi'^2\right\rangle}{a^{2\alpha}} -\frac{\left\langle(\nabla \chi)^2\right\rangle}{a^2}\Big) - 6\xi \left\langle \chi V_{,\chi} \right\rangle 
    + 4\left\langle V \right\rangle - \frac{\left\langle\phi'^2\right\rangle}{a^{2\alpha}} + \frac{\left\langle(\nabla \phi)^2\right\rangle}{a^2}\Big]\over m_p^2+(6\xi-1)\xi{\langle\chi^2\rangle}} \,,
\end{eqnarray}
and where the total potential is given by $V = V_{\rm inf}(\phi) + V_{\rm NMC}(\chi)$. In the following we set $V_{\rm NMC} = 0$, unless stated otherwise (only in subsection~\ref{subsec:selfinteractions} we will restore $V_{\rm NMC} \neq 0$). We note that the Hubble rate can be used as a constraint equation, as it is sourced from the energy density of both the inflaton and the NMC field,
\begin{eqnarray}\label{eq:HubbleSim}
    \mathcal{H}^2 &\equiv& \left(\frac{a'}{a}\right)^2 = \frac{a^{2\alpha}}{3 m_p^2}[E_{\phi} + E_{\chi}] \:,\\
    E_{\phi} &=& \frac{1}{2a^{2\alpha}}\left\langle\phi'^2\right\rangle + \frac{1}{2a^{2}}\left\langle(\nabla \phi)^2\right\rangle + \left\langle V_{\text{inf}}(\phi) \right\rangle\:,  \\
    \label{eq:ChiEnergy}
    E_{\chi} &=& \frac{1}{2a^{2\alpha}}\left\langle\chi'^2\right\rangle + \frac{1}{2a^{2}}\left\langle(\nabla \chi)^2\right\rangle + \left\langle V_{\rm NMC}(\chi) \right\rangle + \frac{3 \xi}{a^{2\alpha}} \mathcal{H}^2\left\langle \chi^2 \right\rangle + \frac{6\xi}{a^{2\alpha}} \mathcal{H} \left\langle \chi \chi'\right\rangle \;,
\end{eqnarray}
so that during evolution, we monitor that the left hand side of Eq.~(\ref{eq:HubbleSim}), $3m_p^2(a'/a)^2$, is well balanced by the right hand side, $a^{2\alpha}(E_\phi+E_\chi)$, with the solution for $\lbrace a(\eta), E_\phi, E_\chi \rbrace$ obtained from solving Eqs.~(\ref{eq:LatticeEoMchi})-(\ref{eq:FriedmannSimulation}). Simulations were run in lattices with $240 - 1000$ sites/dimension, depending on the case, always with lattice ultraviolet scales guarantying  $k_{\rm UV} > k_*$, with $k_*/aH \simeq \sqrt{\xi |R|/H^2}$ the threshold scale that separates tachyonic ($k \lesssim k_*$) from non-excited ($k \gtrsim k_*$). In the following we present our results on lattice simulations for $p = 2$ with $M/m_p = \lbrace 5,0.1\rbrace$ and $\xi = 500$, for $p=4$ with $M/m_p = \{5,1,0.1\}$ and $\xi = \{25, ..., 500\}$, and for $p=6$ with $M/m_p = \{5,1,10^{-1},10^{-2},10^{-3}\}$ and $\xi = \{25, ..., 500\}$. We choose to plot the evolution of physical quantities in {\it e-folds} $N > 0$ after the end of inflation, defined by $N(\eta)\equiv \log(a(\eta)/a_{\rm end})$.

\subsection{Lattice results for $p=2$}\label{subsec:p2}

Here we review the relevant results on geometric (p)reheating for $p=2$, i.e.~when the inflaton potential is quadratic around $\phi = 0$.  By analyzing the linear dynamics in the previous section, we already showed that the inflaton oscillations in this case are such that $R$ remains mostly positive. As the tachyonic instability requires that $R<0$ for as long as possible, we expect the energy transfer to $\chi$ to be very inefficient in this scenario. This can be observed e.g.~in the top right panel of Fig.~\ref{fig:ScalarPSandRicci}, or in the top panels of Fig.~\ref{fig:lowScalePS}, where spectra of $\chi$ do not grow very much compared to the cases with higher $p$. Despite this, we have still run lattice simulations for large values of the coupling $\xi$, to see if this can compensate for the shortness of the stages with $R < 0$. 

\begin{figure*}[th]
    \begin{center}
    \includegraphics[width=0.45\textwidth,height=4.5cm]{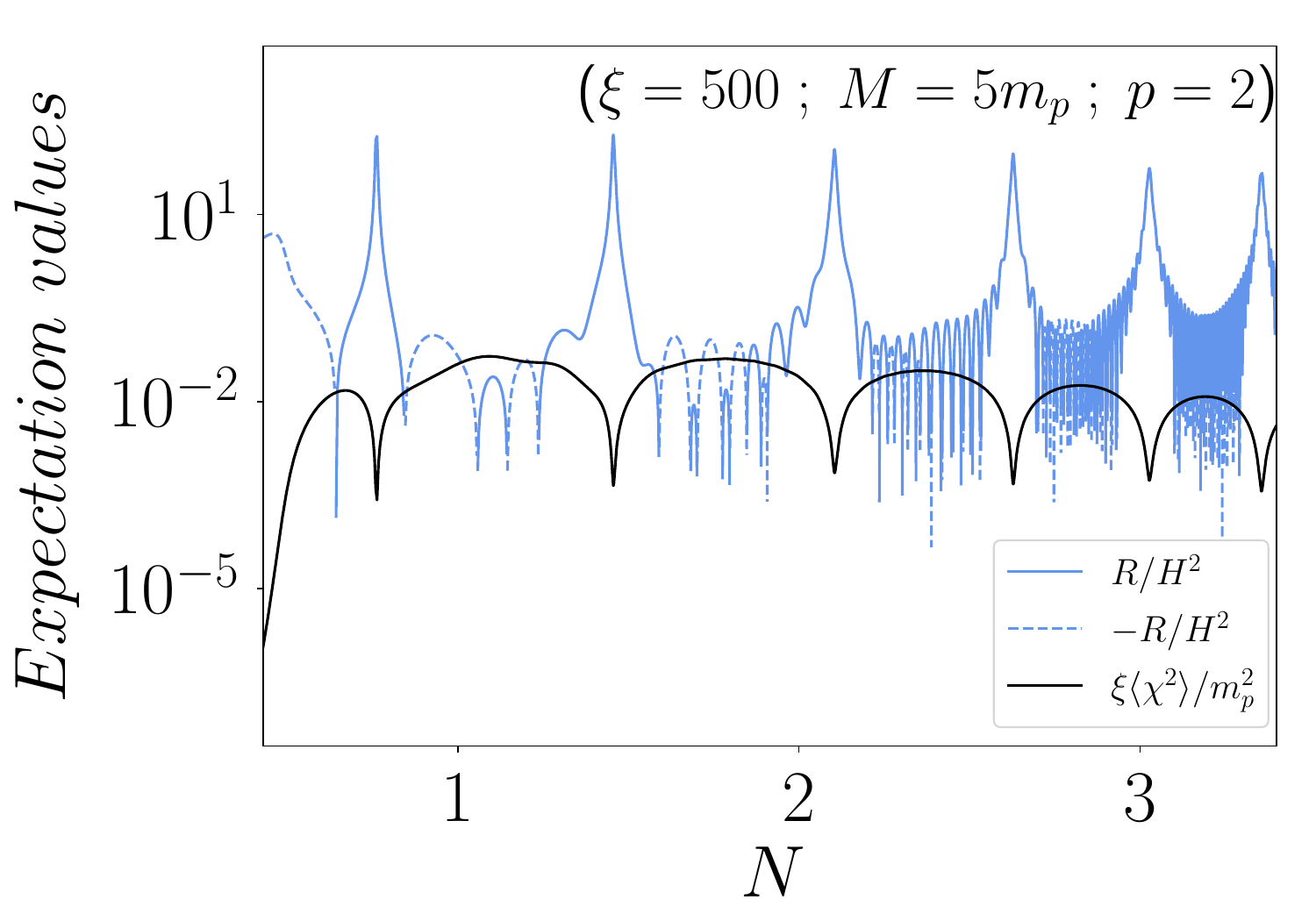} 
    \includegraphics[width=0.45\textwidth,height=4.5cm]{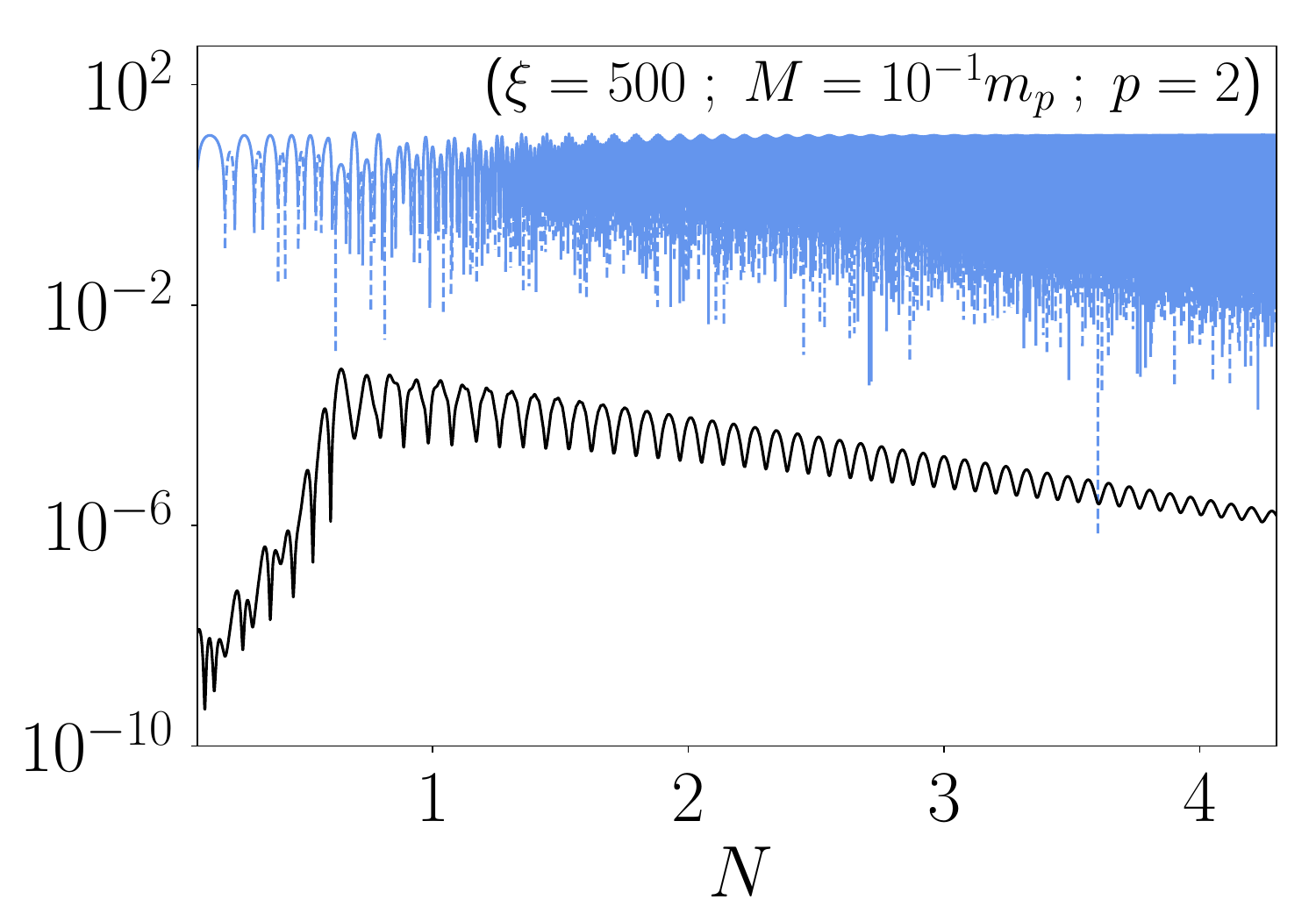} 
    \includegraphics[width=0.45\textwidth,height=4.5cm]{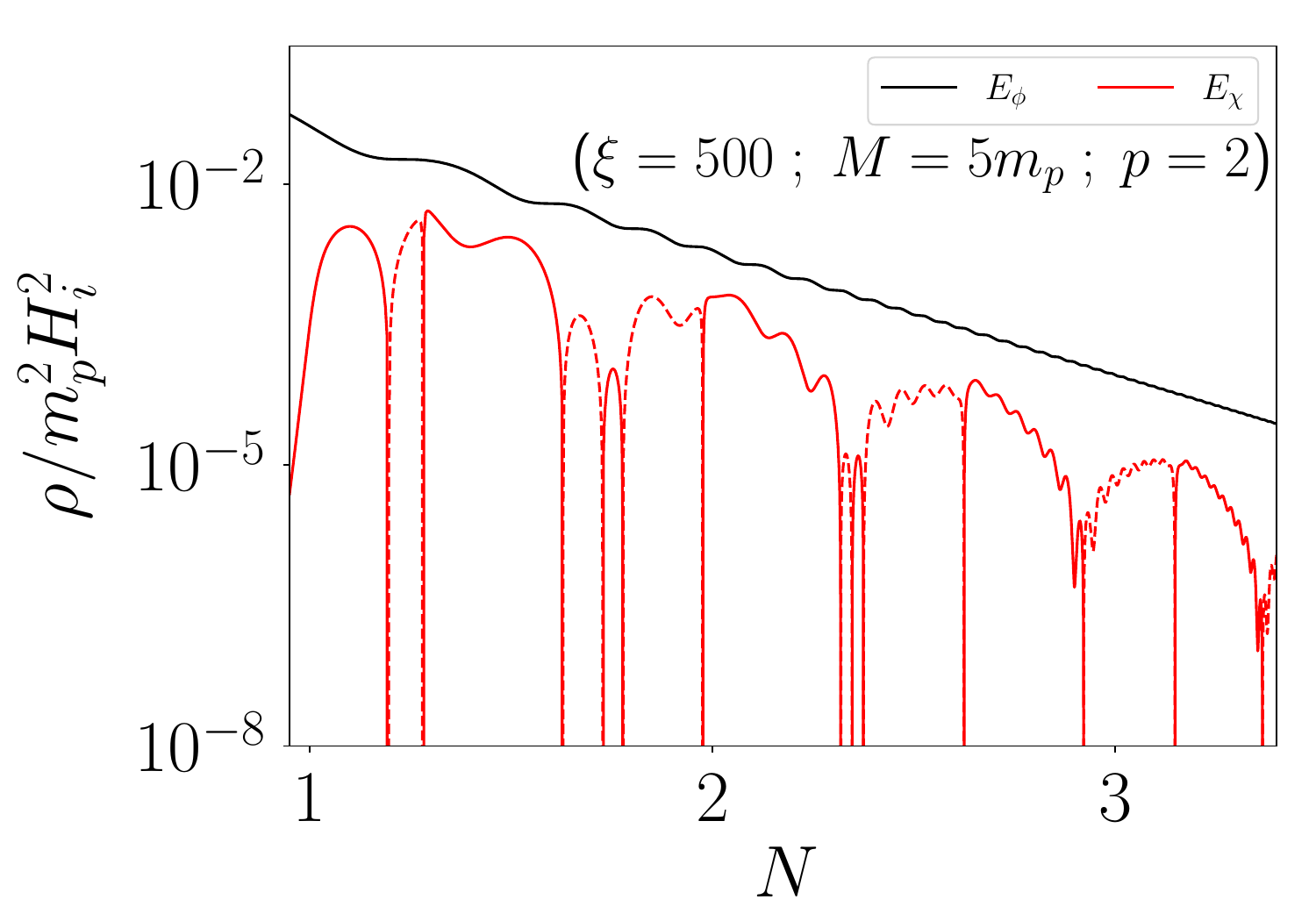} 
    \includegraphics[width=0.45\textwidth,height=4.5cm]{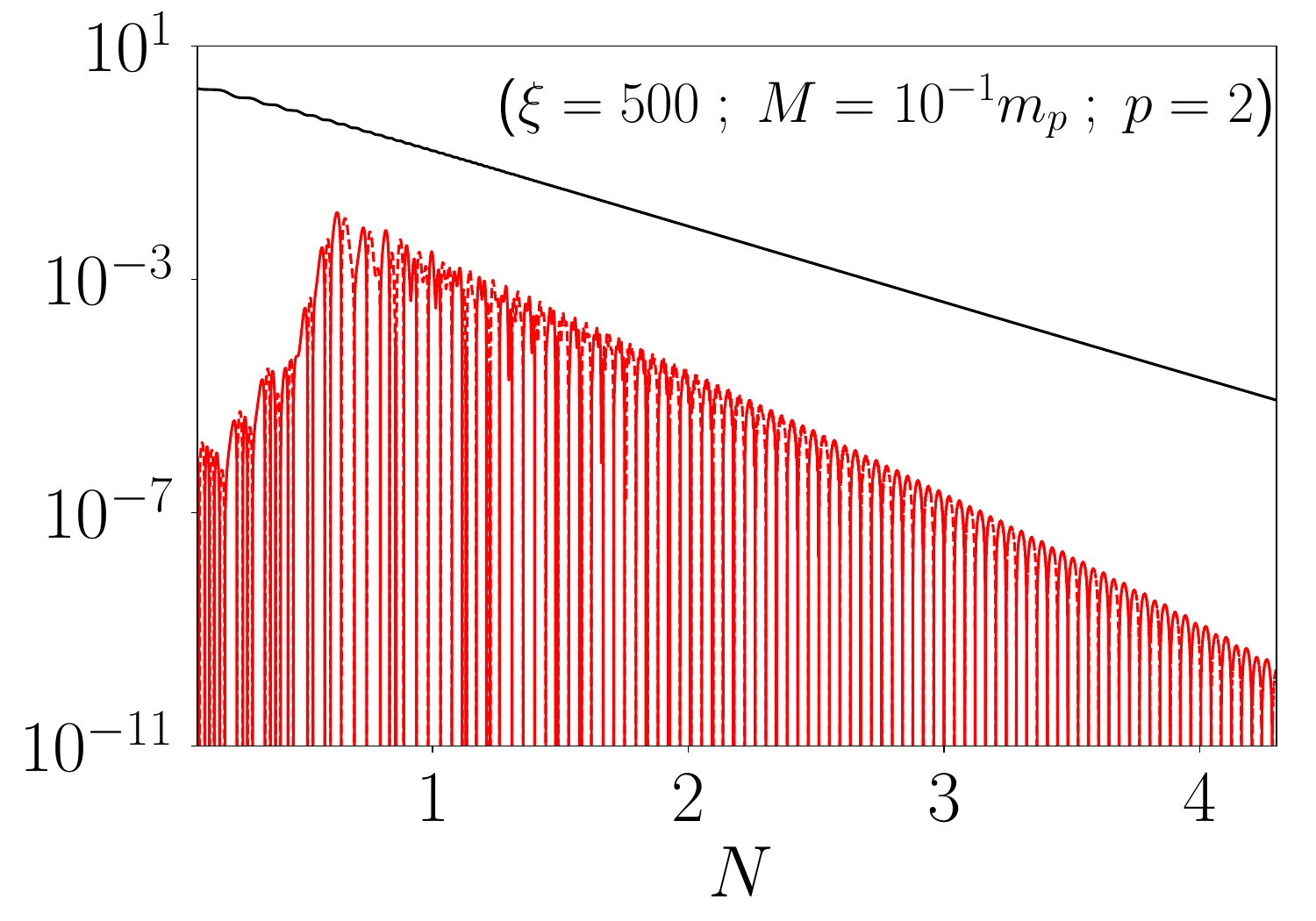} 
    \includegraphics[width=0.45\textwidth,height=4.5cm]{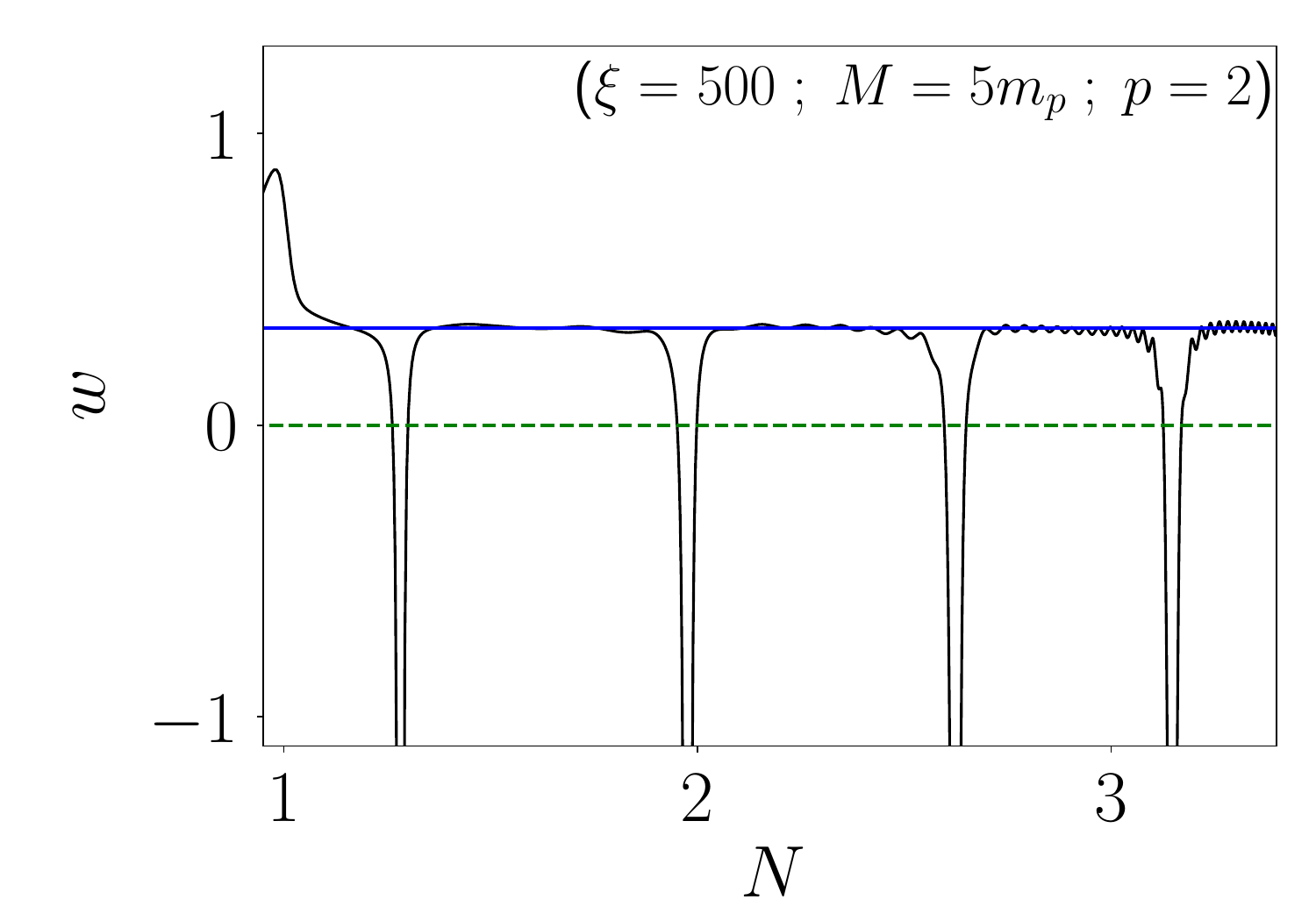} 
    \includegraphics[width=0.45\textwidth,height=4.5cm]{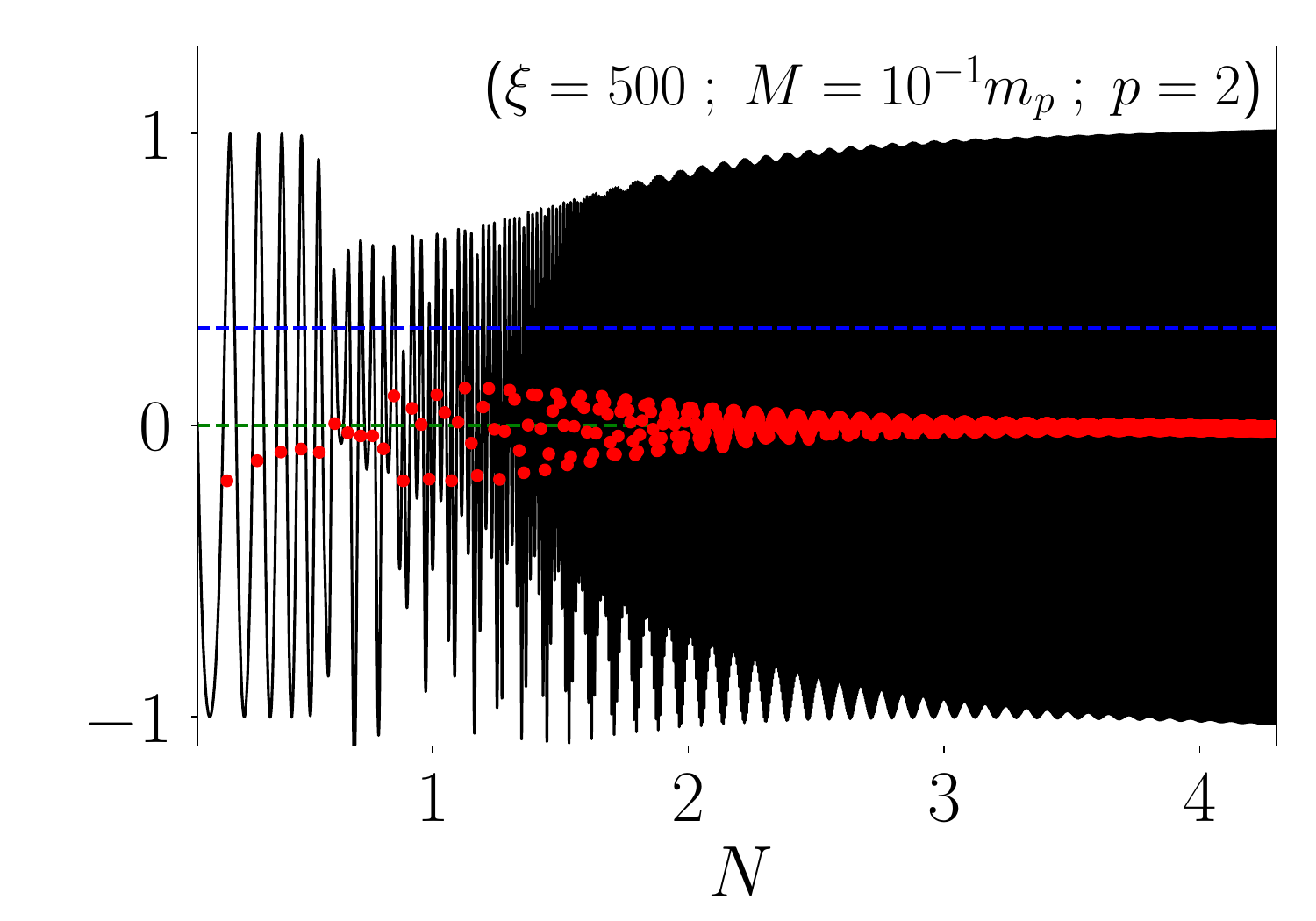} 
    \end{center}
    \vspace*{-0.5cm}
    \caption{Time evolution of various quantities for $p=2$, $\xi=500$,  and $M=5m_p$ (left panels) and $M=10^{-1}m_p$ (right panels). {\it Top Panel:} Evolution of the NMC field variance $\xi \langle \chi^2 \rangle/m_p^2$ and $R/H^2$. {\it Middle Panel}: Evolution of the total energy of the inflaton and of the NMC field. {\it Bottom Panel:} Evolution of the equation of state ($w$) computed from Eq.~(\ref{eq:EoSfromRonH2}), with the green dashed line at $w=0$ indicating matter domination, and red dots depicting $w_{\rm osc}$ per oscillation, c.f.~Eq.~(\ref{eq:RvsW}). We note that in logarithmic plots solid lines stand for positive values and dashed lines for negative values.} \label{fig:Plotsxi500p2} \vspace*{-0.3cm}
\end{figure*}

In Fig.~\ref{fig:Plotsxi500p2} we show the time evolution of the NMC field variance $\langle\chi^2\rangle$  and of the Ricci scalar $R/H^2$, for low scale ($M = 0.1m_p$) and large ($M=5m_p$) scale scenarios with $\xi=500$.  There we observe that $\xi \langle\chi^2\rangle/m_p^2$ grows at most to a value $\mathcal{O}(10^{-1})$ for $M = 5m_p$, or $\mathcal{O}(10^{-3})$ for $M = 0.1m_p$, and then starts decaying. Inspecting the energy densities we see, correspondingly, that the NMC energy reaches at most to a value a factor $\mathcal{O}(10^{-1})$ smaller than the inflaton energy in the large scale scenario, and not more than $\mathcal{O}(10^{-3})$ in the low scale scenario.  Actually, once the NMC energy reaches its maximum (signaling the end of the tachyonic instability during the linear regime), it starts decaying faster than the inflaton energy.  This can be understood by the fact that the inflaton decays as non-relativistic matter, given the exponent $p =2$, while the NMC field decays as radiation,  as there is no other scale in the problem. We plot also the equation of state (EoS) $w$, observing that in the case $M=5m_p$, the effective value obtained from the ratio of oscillation-averaged quantities, c.f.~Eq.~(\ref{eq:HomEoS}), approaches $w_{\rm osc}=1/3$, which reflects the fact that the NMC-field has grown enough to backreact onto the Ricci scalar, dragging it to zero; while in the case $M=10^{-1}m_p$, the oscillations of $w$ are affected only mildly. We note that eventually $w_{\rm osc}$ goes back to oscillate with increasing amplitude. For the case $M=5m_p$, it does not reach back the typical oscillations for a matter dominated EoS during our simulation, but for the case $M=10^{-1}m_p$ it does. We can safely conclude that the case $p=2$ leads to an inefficient preheating energy transfer, even for large scale inflation and couplings as large as $\xi \gtrsim 500$. 

It is perhaps interesting to note that if we had chosen larger couplings, it is likely that, at least in large scale inflationary models (i.e.~$1 \lesssim M/m_p \lesssim 8$), the energy of $\chi$ might reach the energy level of $\phi$, seemingly reheating the Universe. This corresponds, however, only to a temporary achievement of reheating: as the energy densities of the NMC and inflaton fields would scale from that moment onwards as radiation and non-relativistic matter, respectively, soon enough the inflaton would dominate again, with $\chi$ becoming gradually more and more energetically subdominant. In the case of small scale inflationary models, as the initial energy transfer during the linear regime becomes less and less efficient as we reduce the scale $M$, we would not even reach such temporal reheating stage.

\begin{figure*}[tbp]
    \begin{center}
    \includegraphics[width=0.45\textwidth,height=4cm]{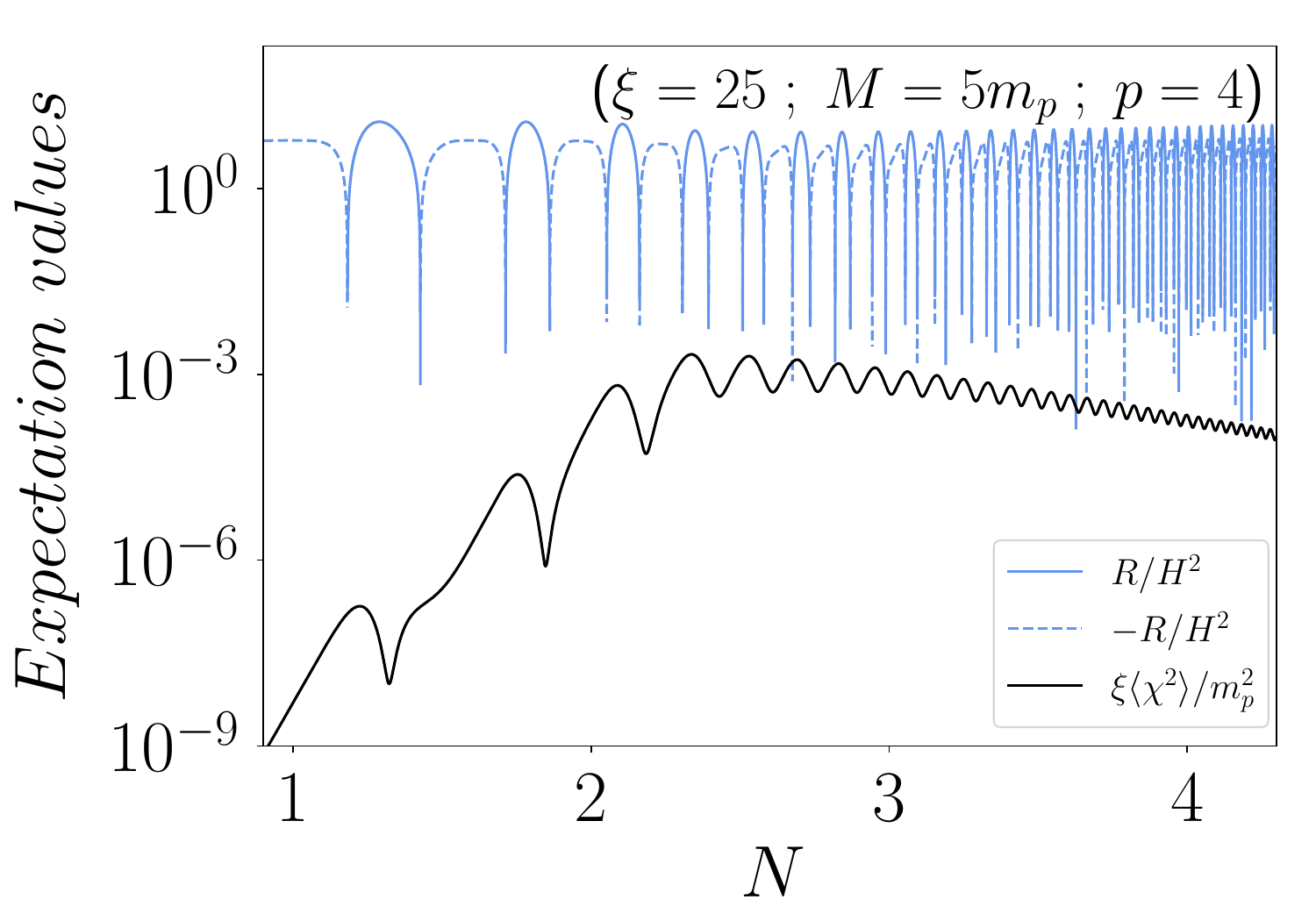} 
    \includegraphics[width=0.45\textwidth,height=4cm]{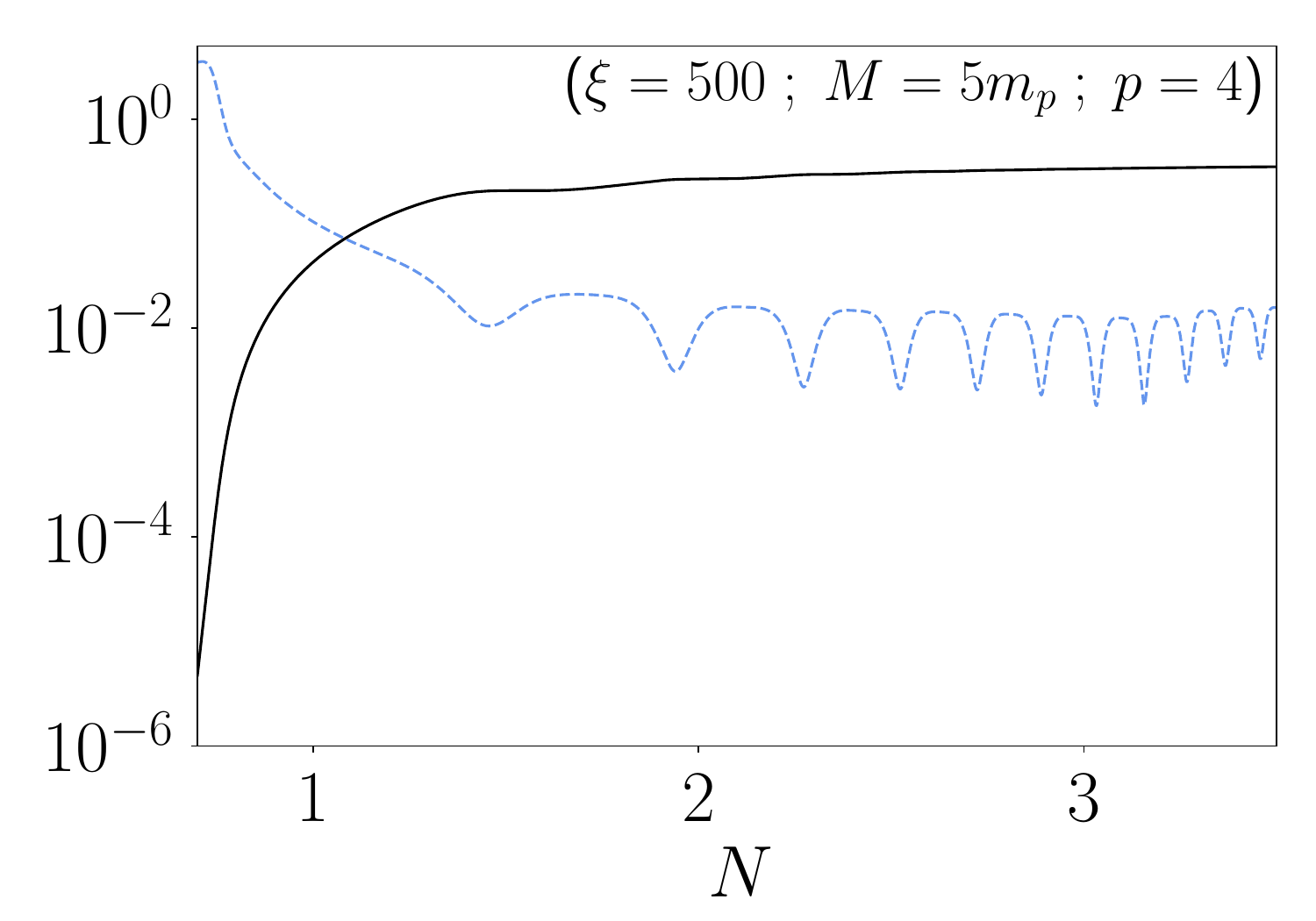} 
    \includegraphics[width=0.45\textwidth,height=4cm]{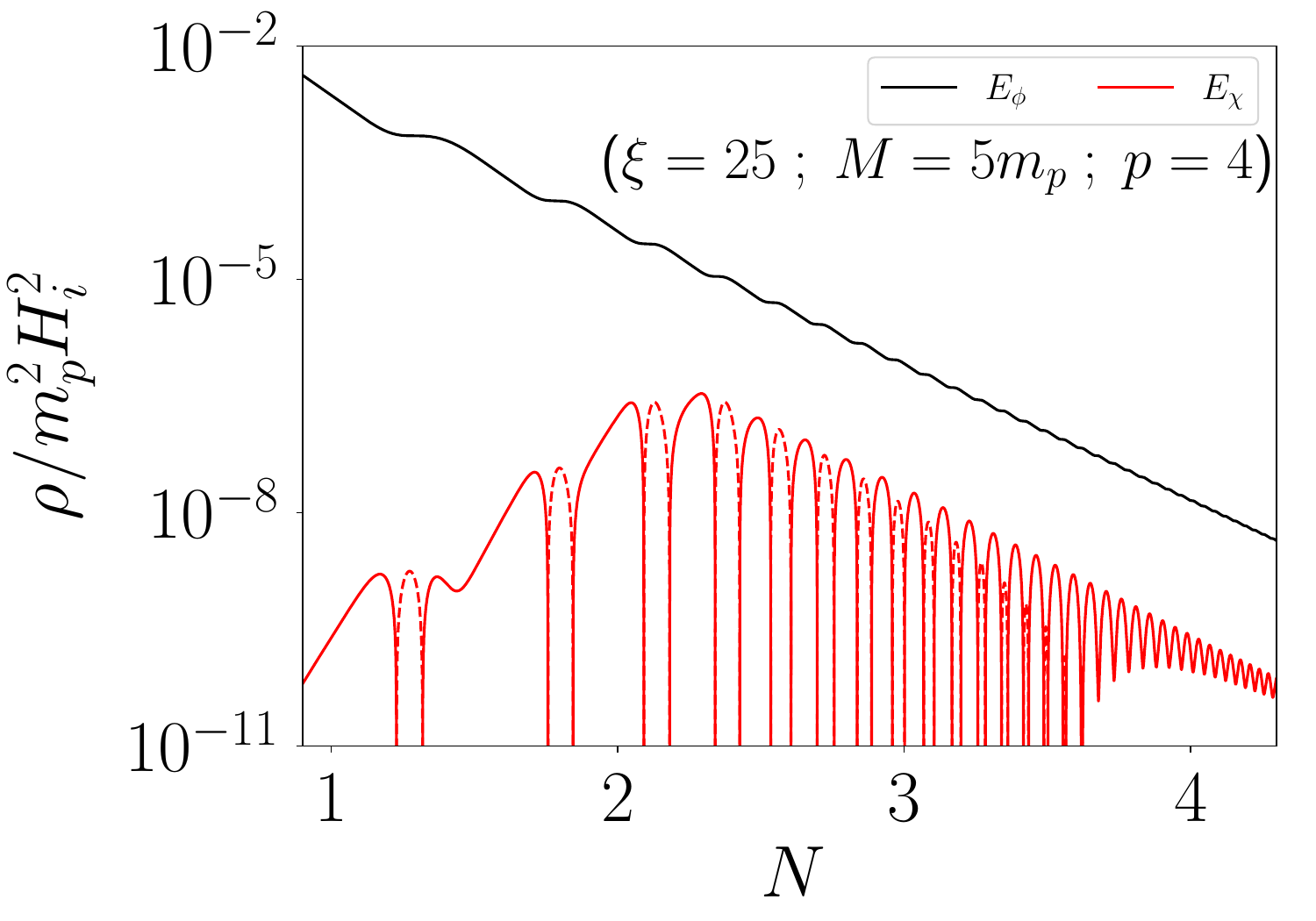} 
    \includegraphics[width=0.45\textwidth,height=4cm]{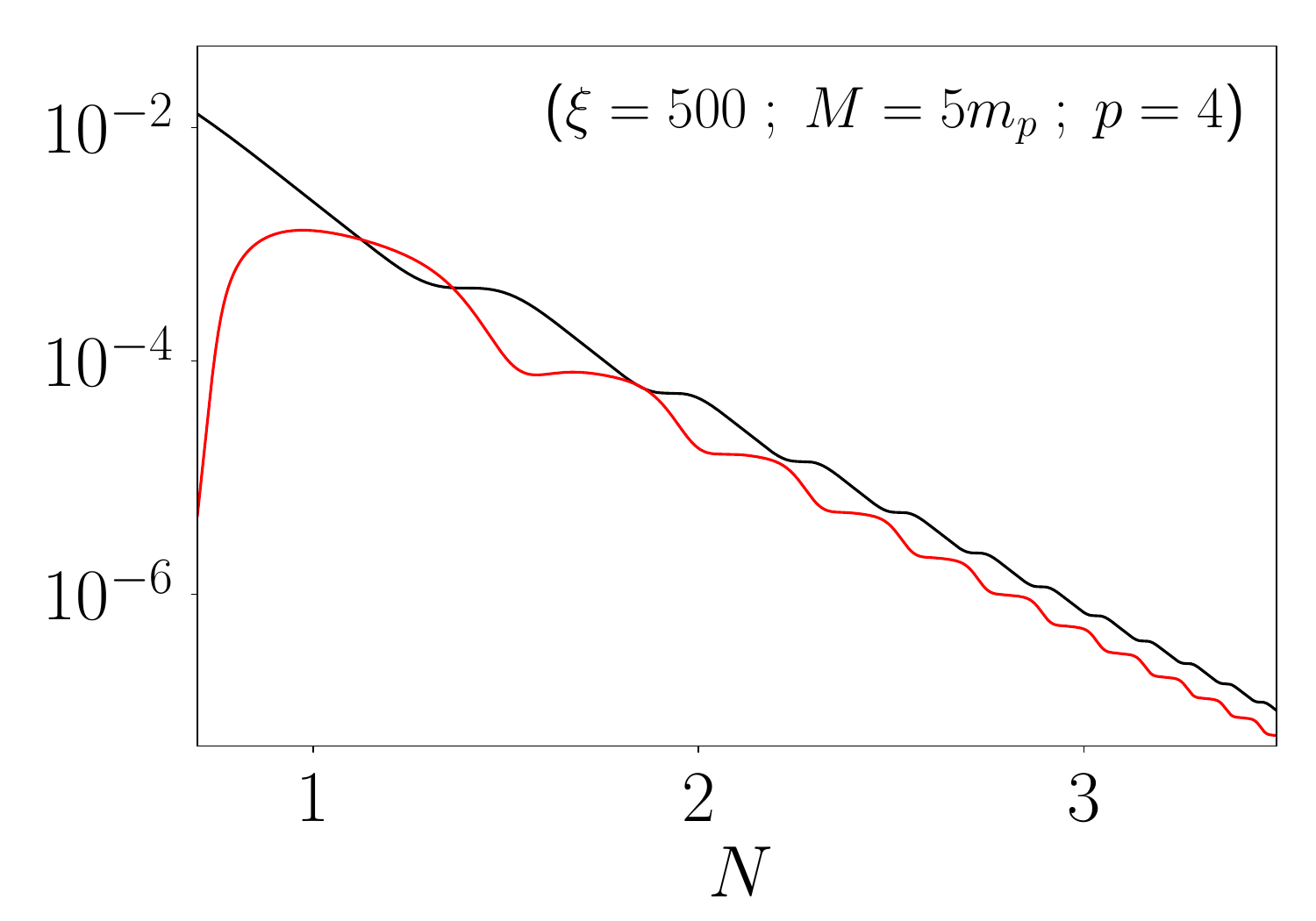} 
    \includegraphics[width=0.45\textwidth,height=4cm]{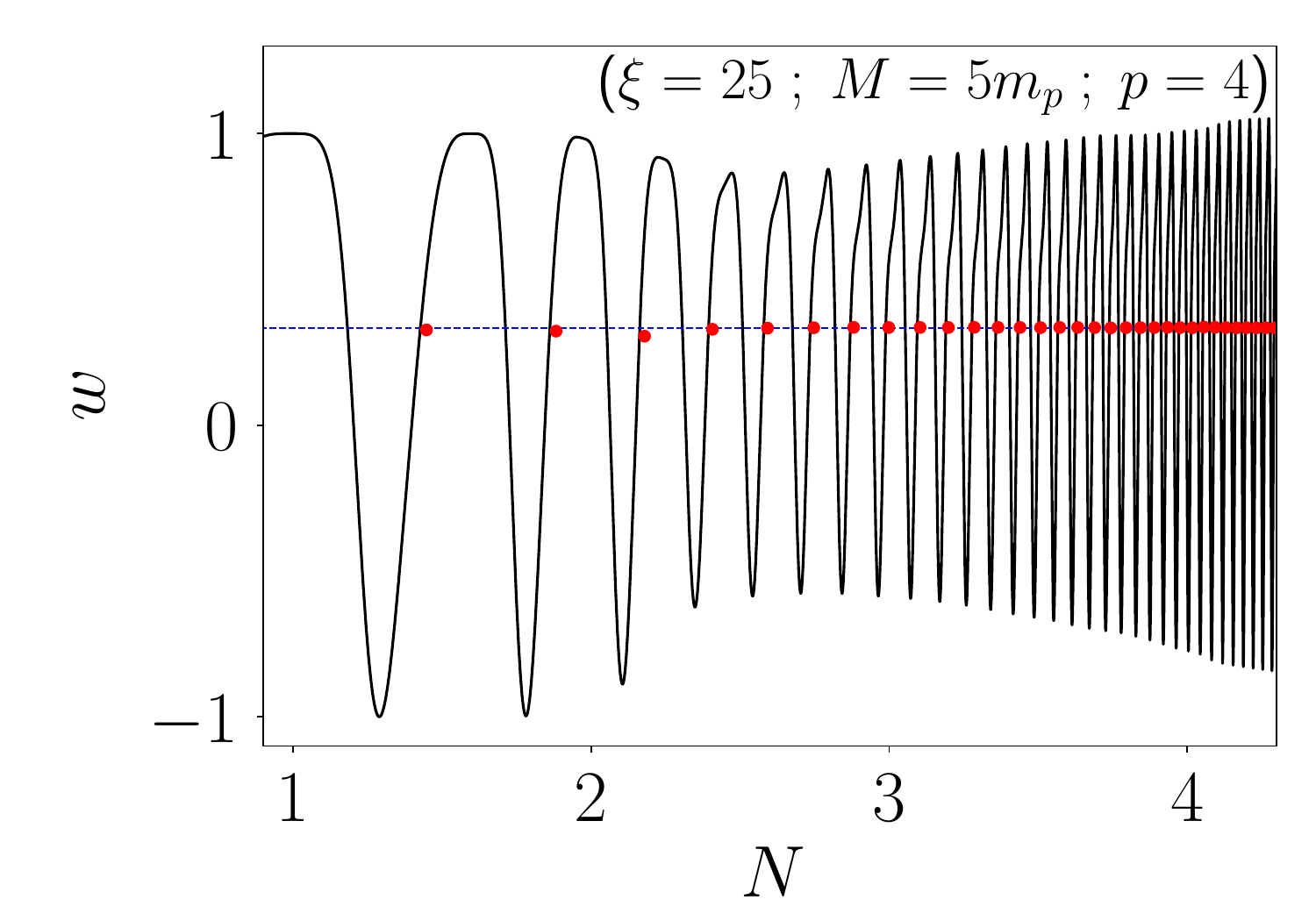} 
    \includegraphics[width=0.45\textwidth,height=4cm]{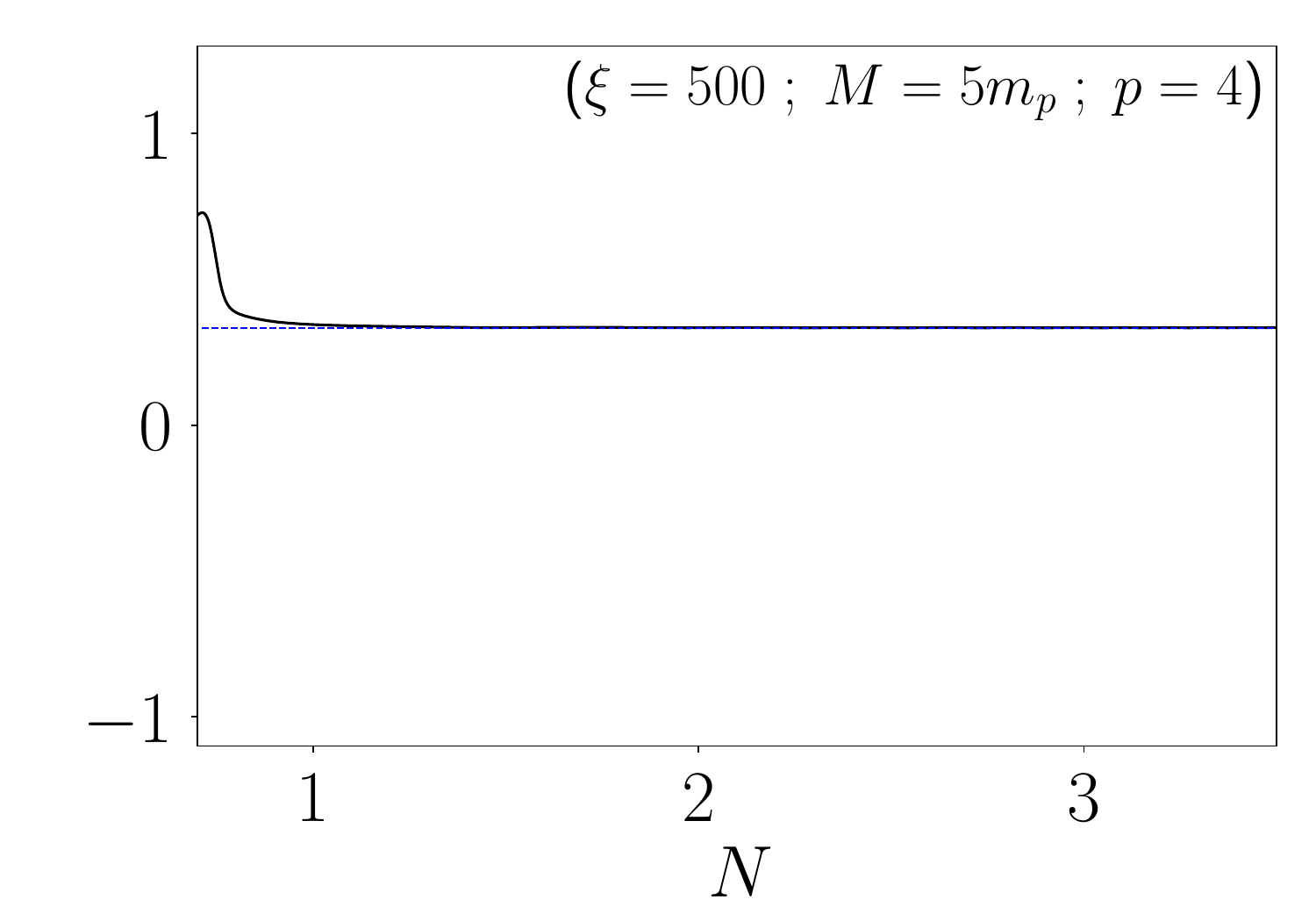} 
    \end{center}
    \vspace*{-0.5cm}
    \caption{Time evolution of various quantities for $p=4$,  $M=5m_p$, $\xi=25$ and (left panels) and $\xi=500$ (right panels). {\it Top Panel:} Evolution of the NMC field variance $\xi \langle \chi^2 \rangle/m_p^2$ and $R/H^2$. {\it Middle Panel}: Evolution of the total energy of the inflaton and of the NMC field. {\it Bottom Panel:} Evolution of the equation of state ($w$) computed from Eq.~\ref{eq:EoSfromRonH2}, blue line is at $w=1/3$ (Radiation Domination), and red dots depicting $w_{\rm osc}$ per oscillation, c.f.~Eq.~(\ref{eq:RvsW}). We note that in logarithmic plots solid lines stand for positive values and dashed lines for negative values.} \label{fig:PlotsM5p4} \vspace*{-0.3cm}
\end{figure*}

\subsection{Lattice results for $p=4$}

Here we review the relevant results on geometric (p)reheating for $p=4$, i.e.~when the inflaton potential is quartic around $\phi = 0$. In the top panels of Fig.~\ref{fig:PlotsM5p4}, we show the evolution of the variance of the NMC field and the Ricci scalar $R/H^2$. In the left-top panel we observe how the variance of the NMC fields grows whenever $R$ is negative, while it oscillates whenever it is positive. For a large scale model ($M=5m_p$) we see that for $\xi=25$, the growth of $\langle \chi^2 \rangle$ eventually stops at $N_{\rm max} \simeq 2.2$ efolds after inflation, during the linear regime. This happens because during a negative semi-oscillation of $R$, the expansion of the universe dilutes the field faster than the tachyonic resonance is capable of making it grow. This is expected for small couplings. However, for couplings as large as $\xi=500$, we see that the variance rapidly grows (in the first negative semi-oscillation of $R$) and saturates to a value $\xi\langle\chi\rangle^2/m_p^2 \simeq 1$. At this point the backreaction of the NMC field onto the background has a major effect, which is taking the value of $R/H^2$ close to zero. This can be understood by analyzing the dependence of $R$ on the different field terms, c.f.~Eq.~\eqref{eq:RiccifromMatter}. Namely, the growth of $\langle \chi^2 \rangle$ in the denominator is the main reason for $R$ to decrease, while the terms in the numerator determine the sign of $R$. In the top-left panel of Fig.~\ref{fig:PlotsM5p4} we observe how $R/H^2$ decreases as $\langle\chi^2\rangle$ grows, and once the variance saturates, then the value of $R/H^2$ remains negative, as determined by the fact that the NMC field's  gradient energy dominates over its kinetic energy. In the middle panels we show the evolution of the field energies. For the case $\xi=25$, we see how the energy of $\chi$ grows coinciding with the growth of the variance $\langle\chi^2\rangle$. The energy of the NMC field becomes negative at some point, coinciding with $R$ being positive, which causes a rapid flip in the sign of the NMC velocity. During this stage the energy of the NMC field is dominated by the term $\propto \langle \chi \chi'\rangle$, a situation that lasts for a short time, and later on the field oscillates as usual. In the right panels of the same figure, we see that for $\xi=500$, the energy of the NMC field grows to a maximal value within the first negative semi-oscillation of $R$, and at this point already half of the total energy is stored in the NMC field. The NMC field becomes a massless free field as $R$ decays due to the backreaction, and hence behaves as radiation. This is the reason why the energy density of both the inflaton and the NMC fields decay at the same rate\footnote{We recall that for $p = 4$, the energy density of the inflaton decays as radiation as well (at least once the inflaton is confined to oscillate within the positive curvature part of its potential)}. In the bottom panels we show the equation of state (EoS) as computed from Eq.~(\ref{eq:EoSfromRonH2}). For $\xi=25$ the EoS starts oscillating between $1$ and $-1$, with an average value of $1/3$, and eventually the backreaction squeezes the oscillations' envelope, while maintaining its average value. Backreaction is not sufficiently strong to send $R$ to a small value and by the time tachyonic resonance has become inefficient, the EoS turns back to oscillate between $1$ and $-1$, with average value $1/3$. For $\xi=500$, backreaction happens so fast, within the first semi-oscillation of $R$, that $w$ does not have time to oscillate and it goes directly to the fixed value $w=1/3$. 

\begin{figure*}[tbp]
    \begin{center}
    \includegraphics[width=0.55\textwidth,height=5.0cm]{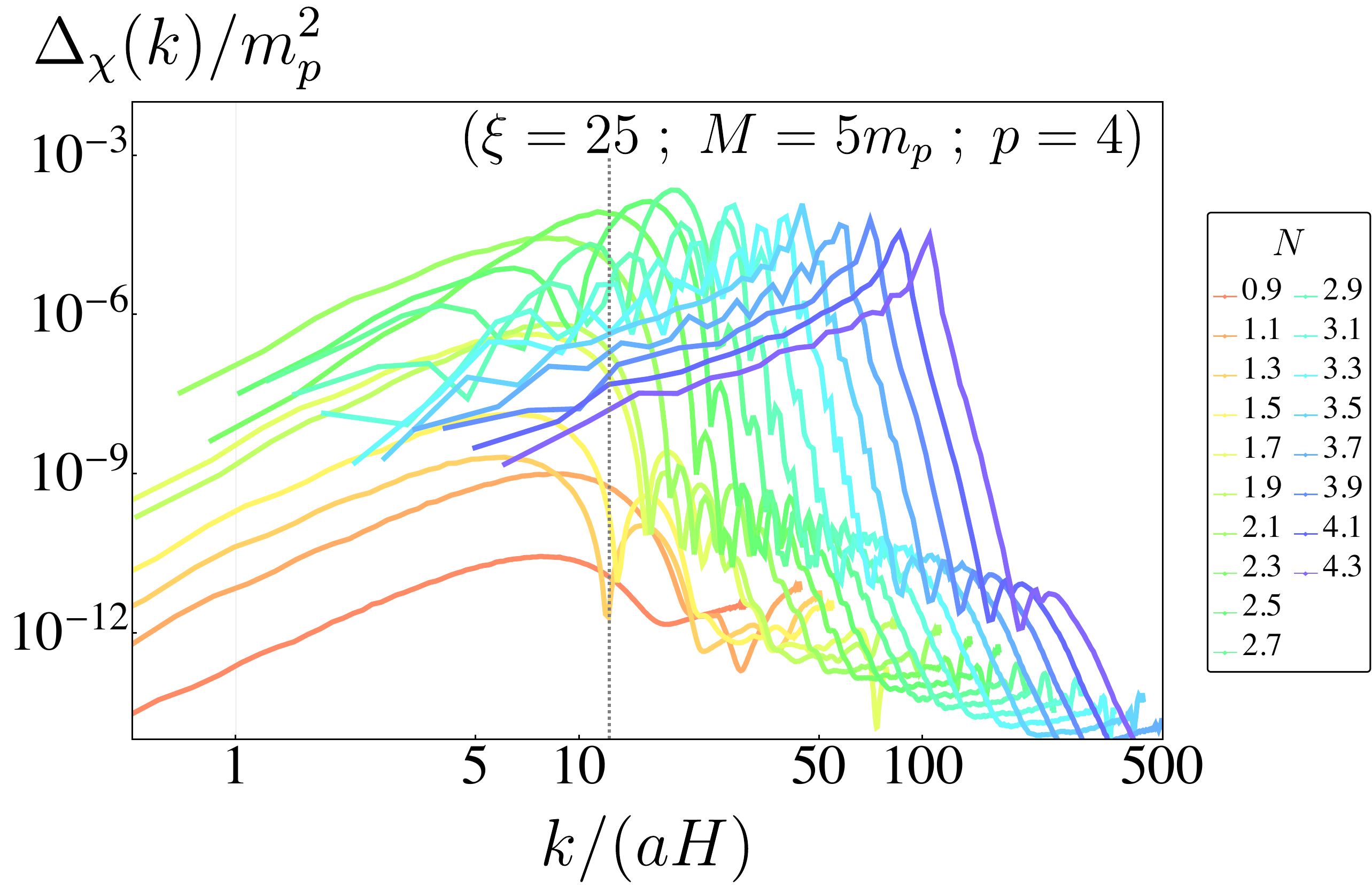}
    \includegraphics[width=0.55\textwidth,height=5.0cm]{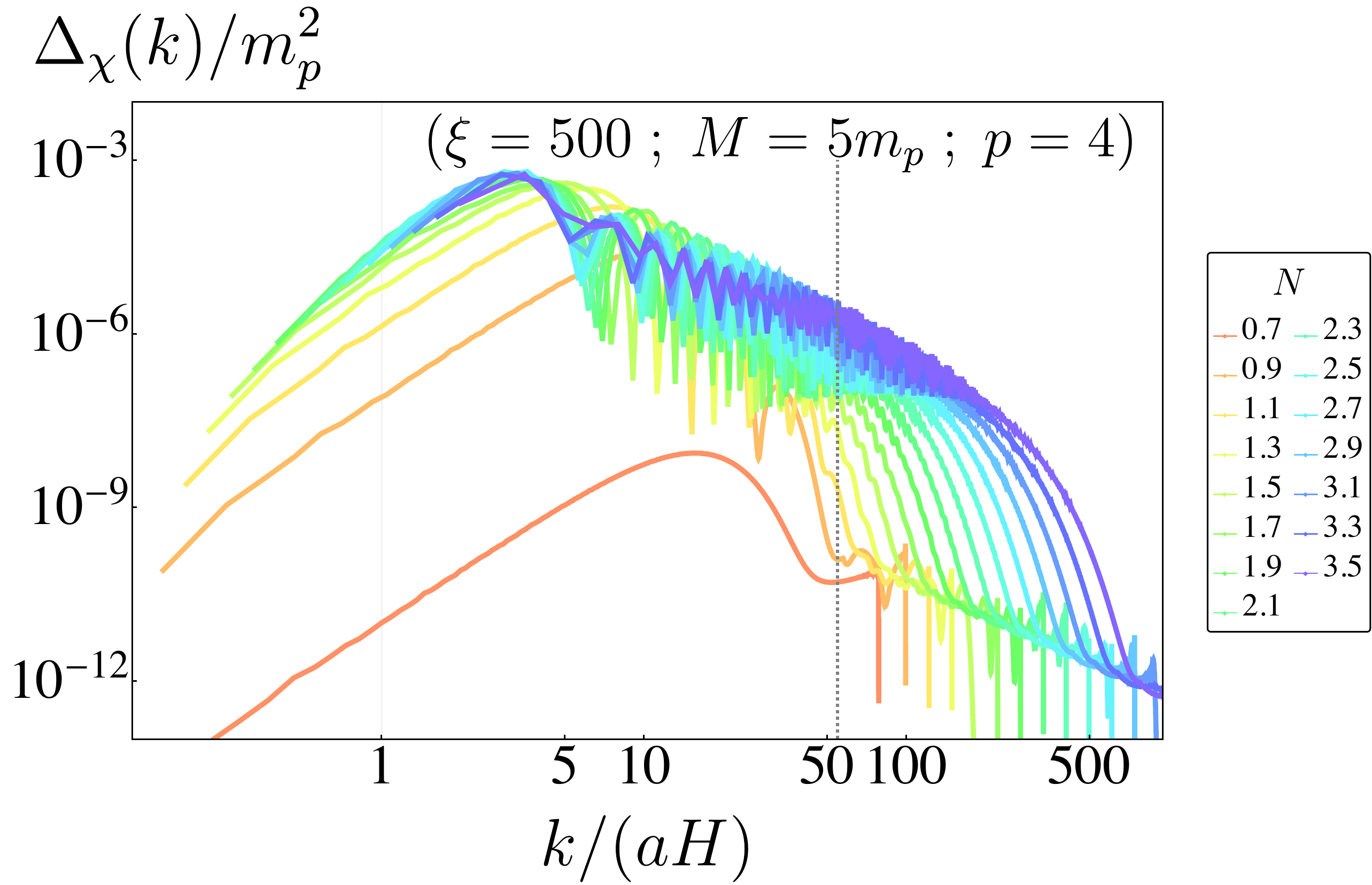} 
    \end{center}
    \vspace*{-0.5cm}
    \caption{Power spectrum of NMC field versus $k/aH$ for $M=5m_p$ and $p=4$. {\it Left:} $\xi=25$ and {\it Right:} $\xi=500$. Dashed line shows threshold scale $k_*/aH = \sqrt{6\xi}$} \label{fig:MatterspectraM5p4} \vspace*{-0.3cm}
\end{figure*}

In Fig.~\ref{fig:MatterspectraM5p4} we see power spectra for a large scale scenario with $M=5m_p$, $p=4$, and two different couplings, $\xi=25$ and $\xi=500$. We see that for $\xi=25$ the initially growing spectrum of $\chi$ is peaked at a fixed scale $k_*/(aH) \lesssim 8$. Whenever $R/H^2$ is maximally negative, the modes below the cutoff scale $k \lesssim k_*$ grow, while whenever $R/H^2$ is positive, modes oscillate. The power spectrum of $\chi$ grows until the expansion of the universe shuts down the tachyonic resonance during the linear regime, and from this point onward the power spectrum range simply redshifts, while its amplitude decays. The large coupling case $\xi=500$ shows, on the contrary, how the resonance is ended by $\chi$'s own backreaction on the Ricci scalar. First, we note that the NMC field spectrum grows while peaking at the scale $k_*/(aH) \lesssim  30$, but contrary to the weak coupling case, the expansion of the universe does not shut down the resonance now. Instead, the backreaction on the Ricci scalar happens so fast, that within its first semi oscillation $R$ is damped to a small values. This in turn forces a shift of the NMC power spectrum towards larges scales, as noted in the bottom panel, where we observe how the spectra drifts towards the Hubble scale. This is a consequence of the fact that the tachyonic threshold scale is proportional to the Ricci scalar as $k_* \propto \sqrt{|R|}$. When the smallest excited mode leaves the lattice, our simulation cannot be trusted anymore, so we stop the simulation typically slightly before this, when the infrared tail of the spectrum still has some extension left to the peak $(k_*/aH)$ itself.

\begin{figure*}[tbp]
    \begin{center}
     \includegraphics[width=0.32\textwidth,height=3.2cm]{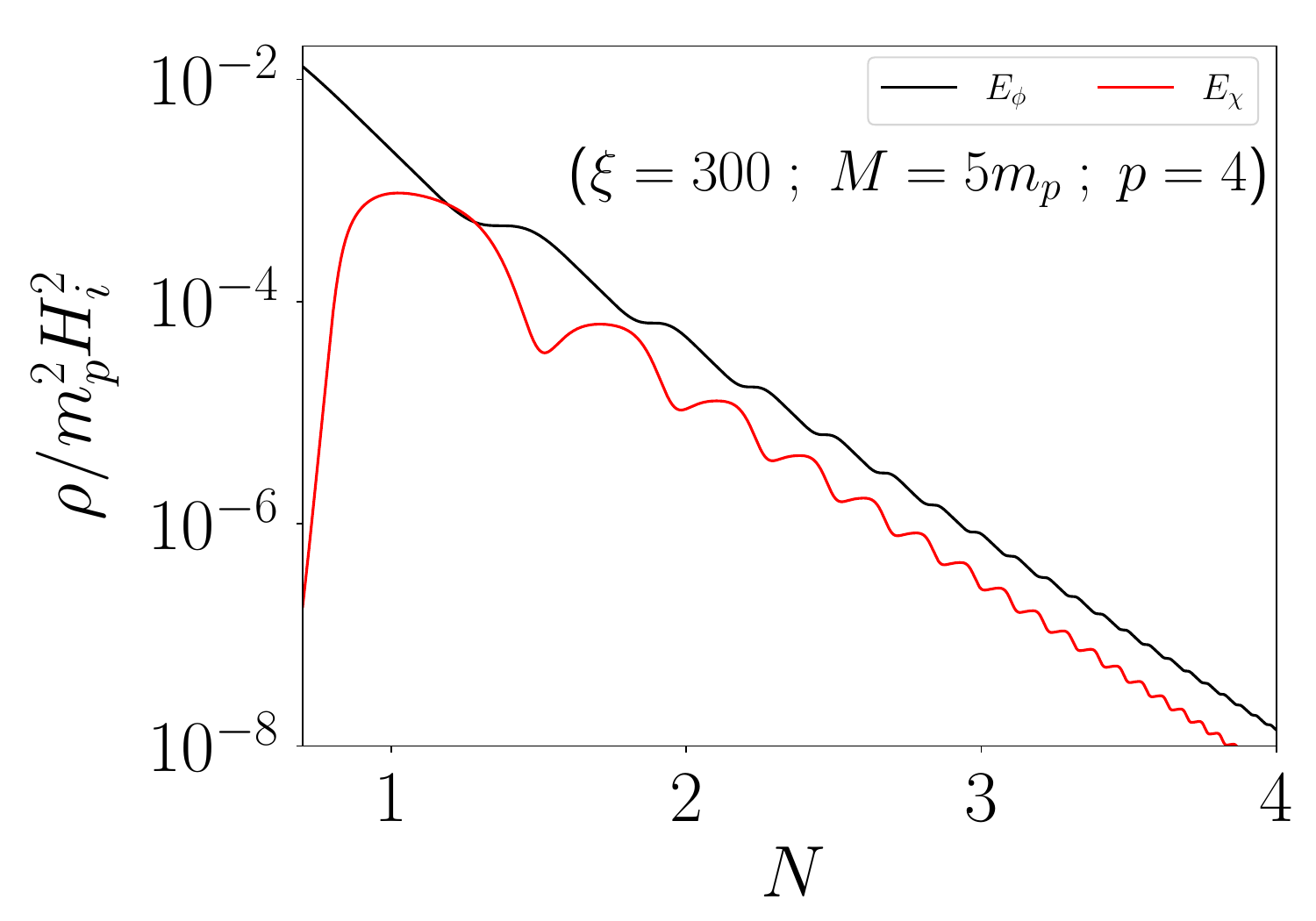} 
    \includegraphics[width=0.32\textwidth,height=3.2cm]{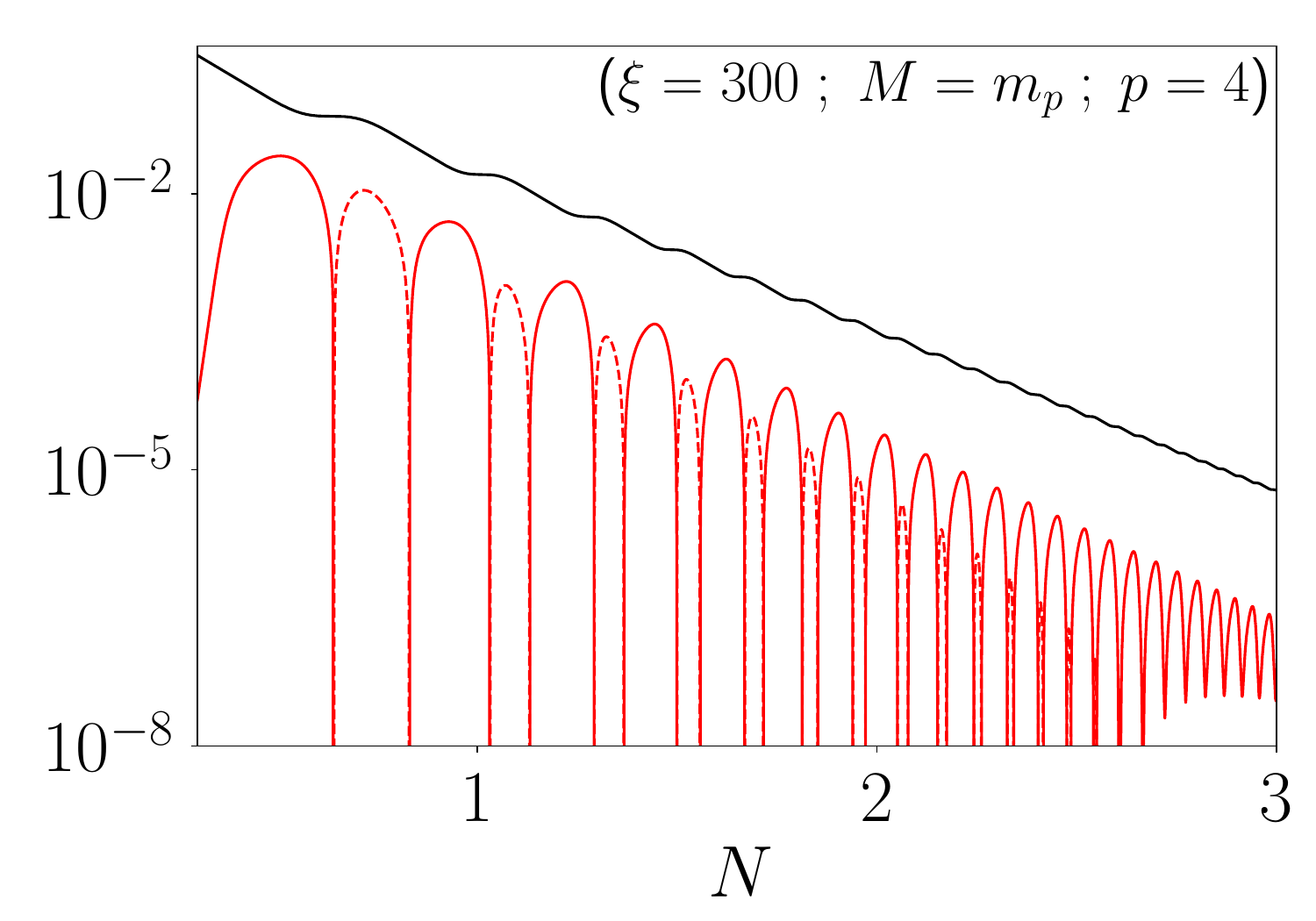} 
    \includegraphics[width=0.32\textwidth,height=3.2cm]{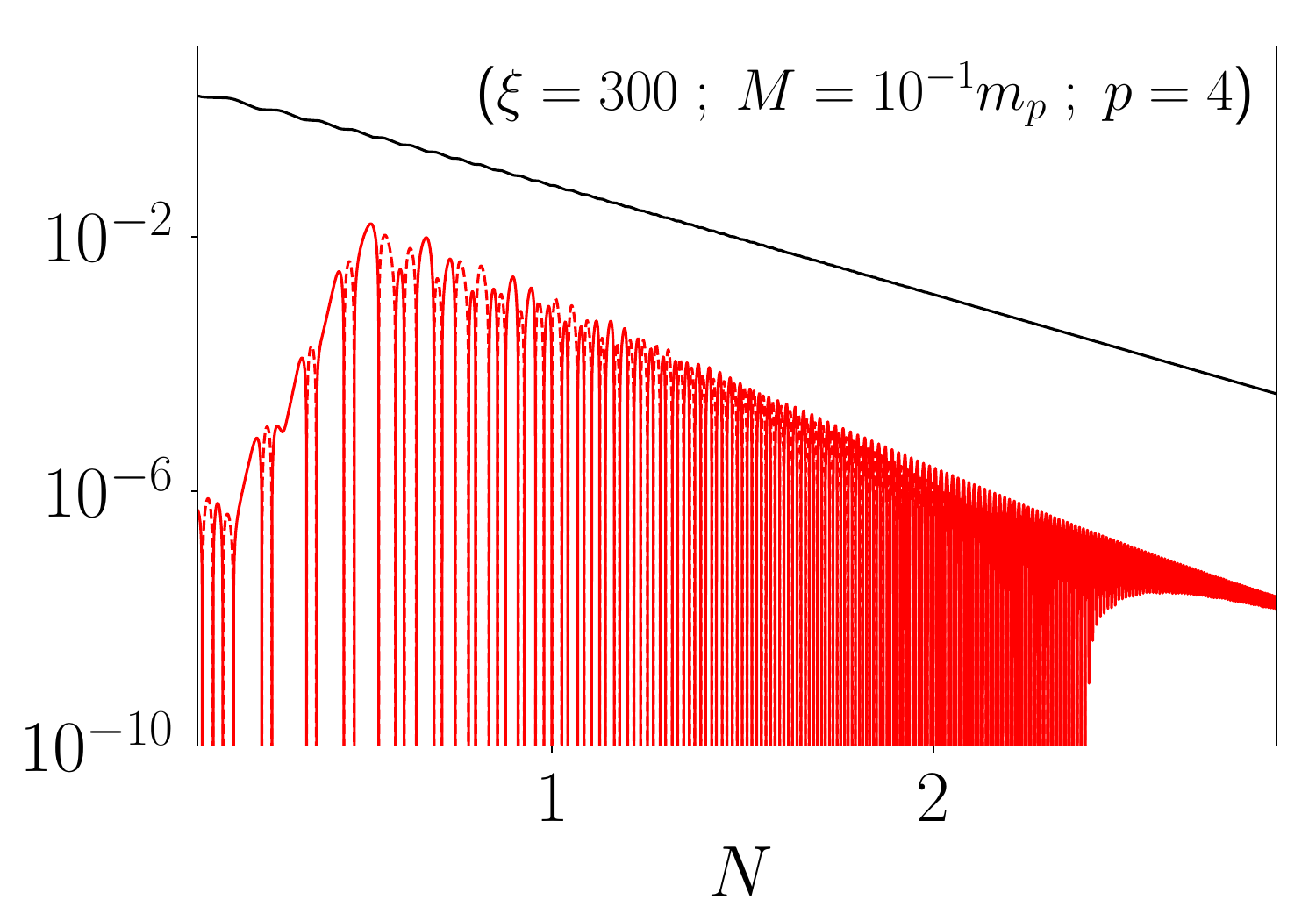}
     \includegraphics[width=0.32\textwidth,height=3.2cm]{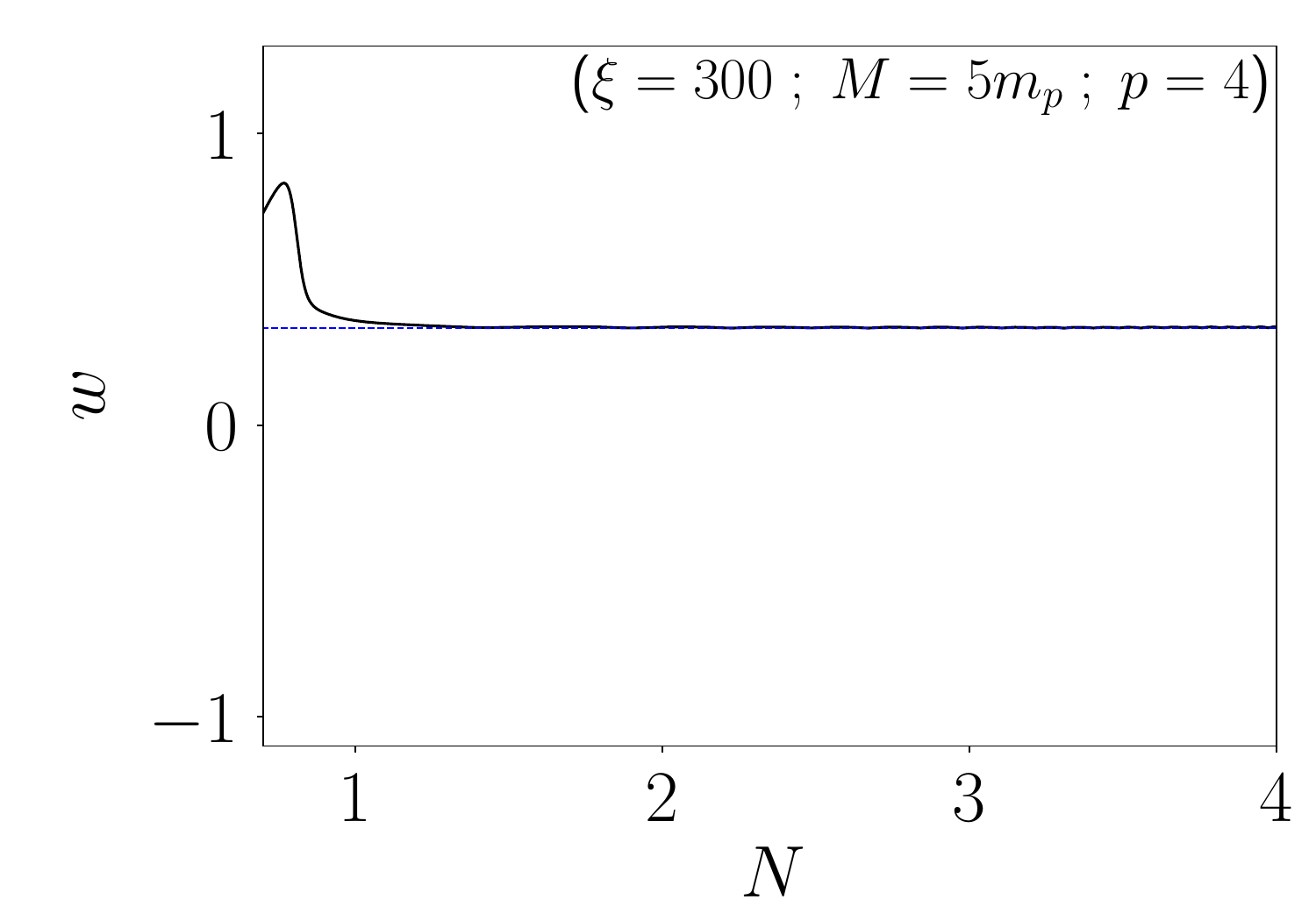} 
    \includegraphics[width=0.32\textwidth,height=3.2cm]{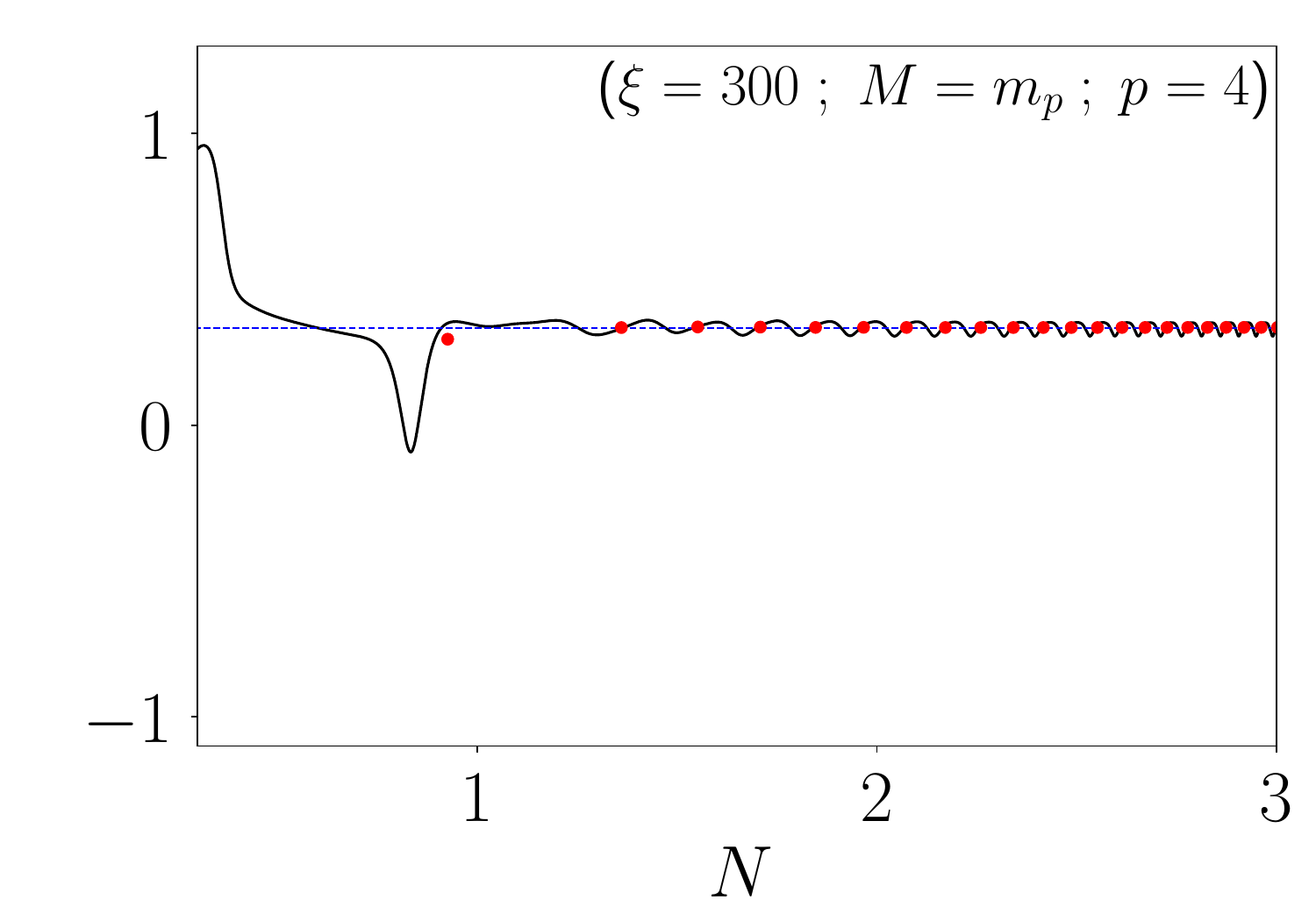} 
    \includegraphics[width=0.32\textwidth,height=3.2cm]{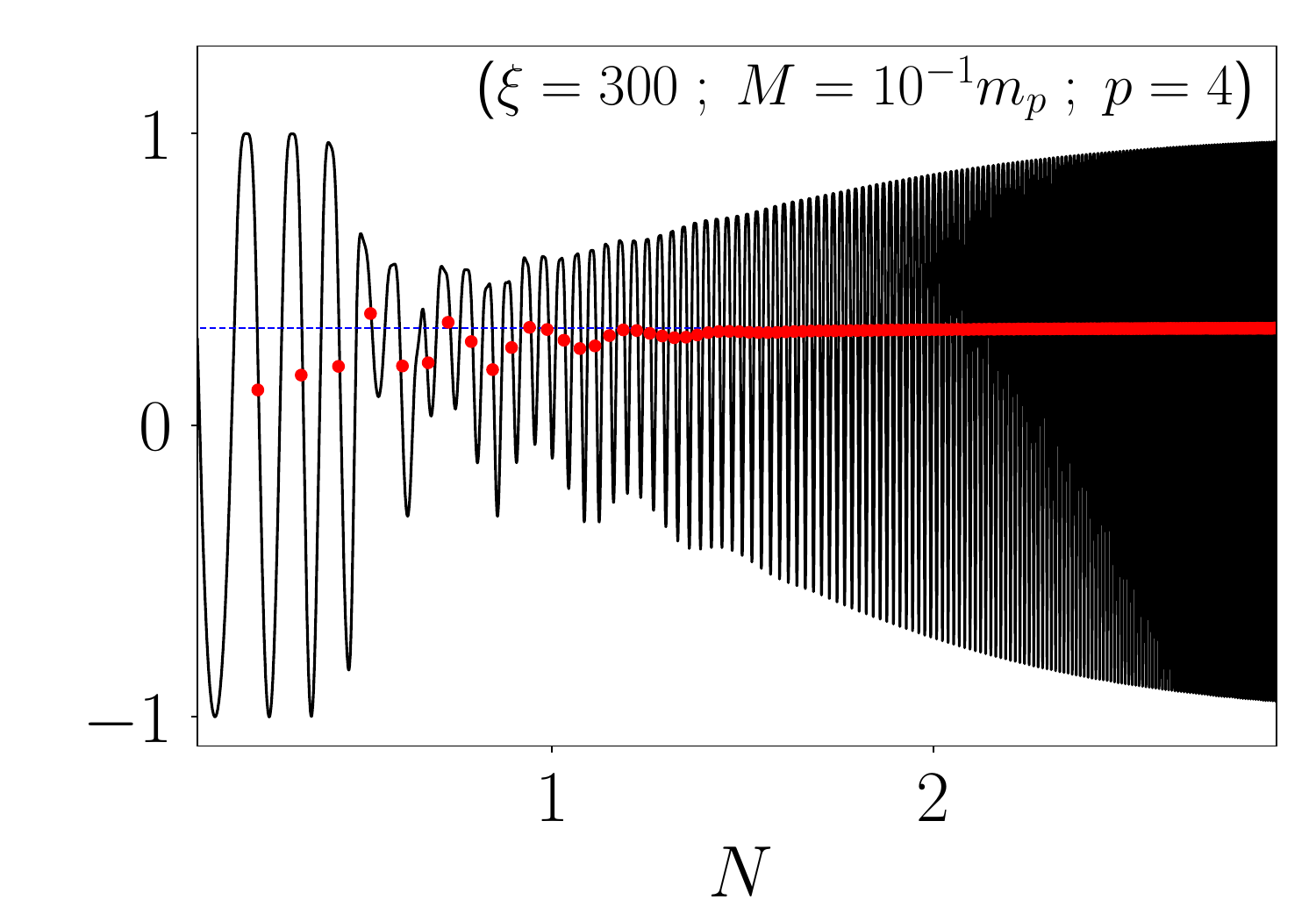}
    \end{center}
    \vspace*{-0.5cm}
    \caption{Energy densities and equation of state (EoS) for $p = 4$ and fixed coupling $\xi=300$, with $M=5m_p$ (left panels), $M=m_p$ (central panels) and $M=0.1m_p$ (right panels). {\it Top:} Evolution of the total energy density of the inflaton and of the NMC field. Solid lines denote positive values and dashed lines negative values. {\it Bottom:} Evolution of the EoS  $w$ computed from Eq.~(\ref{eq:EoSfromRonH2}), where the blue solid line at $w=1/3$ denotes RD, and the red dots depict the effective value $w_{\rm osc}$ computed from Eq.~(\ref{eq:RvsW}).} \label{fig:Plotsxi300p4} \vspace*{-0.3cm}
\end{figure*}

We now focus on the dependency of energy transfer on $M$. We basically consider two variables in order to quantify the effectiveness of reheating: the fraction of energy transferred into the NMC field, and the time scale to reach an equation of state as $\omega \simeq 1/3$ (i.e.~radiation domination). If the NMC field energy accounts for at least 90\% of the total energy of the two-field system, we will consider the Universe to have been properly reheated. In Fig.~\ref{fig:Plotsxi300p4} we show in a series of panels with decreasing $M$, the evolution of the energy components and of the equation of state (EoS) for a fixed coupling $\xi=300$. For the large scale case $M=5m_p$ (left panels), almost half of the energy of the inflaton has been transferred into the NMC field by $N_{\rm max} \lesssim 1$, and since the NMC field and the inflaton both behave as radiation in this case, the EoS settles down, from then on, to the value $w=1/3$. In the central panels we consider the case $M=m_p$. We first note that reheating is not fully achieved, as the NMC field never reaches more than approximately $5\%$ of the total energy budget. While the initial dynamics $\chi$ makes the EoS to oscillate, it eventually settles down to a fixed value $w=1/3$, as dictated by the dominant inflaton which scales as radiation. We note that as the energy of the NMC field remains subdominant and it is still affected by its interaction with $R$, $\chi$ does not behave as radiation, and the oscillation envelope of its energy density decays faster than $1/a^{4}$. The right panels consider finally the case for $M=10^{-1}m_p$. Reheating is then never achieved and the tachyonic resonance is ended in the linear regime, when the energy of the NMC field accounts for less than 1~\% of the total budget. The effective EoS starts as $w_{\rm osc} < 1/3$, which is characteristic of low scale models, as the first inflaton oscillations are dominated by the potential while the field samples values above the inflection point. The effective EoS eventually saturates to the value $w_{\rm osc}=1/3$, as expected, when oscillations become confined only to the (positive curvature) part of the potential $V_{\rm inf} \propto \phi^4$.  

The energy transfer for $p  =4$ looses efficiency as the scale $M$ decreases. The reason lies on two facts: 

$i)$ lowering the scale $M$ forces the inflaton to end inflation above the inflection point of the potential. This means that the first oscillations will be mostly dominated by the potential energy, resulting in the Ricci scalar being mostly positive. 

$ii)$ lowering the scale $M$ makes the inflaton to oscillate faster as compared to the Hubble rate. This means that the negative semi-oscillations of $R$ lasts for very short time intervals, which makes the tachyonic growth inefficient compared to the dilution due to the expansion of the universe. 

\begin{figure*}[tbp]
\begin{center}
\includegraphics[width=0.6\textwidth,height=5.5cm]{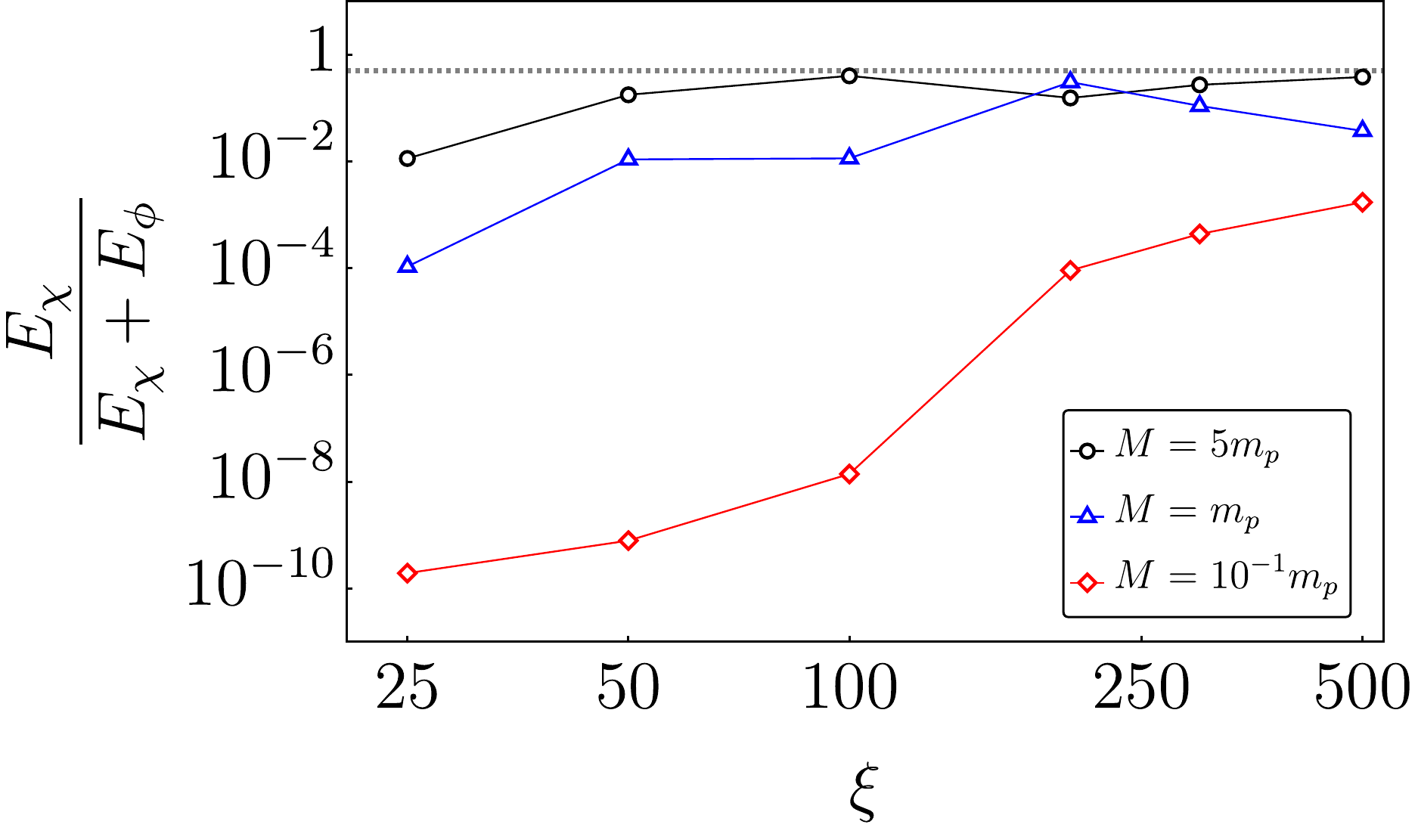} 
    \end{center}
    \vspace*{-0.5cm}
    \caption{Final energy fraction of the NMC field compared to the total energy, for $p=4$. The horizontal grey line denotes $E_{\chi} = E_{\phi}$.} \label{fig:ENMCp4plot} \vspace*{-0.3cm}
\end{figure*}

In Fig.~\ref{fig:ENMCp4plot} we plot the fraction of energy of the NMC field compared to the total energy. The parameter dependence of the energy fraction of the NMC field shows the two effects described above. Raising the coupling $\xi$ makes more effective the transfer of energy and, eventually, for large enough couplings, energy can be equally distributed between both fields, achieving a partial reheating of the Universe. Decreasing the mass scale $M$ affects the efficiency of the energy transfer during preheating to the point that for the largest coupling considered, $\xi=500$, and only a moderately small mass scale, $M=10^{-1}m_p$, the energy fraction of the NMC field already accounts only for 0.1~\% of the total energy.

\subsection{Lattice results for $p=6$}
\label{subsec:p6}

\begin{figure*}[tbp]
    \begin{center}
     \includegraphics[width=0.32\textwidth,height=3.5cm]{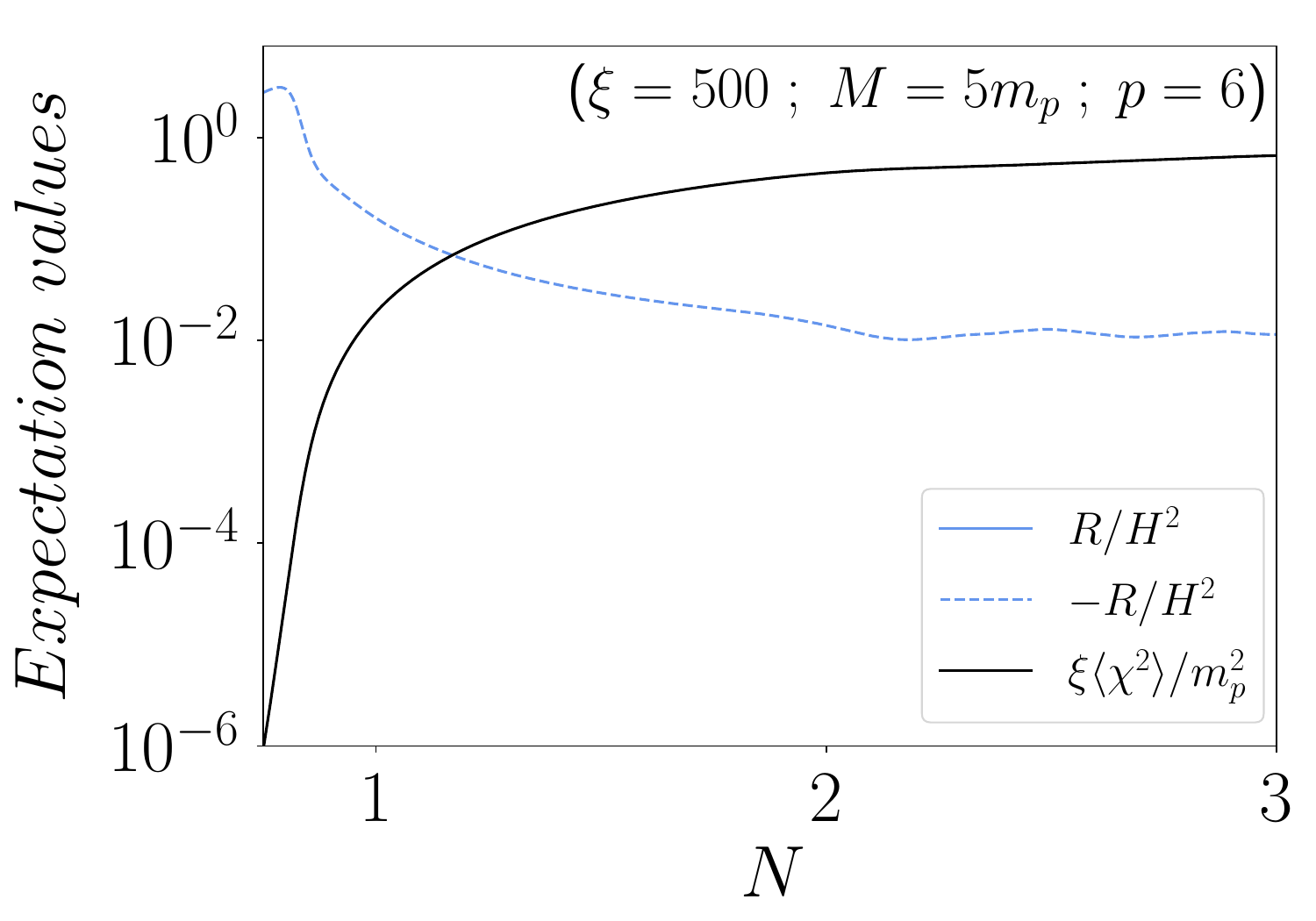} 
    \includegraphics[width=0.32\textwidth,height=3.5cm]{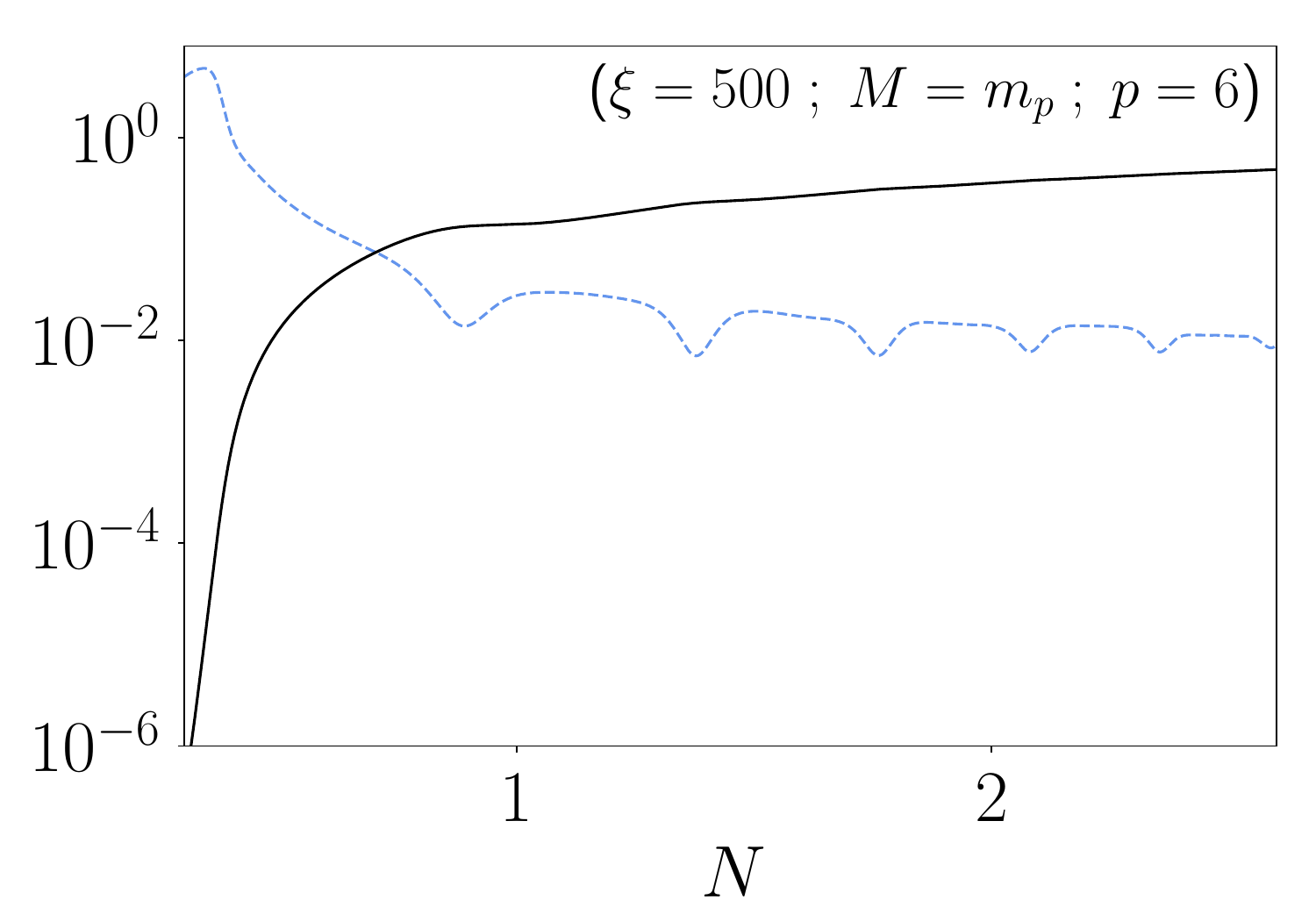} 
    \includegraphics[width=0.32\textwidth,height=3.5cm]{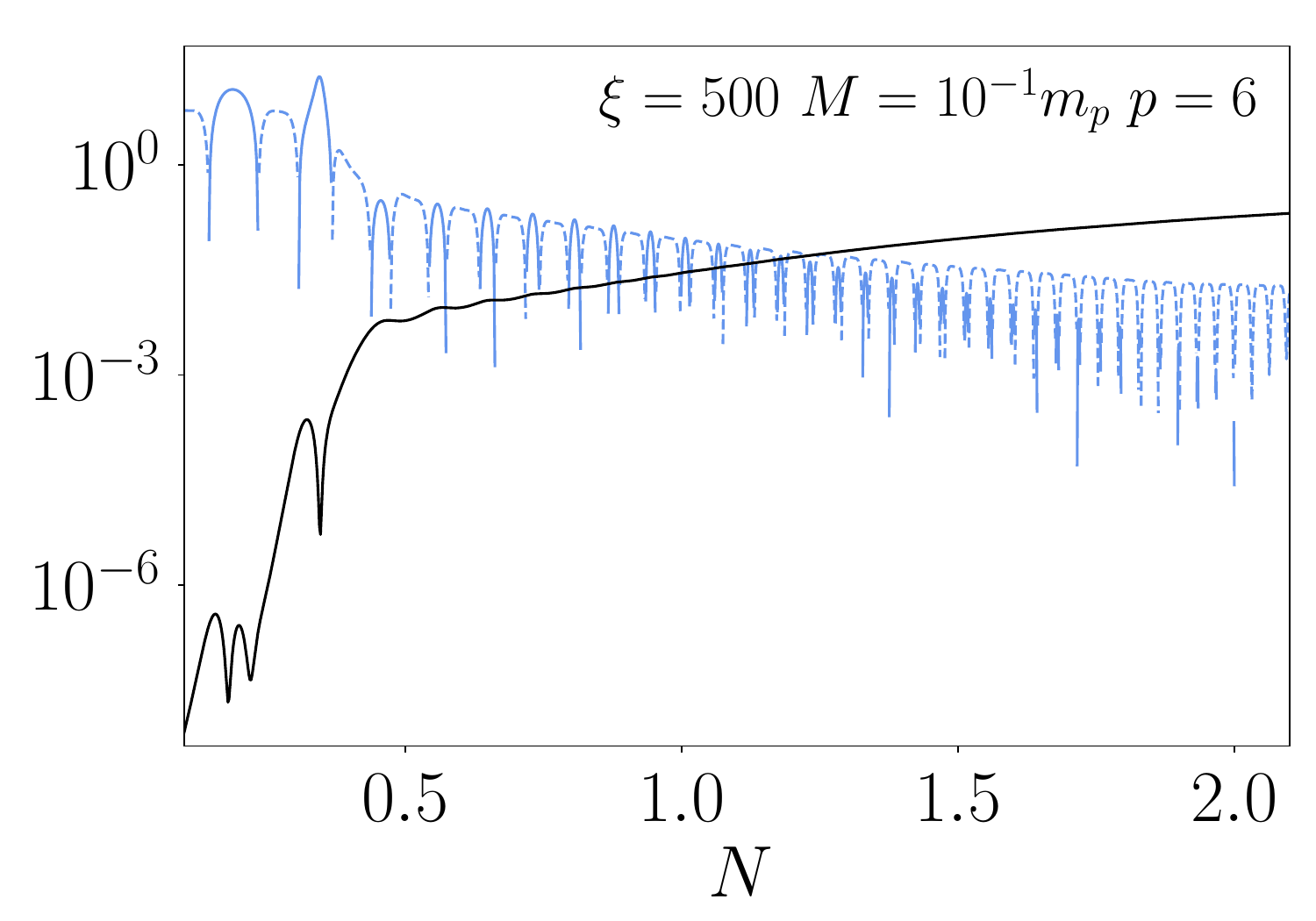}
     \includegraphics[width=0.35\textwidth,height=3.5cm]{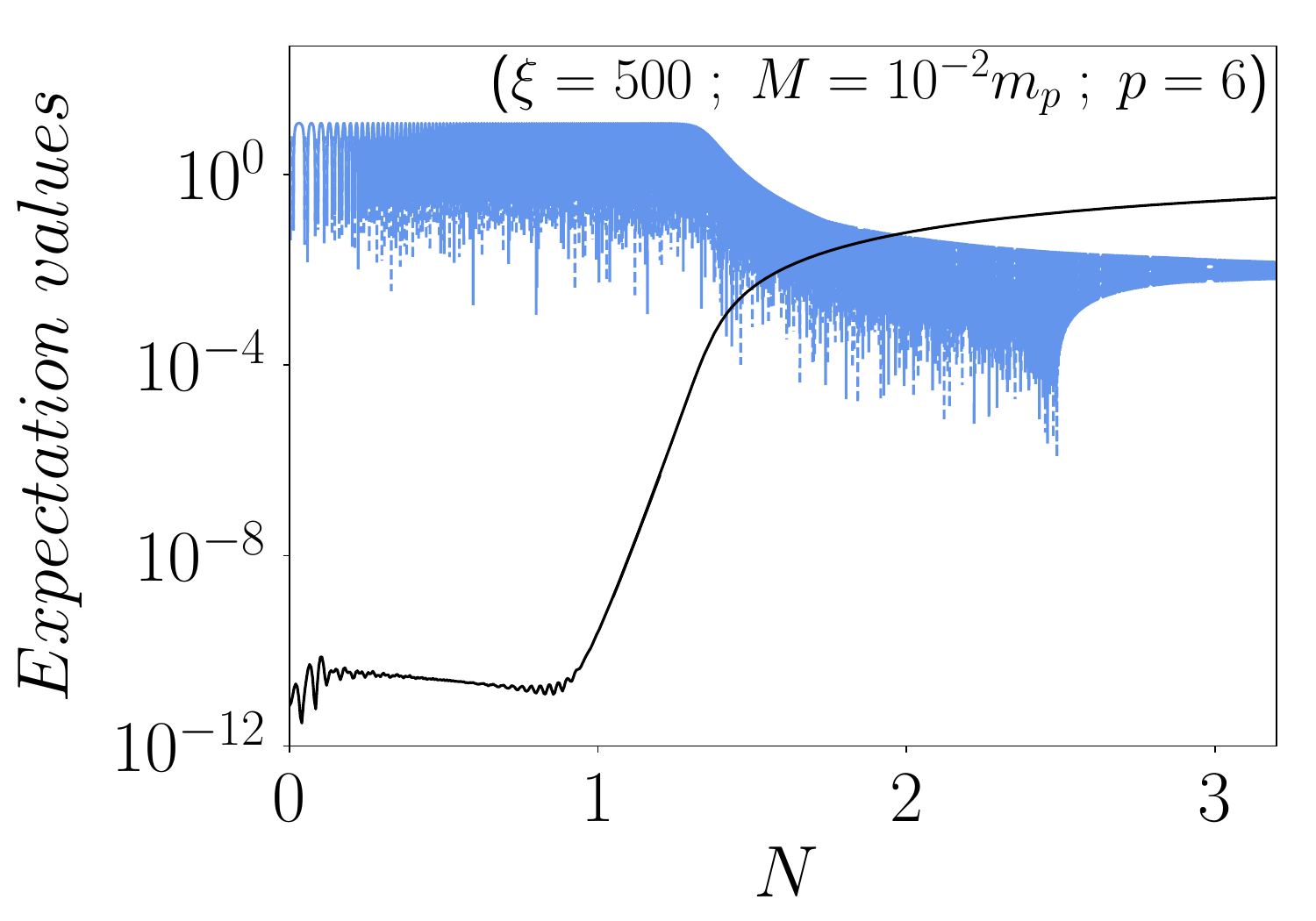} 
    \includegraphics[width=0.35\textwidth,height=3.5cm]{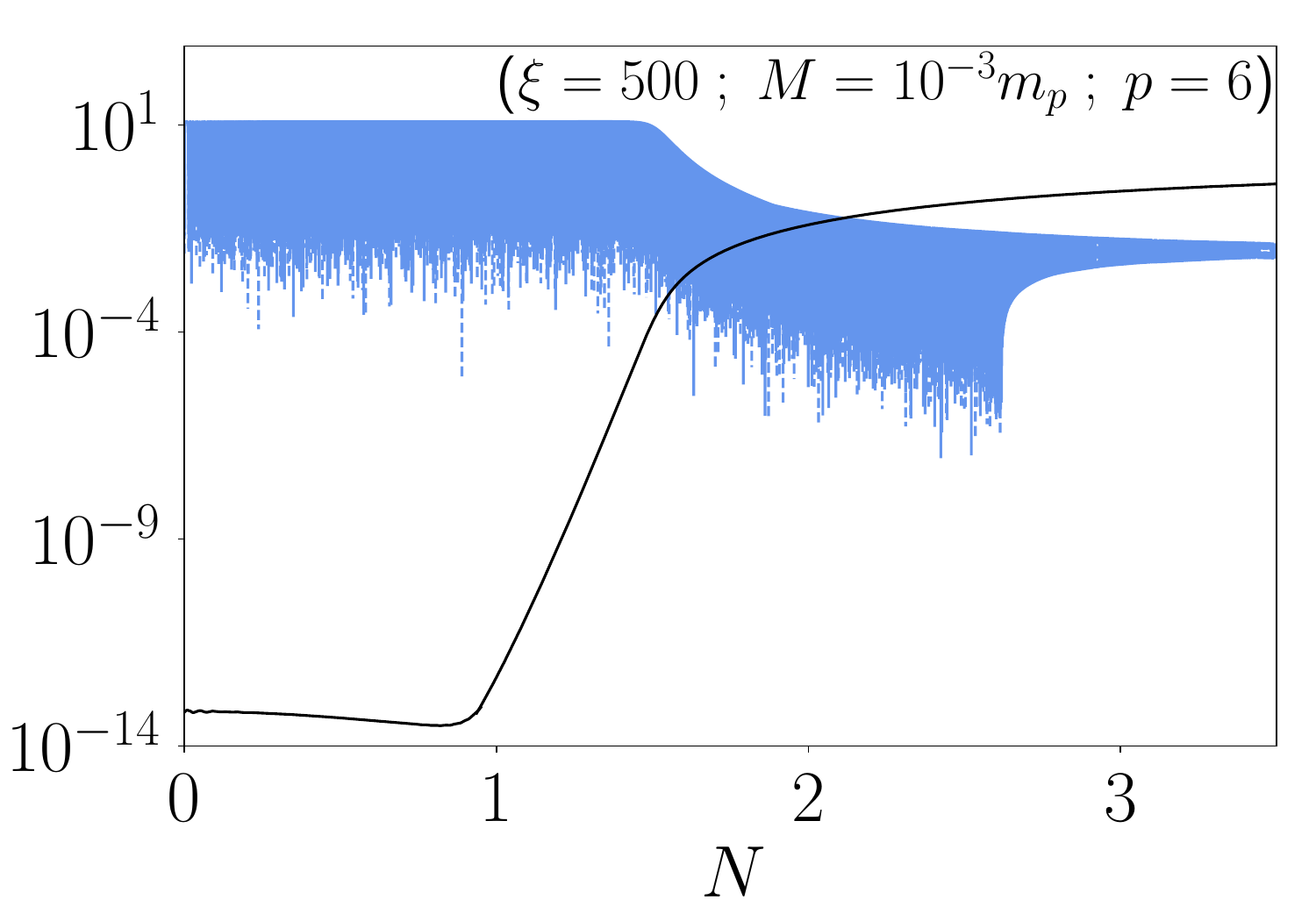} 
    \end{center}
    \vspace*{-0.5cm}
    \caption{Time evolution of the variance of the NMC field times the coupling, $\xi \langle \chi^2 \rangle/m_p^2$, and $R/H^2$ for $\xi=500$ and (from left to right) $M=5\,m_p$, $M=m_p$, $M=10^{-1}\,m_p$, $M=10^{-2}\,m_p$, $M=10^{-3}\,m_p$. Solid lines denote positive values and dashed lines negative values.} \label{fig:Plotsxi500p6M} \vspace*{-0.3cm}
\end{figure*}

In this section we review the most relevant results on geometric (p)reheating for potential~(\ref{eq:potential}) with $p = 6$. In Fig.~\ref{fig:Plotsxi500p6M} we show the time evolution of the NMC field variance $\xi\langle\chi^2\rangle/m_p^2$ for different values of $M$, together with the Hubble-normalized Ricci curvature $R/H^2$. We notice that for a given coupling, the field variance starts growing at different moments depending on $M$, and eventually saturates to $\xi\langle\chi^2\rangle/m_p^2 \lesssim 1$.  The backreaction onto the Ricci scalar becomes noticeable at some point in all cases, as $R/H^2$ starts decaying once the variance of $\chi$ grows enough. Remarkably,  this effect happens for all choices of $M$, i.e.~independently of how fast $R$ oscillates compared to the expansion of the universe. The delay on the initial growth of $\langle \chi^2 \rangle$ depends on the inflationary energy scale, with the delay being longer the smaller the value of $M$, see e.g.~bottom panels in Fig.~\ref{fig:Plotsxi500p6M} for $M/m_p=\{10^{-2},10^{-3}\}$. This is due to the initial oscillations of the inflaton sampling the negative curvature part of its potential, leading to an initially positive oscillation-averaged curvature $\overline R > 0$, recall Sect.~\ref{sec:ScaleofInflation}. Depending on the smallness of $M$ it takes longer to decay into the asymptotic value given by Eq.~\eqref{eq:EoSfromRonH2}, as it takes longer for the inflaton to reach an oscillation regime where only positive curvature parts of the potential are explored. Once the latter regime is established, the effective equation of state EoS settles down to  $\overline{w} = 0.5$, {\it c.f.~}Eq.~\eqref{eq:HomEoS} for $p=6$, which leads to an oscillation-averaged Ricci curvature as $\overline{R}/\overline{H^2}=-1.5$, {\it c.f.~}Eq.~\eqref{eq:EoSfromRonH2}. 

The novelty for $p = 6$ is that, contrary to $p \leq 4$, the averaged Ricci curvature becomes always eventually negative. The tachyonic instability is therefore guaranteed to be triggered at some point, the later the smaller the energy scale $M$, as long as the inflaton condensate remains homogeneous\footnote{We notice that for the time being we are ignoring inflaton fragmentation, see subsection~\ref{subsec:InflatonFragmentation}. Once we take it into account, homogeneity of the inflaton will be broken very rapidly, particularly for small mass scales $M$. For clarity of the discussion, we ignore first this effect, and later on we quantify how the achievement of proper reheating can be affected by this.}. Furthermore, as the energy of $\chi$ decays as radiation once the condition $\xi\langle\chi^2\rangle/m_p^2 \lesssim 1$ is reached, whereas the inflaton energy decays faster than radiation for $p = 6$, the field $\chi$ will eventually dominate the energy budget, with the contribution of $\phi$ becoming gradually more and more negligible. We consider that proper reheating has been achieved when $\rho_{\chi} = \rho_{\phi}$, understanding that soon enough the system will naturally flow to $\rho_{\chi} \gg  \rho_{\phi}$.

\begin{figure*}[tbp]
    \begin{center}
     \includegraphics[width=0.32\textwidth,height=3.5cm]{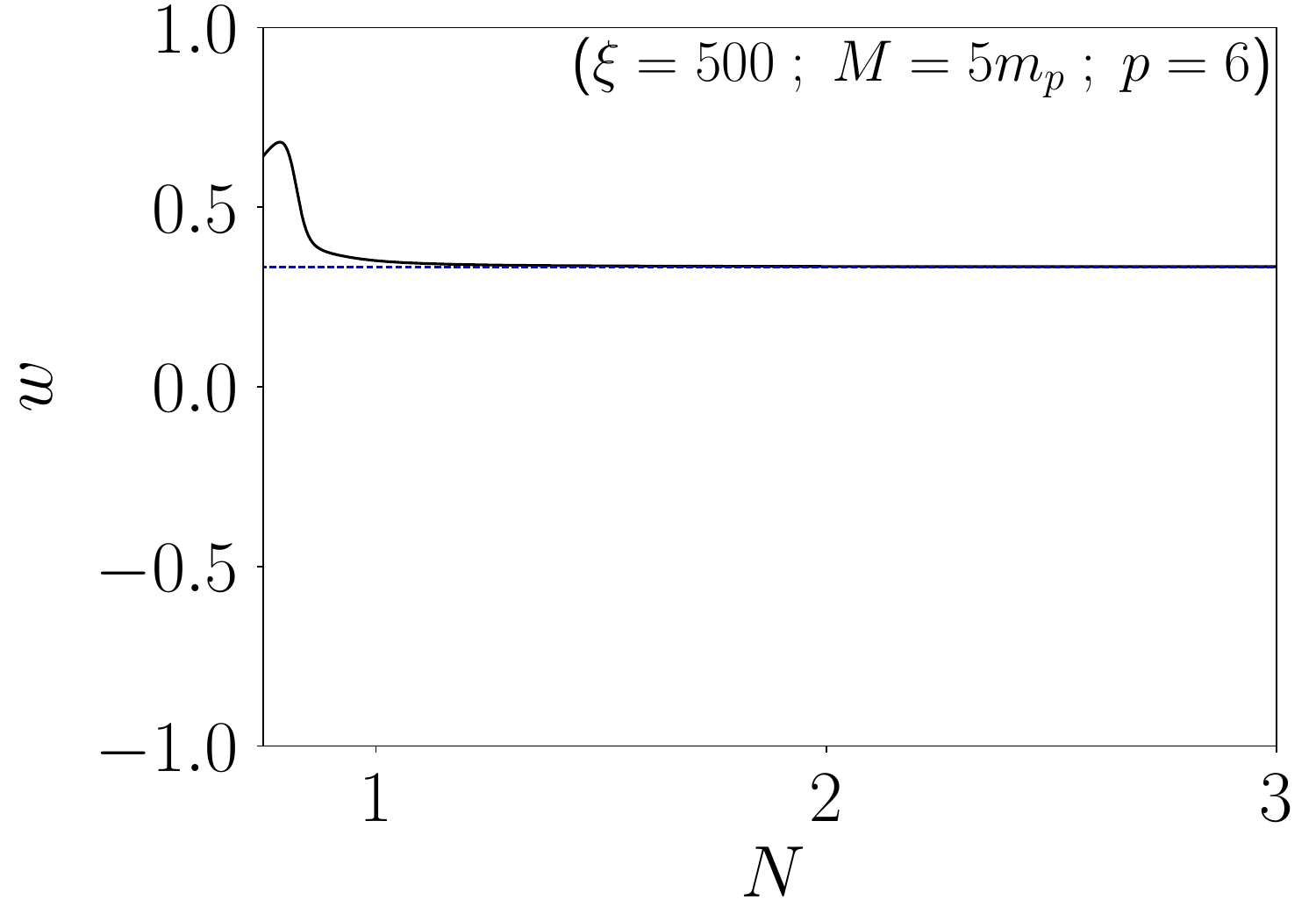} 
    \includegraphics[width=0.32\textwidth,height=3.5cm]{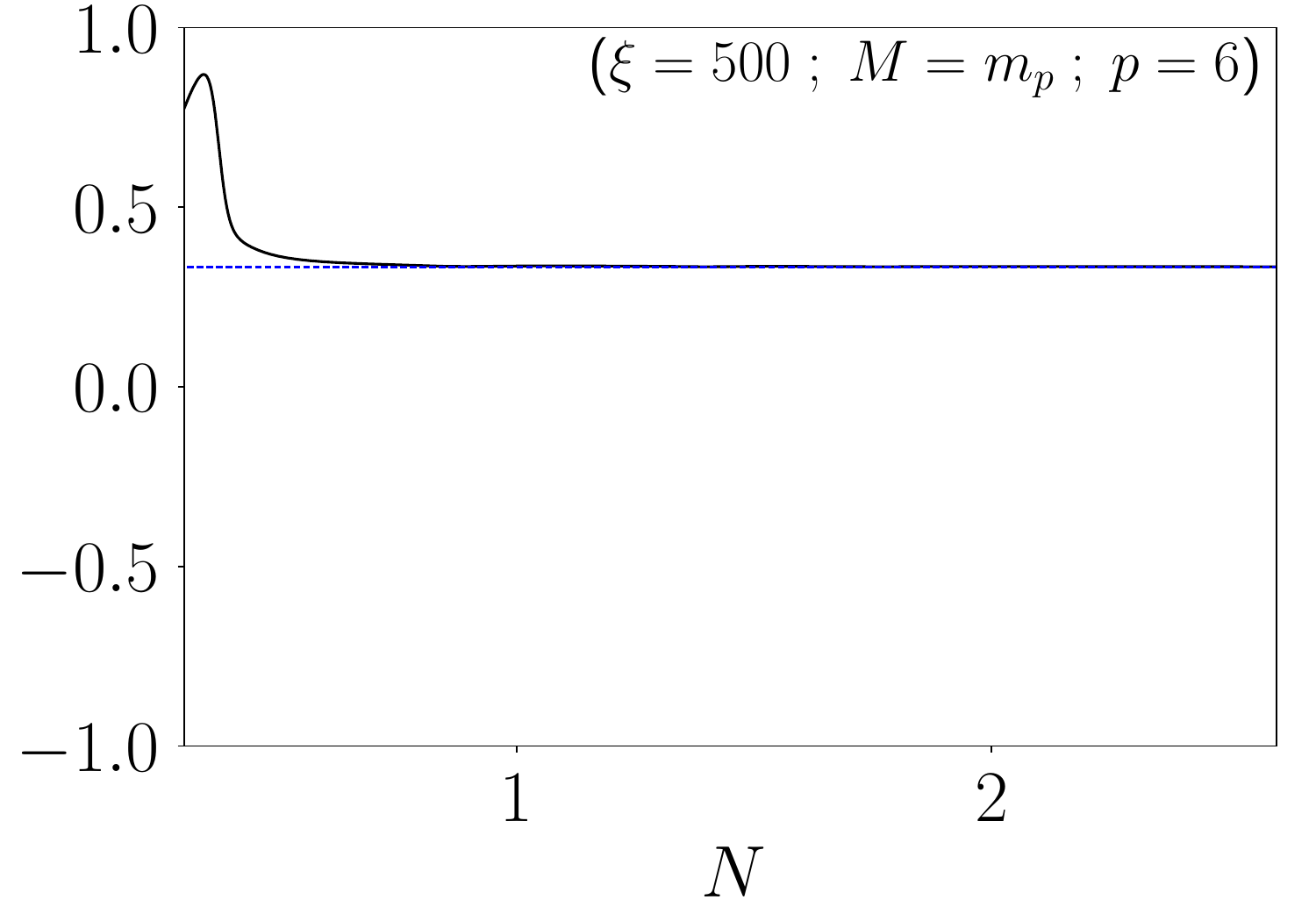} 
    \includegraphics[width=0.32\textwidth,height=3.5cm]{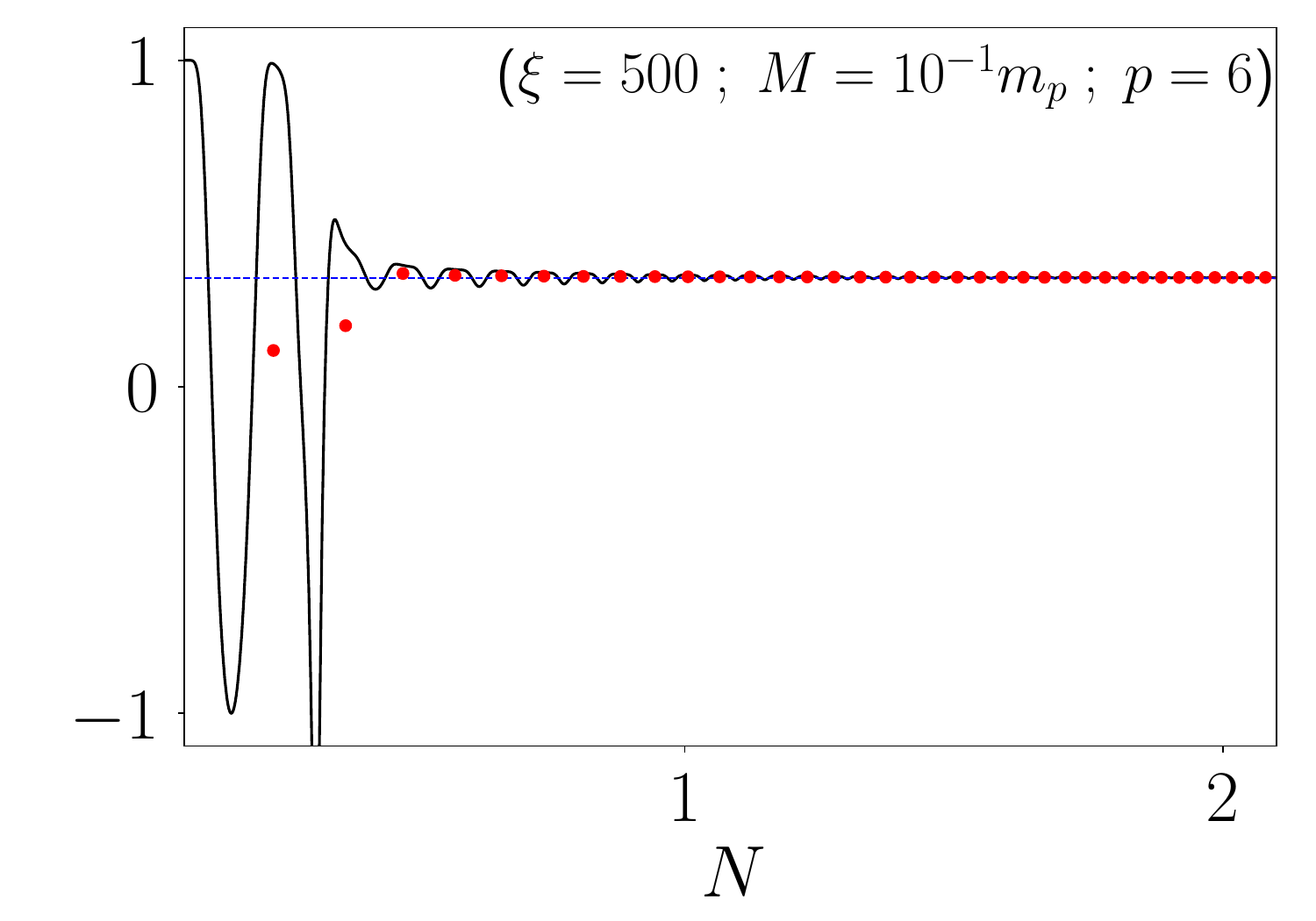}
     \includegraphics[width=0.32\textwidth,height=3.5cm]{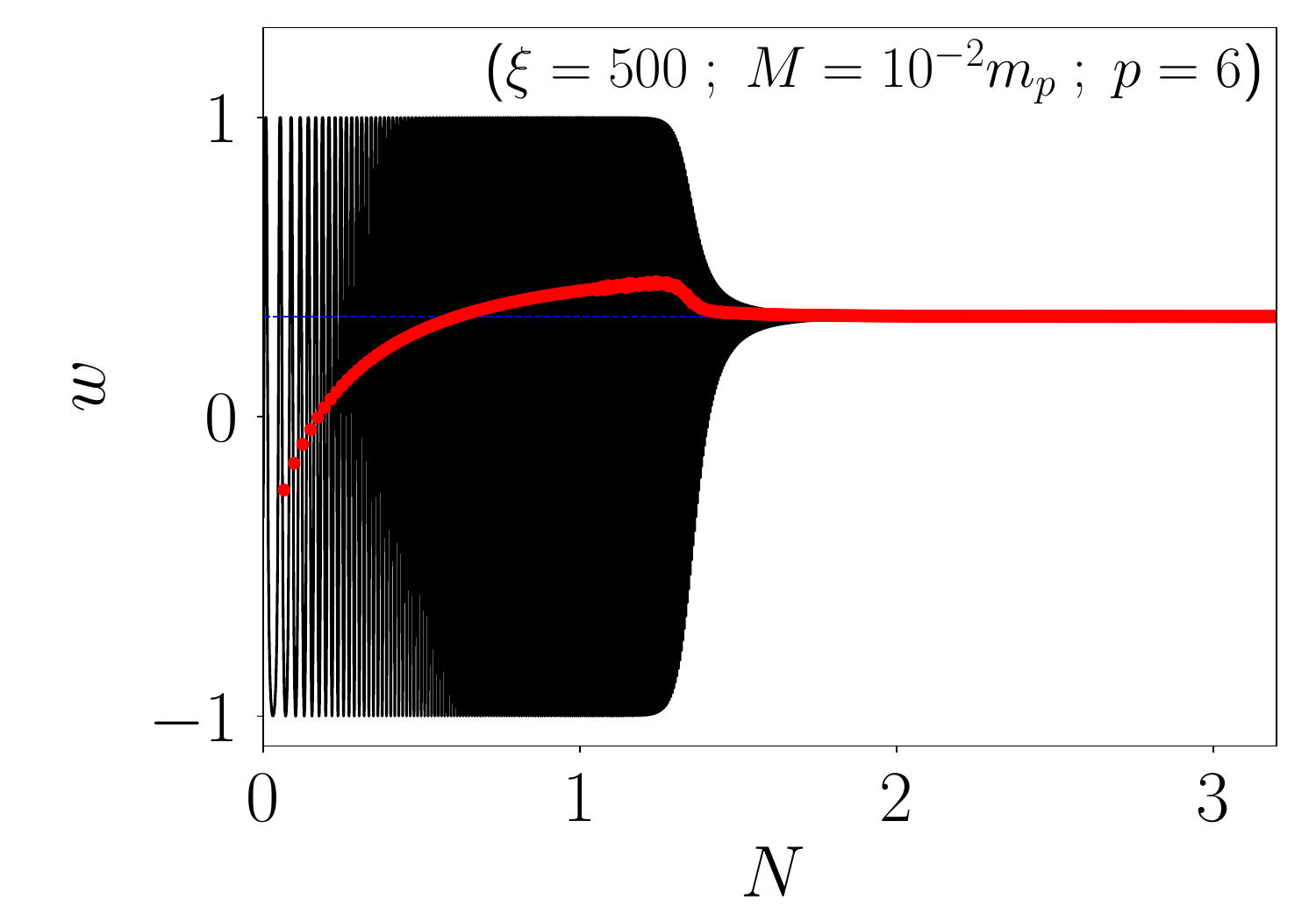} 
    \includegraphics[width=0.32\textwidth,height=3.5cm]{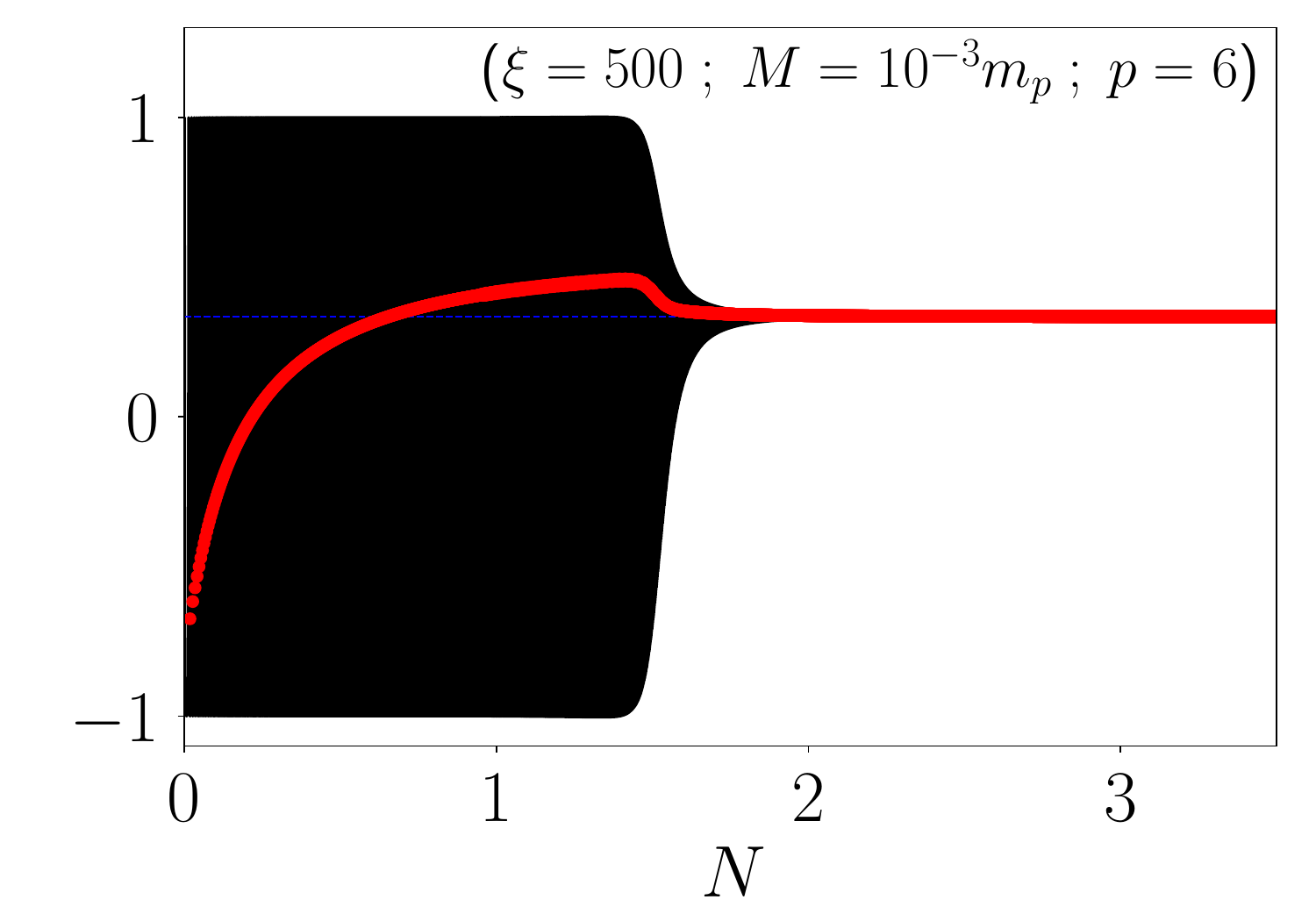} 
    \end{center}
    \vspace*{-0.5cm}
    \caption{Time evolution of the equation of state ($w$) as computed from Eq.~\ref{eq:EoSfromRonH2}, blue line is at $w=1/3$ (Radiation Domination), and red dots depict average value of $w$ per oscillation for $\xi=500$  and and (from left to right) $M=5\,m_p$, $M=m_p$, $M=10^{-1}\,m_p$, $M=10^{-2}\,m_p$, $M=10^{-3}\,m_p$.} \label{fig:Plotsxi500p6EoS} \vspace*{-0.3cm}
\end{figure*}

\begin{figure*}[tbp]
    \begin{center}
     \includegraphics[width=0.35\textwidth,height=3.5cm]{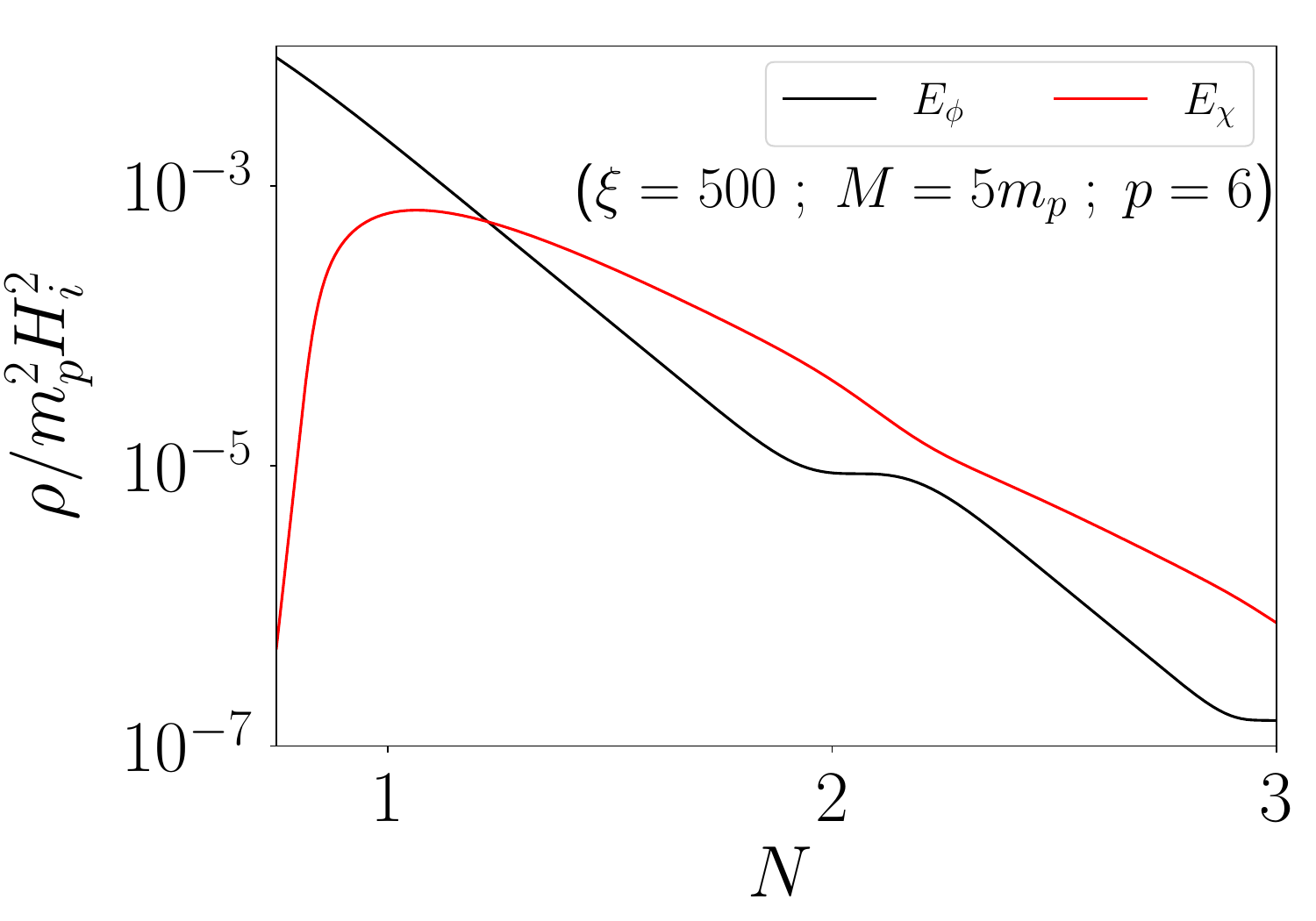} 
    \includegraphics[width=0.31\textwidth,height=3.5cm]{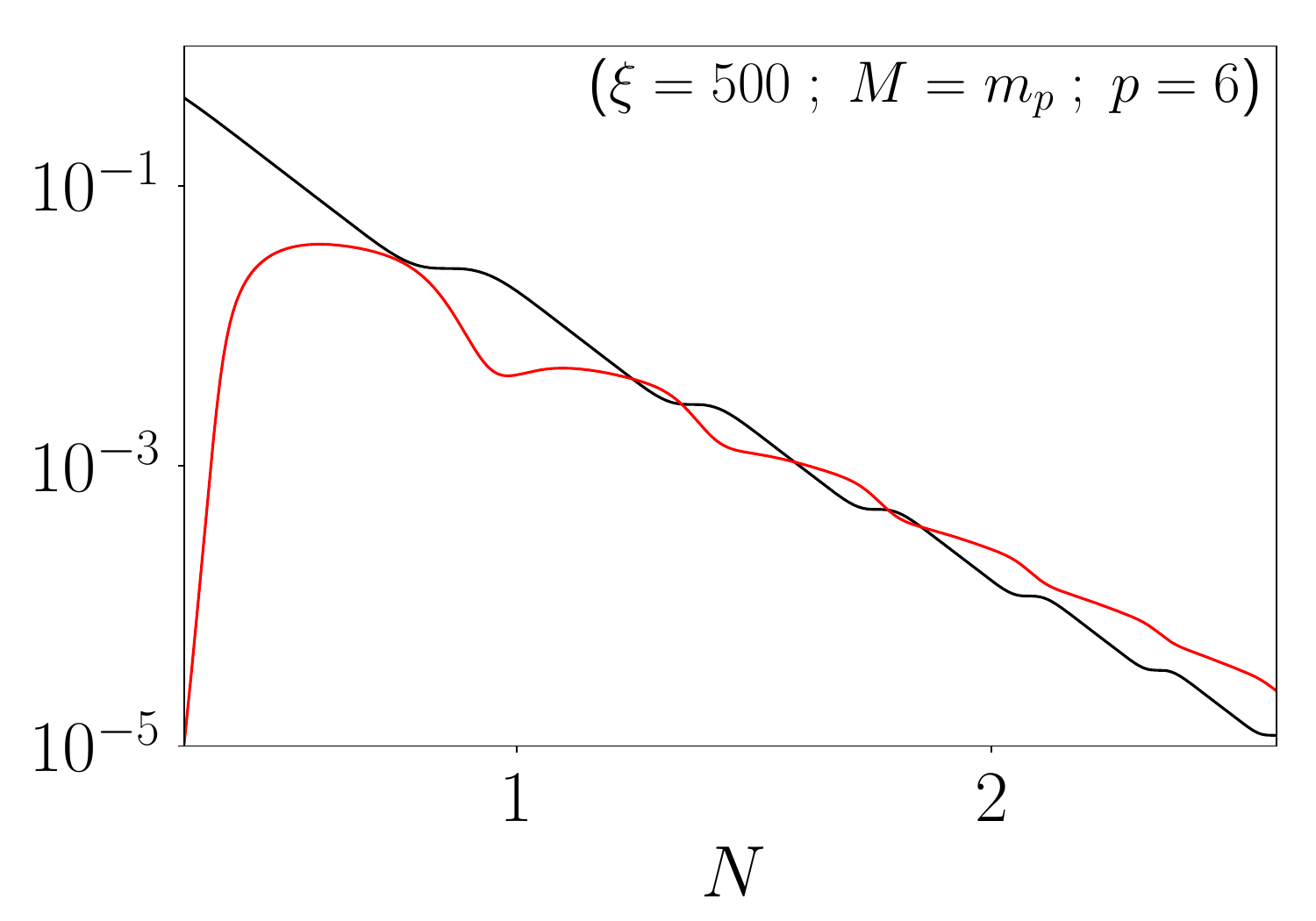} 
    \includegraphics[width=0.31\textwidth,height=3.5cm]{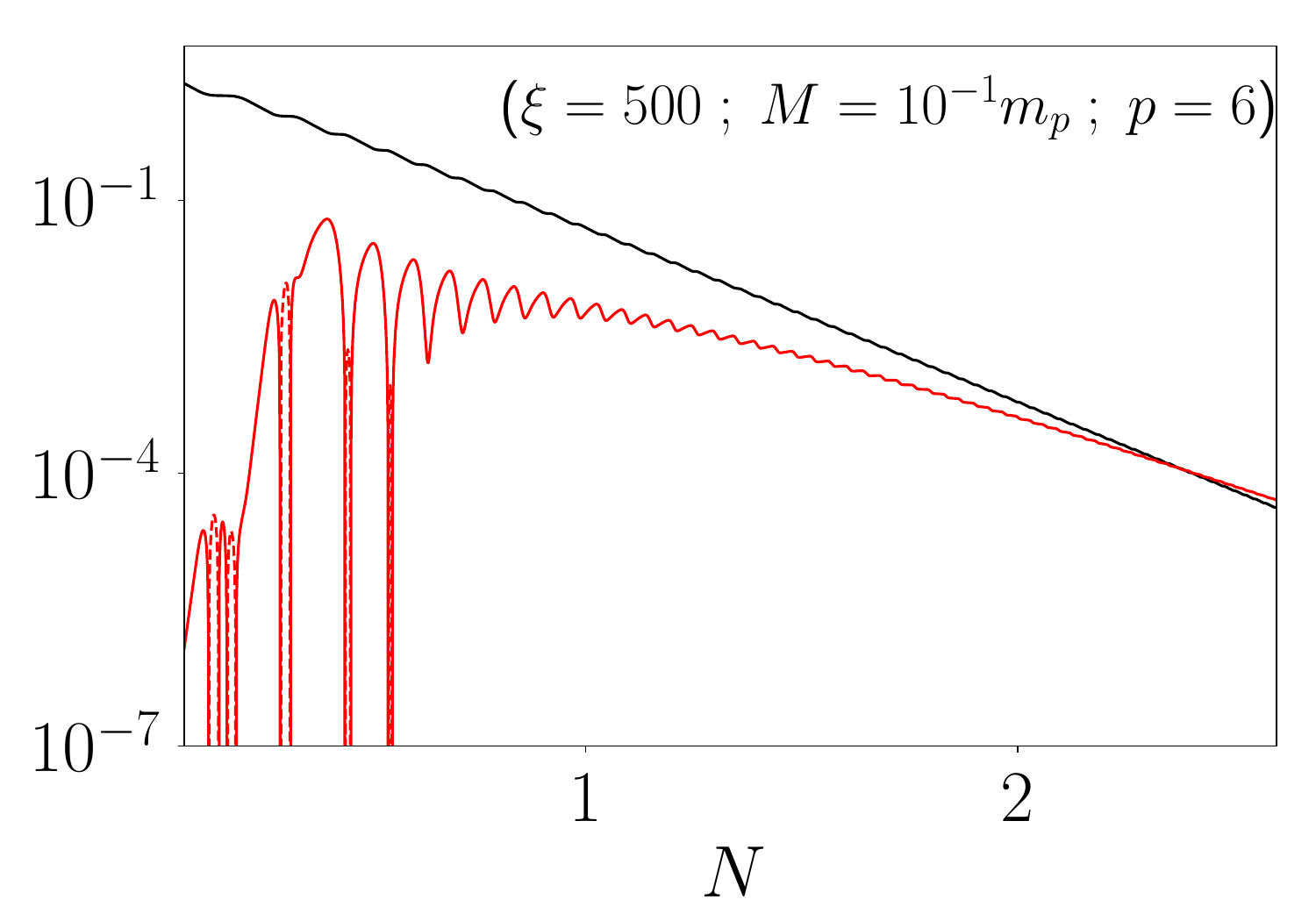}
     \includegraphics[width=0.38\textwidth,height=3.5cm]{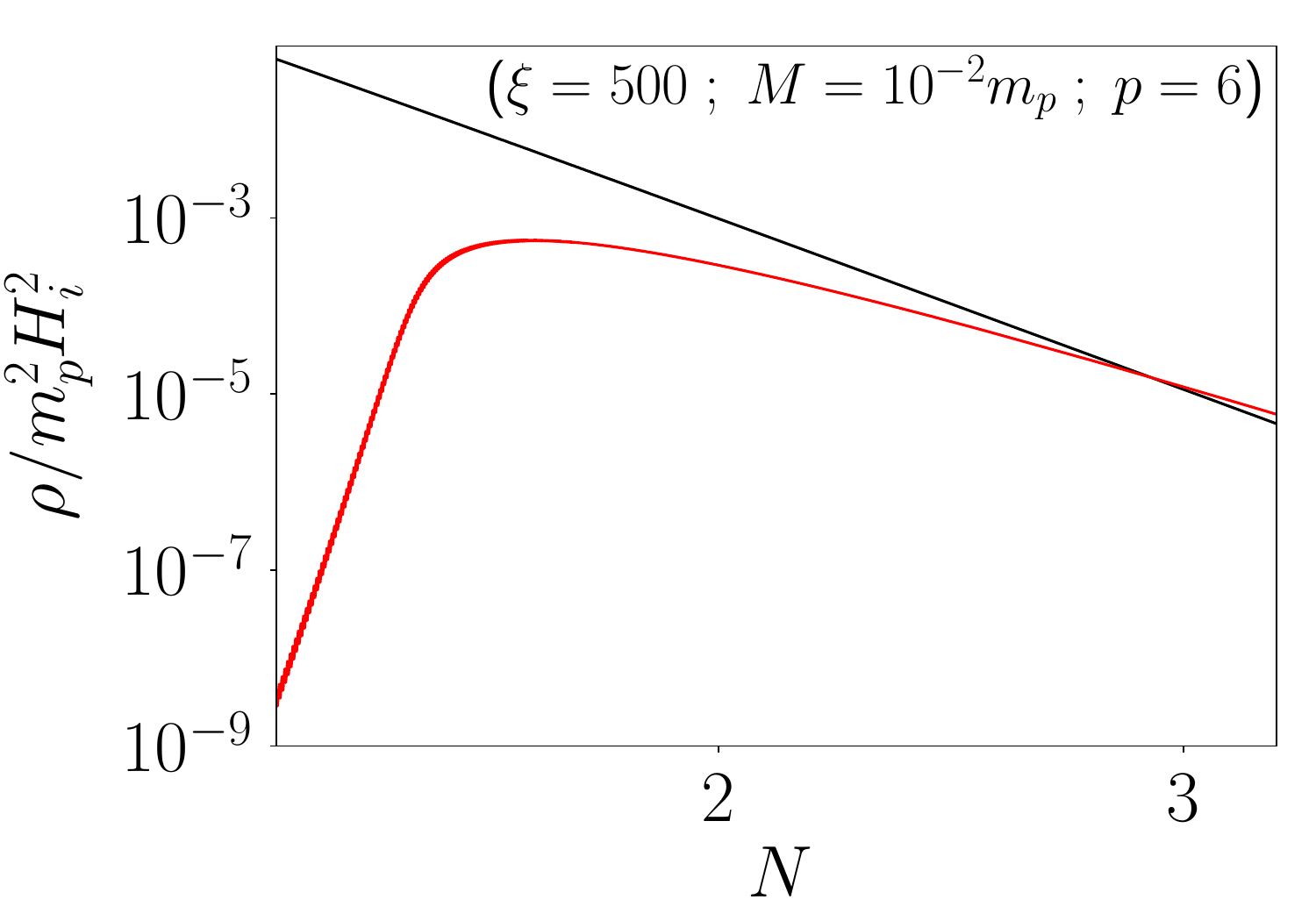} 
    \includegraphics[width=0.35\textwidth,height=3.5cm]{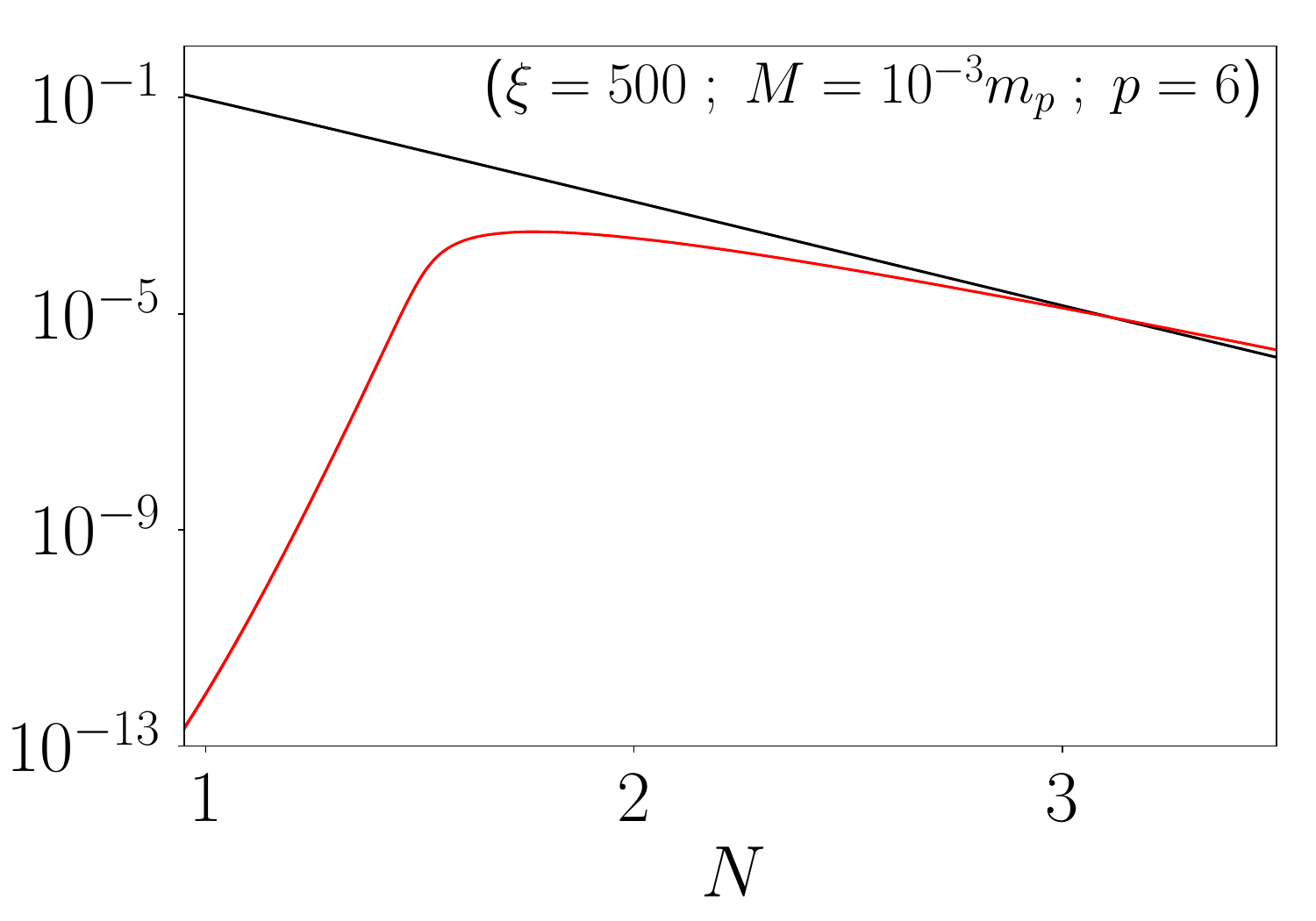} 
    \end{center}
    \vspace*{-0.5cm}
    \caption{Time evolution of the total Energy of the inflaton and the NMC field for $\xi=500$ and (from left to right) $M=5\,m_p$, $M=m_p$, $M=10^{-1}\,m_p$, $M=10^{-2}\,m_p$, $M=10^{-3}\,m_p$.} \label{fig:Plotsxi500p6Energies} \vspace*{-0.3cm}
\end{figure*}

The evolution of the equation of state (EoS) for $p  = 6$ is shown in Fig.~\ref{fig:Plotsxi500p6EoS}. After an initial stage with the EoS remaining positive, oscillatory, or negative, for large ($M \geq m_p$), intermediate ($M = 0.1m_p$) and small ($M = 0.01m_p,0.001m_p$) scale models, respectively, backreaction eventually kicks in (the later the larger the value of $M$), so that $R/H^2$ settles close to zero, driving the EoS down to $w \simeq 1/3$ in all cases.
The evolution of the energy components is also interesting. To begin with, we note that once the tachyonic instability is triggered, the total energy of the NMC field grows as fast as the variance of $\chi$, as it is the term $\propto \langle \chi^2 \rangle$ in Eq.~(\ref{eq:ChiEnergy}) that dominates the energy of NMC scalar in those moments. In Fig.~\ref{fig:Plotsxi500p6Energies} we show the evolution of the total energy densities of the inflaton and of the NMC field, once the tachyonic instability has been established.  We see that the energy of $\chi$ grows exponentially fast due to the tachyonic instability, and eventually backreacts  onto $R$, switching off the instability in that moment, thus stopping the exponential growth. There is still further growth of the NMC energy above the inflaton's level, as the inflaton decays faster than radiation for $p=6$, while the NMC field, once $R$ has become very small compare to $H^2$ (due to backreaction), evolves as free massless field, with its energy decaying as radiation. The energy of the NMC field ends therefore, necessarily, dominating the energy budget, leading to a proper reheating of the Universe.
\begin{figure*}[tbp]
    \begin{center}
    \includegraphics[width=0.6\textwidth,height=5cm]{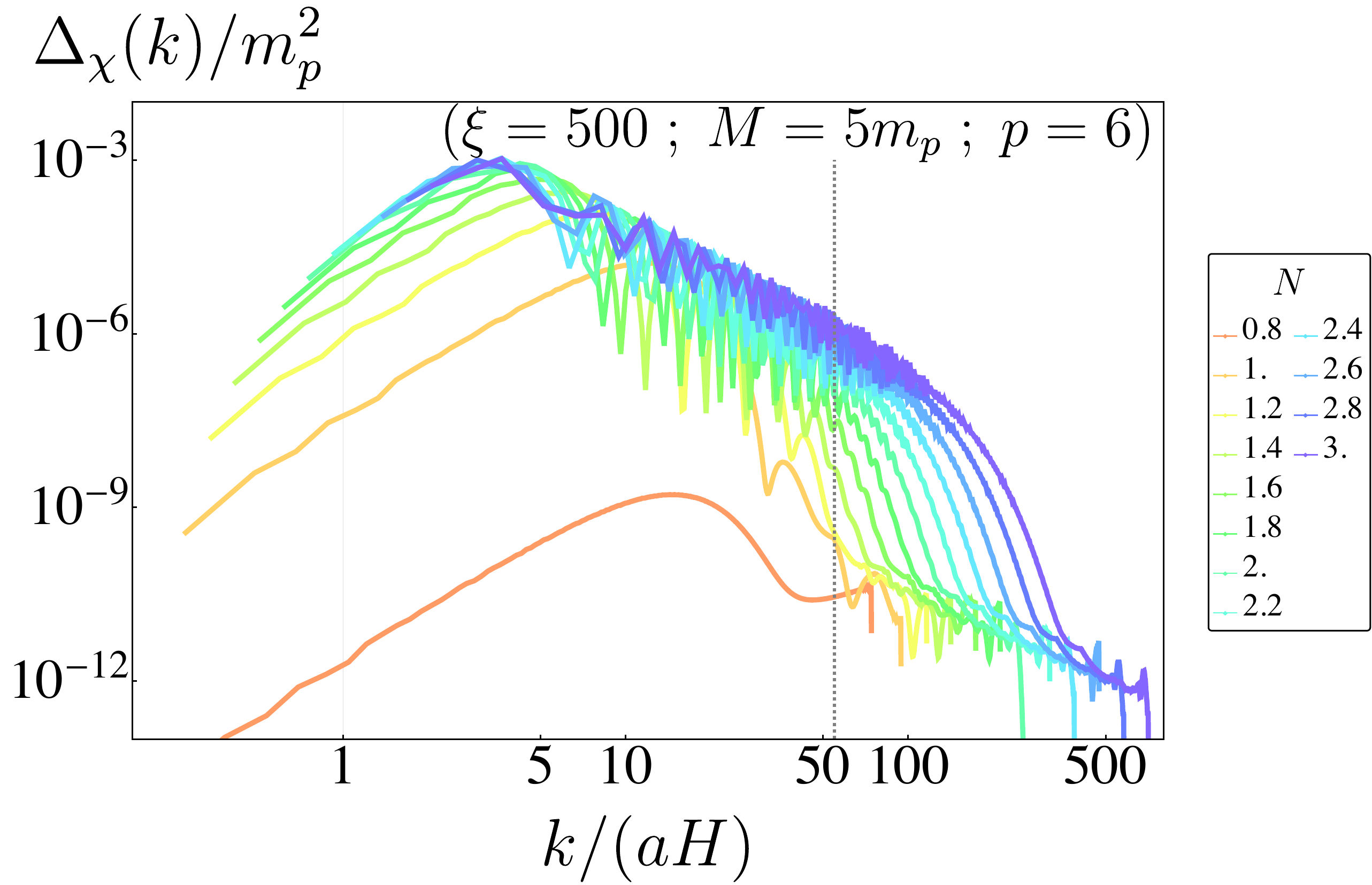} 
    \includegraphics[width=0.6\textwidth,height=5cm]{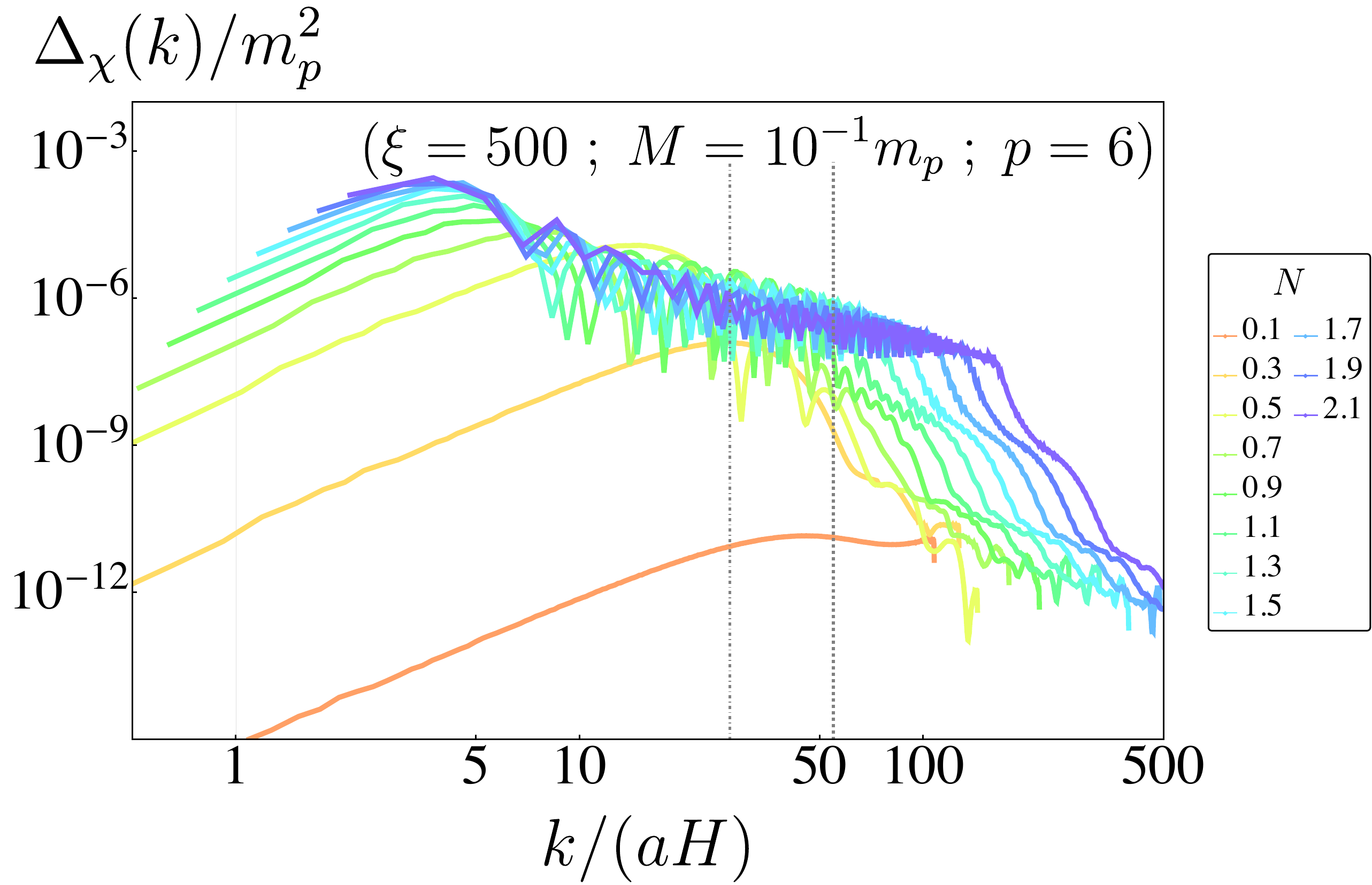} 
    \includegraphics[width=0.6\textwidth,height=5cm]{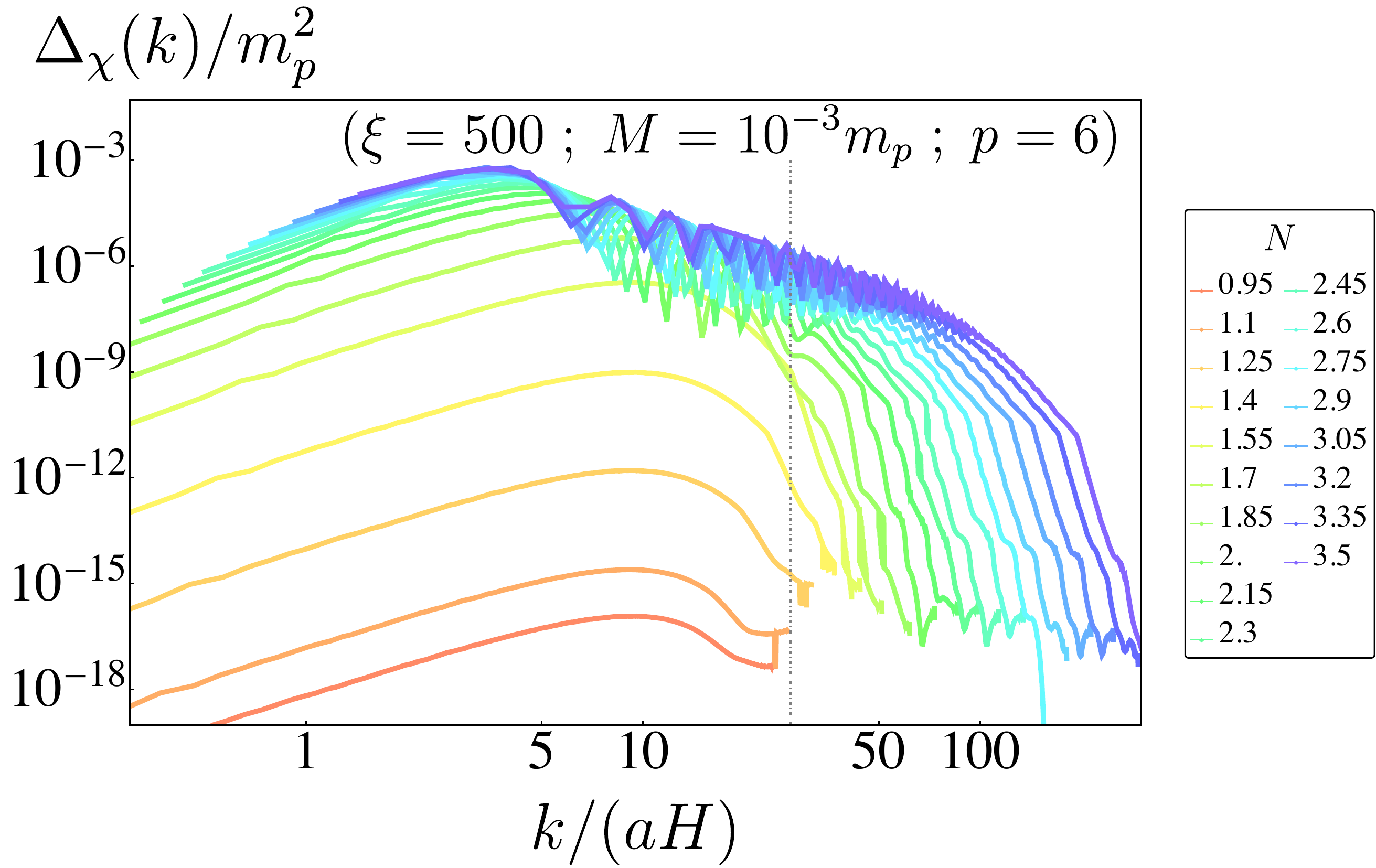} 
    \end{center}
    \vspace*{-0.5cm}
    \caption{Power spectrum of NMC field versus $k/aH$ for $p=6$ and $\xi=500$ for {\it Top:} $M=5 m_p$,  {\it Mid:} $M=10^{-1}m_p$ and {\it Bottom: } $M=10^{-3}m_p$. Dashed line shows threshold scale $k_*/aH = \sqrt{6\xi}$, while dot-dashed line shows threshold scale $k_*/aH = \sqrt{3/2 \xi}$} \label{fig:Matterspectraxi500p6} \vspace*{-0.3cm}
\end{figure*}

In Fig.~\ref{fig:Matterspectraxi500p6} we show the evolution of the power spectrum $\Delta_{\chi}(k,N)$ of the NMC field for different choices of $M$. The growth and saturation of the spectra in a large scale scenario (e.g.~top panel of Fig.~\ref{fig:Matterspectraxi500p6} for $M=5m_p$), takes place within the first negative semi-oscillations of $R$. The initial growth happens up to a threshold momenta $k_*/(aH) \lesssim \sqrt{\xi} |\overline R|^{1/2}/H \sim \sqrt{6\xi}\sqrt{|4-p|/(p+2)} \simeq 27$, {\it c.f.~}Eq.~(\ref{eq:kStarApprox}). Once $\chi$ starts  backreacting onto the Ricci scalar, $R/H^2$ is forced to decrease, and hence the threshold scale $k_*$ that separates tachyonic ($k \lesssim k_*$) vs non-tachyonic ($k \gtrsim k_*$) modes, drifts to smaller values, approaching gradually $k_*/(aH) \simeq 1$. In a small scale scenario, as shown in e.g.~the middle panel of Fig.~\ref{fig:Matterspectraxi500p6} for $M=10^{-1}m_p$, the growth of the $\chi$-spectrum happens in two stages. First, the modes grow every time $R$ becomes negative with, up to a maximum cutoff $k_*/(aH)\sim \sqrt{6\xi}$. Gradually, as times goes by, the negative value of $R$ within a single oscillation, becomes weaker and weaker for the tachyonic instability of the modes to overtake the expansion of the universe. However, a net growth of the modes still takes place when the oscillation averaged curvature $\overline R$ becomes eventually negative, when the inflaton oscillations become dominated by kinetic energy. The tachyonic growth ends when $R/H^2$ becomes sufficiently small due to backreaction, henceforth displacing the peak $k_*/(aH)$ of $\chi$'s power spectrum towards larger scales as $k_*/(aH) \sim 1$. For even smaller scale scenarios, see e.g.~the bottom panel of Fig.~\ref{fig:Matterspectraxi500p6} for $M=10^{-3}m_p$, the growth of unstable modes does not even happen for a single negative semi-oscillation initially (as the dilution due to the expansion of the universe is stronger), but once we enter into the regime $\overline R < 0$, the curvature then tends to  $\bar{R}/\bar{H}^2 = -3/2$, triggering the unstable growth $\chi$'s infrared modes $k \lesssim k_*$. The threshold scale in that moment can be estimated again with Eq.~(\ref{eq:kStarApprox}), as $k_*/(aH) \lesssim \sqrt{6\xi}\sqrt{|4-p|/(p+2)} \simeq 27$. Eventually, the growth of the variance cease to be exponential and saturates to a value $\xi\langle\chi^2\rangle/m_p^2\lesssim 1$, shifting again the peak of the NMC field's power spectrum to scales closer to the horizon $k_*/aH \simeq 1$. 
\begin{figure*}[tbp]
    \begin{center}
    \includegraphics[width=0.65\textwidth,height=6cm]{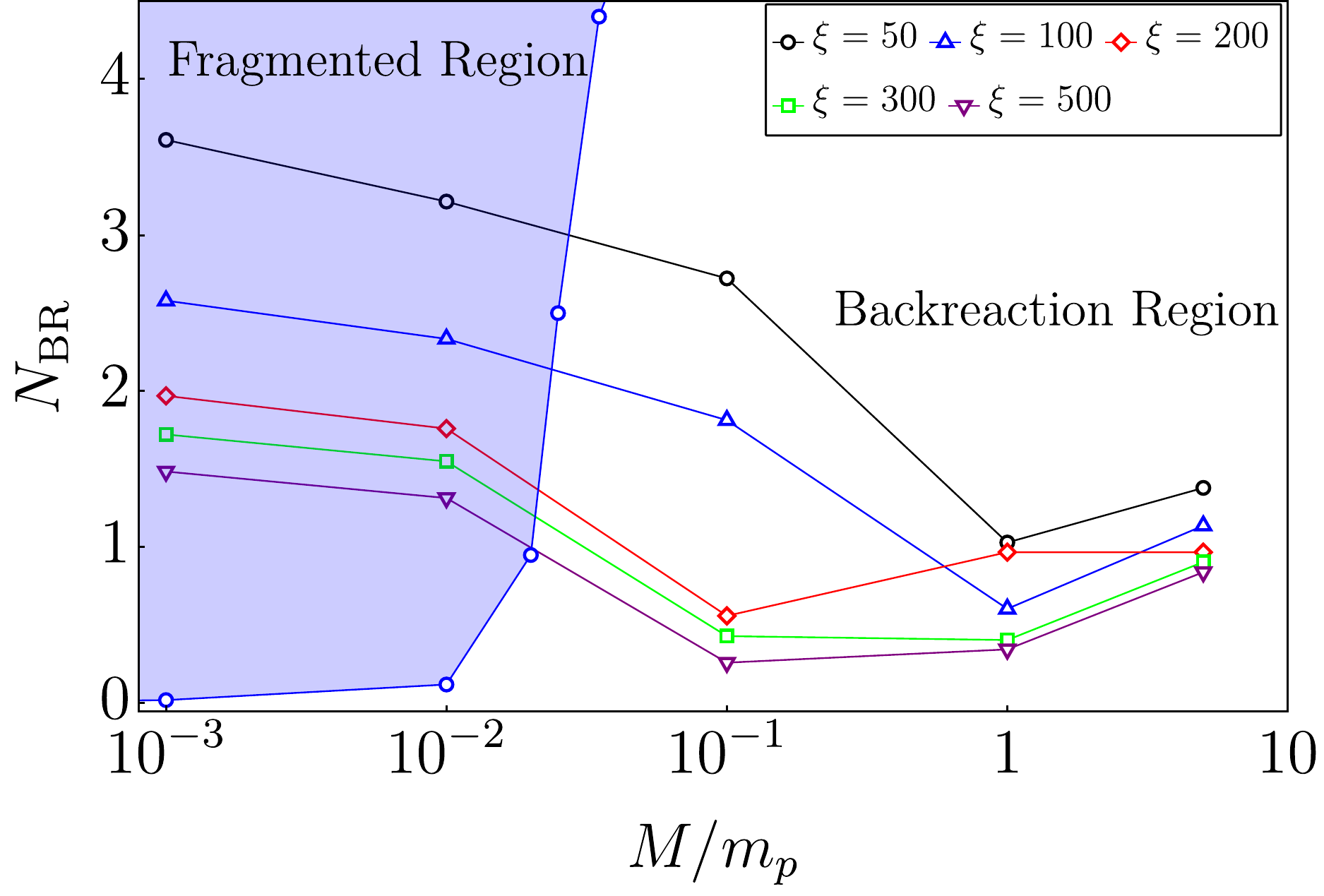} 
    \includegraphics[width=0.65\textwidth,height=6cm]{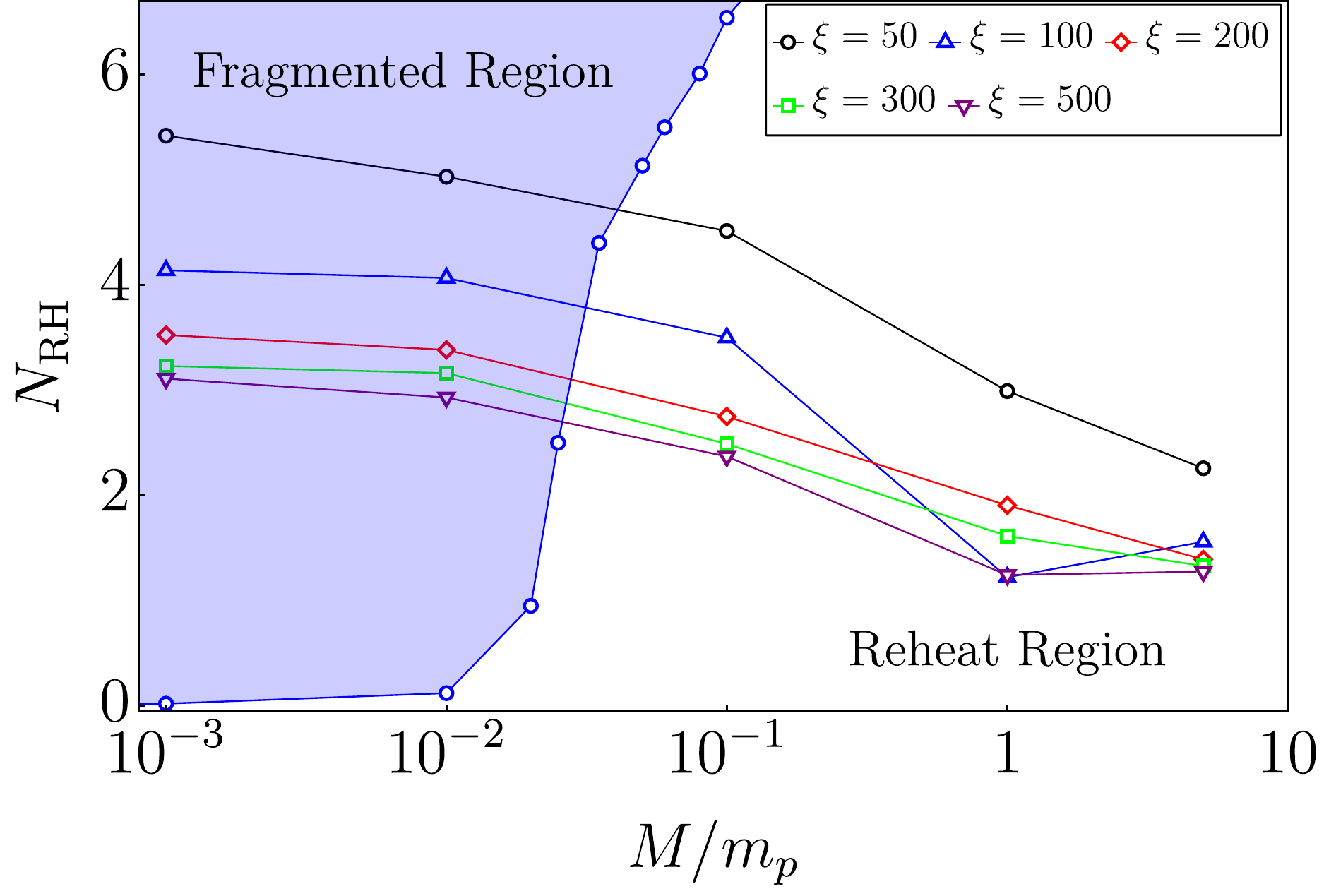}
    \end{center}
    \vspace*{-0.5cm}
    \caption{{\it Top:} Backreaction time as a function of scale $M$ and coupling $\xi$. {\it Bottom:} (P)reheating time as a function of scale $M$ and coupling $\xi$. Blue region the shows parameter space ruled out to complete (p)reheating as inflaton fragments first.} \label{fig:BRandPRtime} \vspace*{-0.3cm}
\end{figure*}

In order to parameterize the the (p)reheating dynamics as a function of $M$ and $\xi$, we have characterized both the time scales when $\chi$ backreacts on $R$, $N_{\rm BR}$, and when $\chi$ starts dominating the energy budget and proper reheating becomes effective, $N_{\rm RH}$. We define the time scale $N_{\rm BR}$ as the moment when the envelope of effective EoS $w_{\rm osc}$ decays below 90~\% of its maximum value. This corresponds also to the moment when the Universe transits from kination domination (KD) to radiation domination (RD), see Fig.~\ref{fig:Plotsxi500p6EoS}. In the {\it top} panel of Fig.~\ref{fig:BRandPRtime}, we see the backreaction time ($N_{\rm BR}$) as a function of $M$ and $\xi$. We notice that increasing the coupling decreases the time of backreaction, as expected, as the growth rate of the tachyonic modes increases monotonically with $\xi$. At the same time, decreasing $M$ makes the backreaction time larger. This is due to the time delay for the effective EoS to reach the value $w_{\rm eff}=0.5$. Another reason comes from the fact that lowering the scale $M$ makes the effective EoS $w_{\rm eff}<1/3$ during the first oscillations, which means that $\overline R$ is positive on average. 

The time scale for the NMC field to become the dominant species in the total energy budget, defines of course the moment when proper reheating is achieved in the model. For practical reasons we define $N_{\rm RH}$ as the time when $E_{\chi}$ becomes equal to $E_{\phi}$. In the {\it bottom} panel of Fig.~\ref{fig:BRandPRtime} we show the parameter dependence of $N_{\rm RH}$ on $M$ and $\xi$. As expected, we obtain that $N_{\rm RH}$ is larger than $N_{\rm BR}$, as the latter measures the moment when the NMC field becomes a relevant contributor to the Ricci scalar $R$, while the former measures the time when the NMC field becomes a relevant contributor of the total energy budget. 

Our analysis so far has ignored the fact that the inflaton condensate may fragment. If we look at Fig.~\ref{fig:lowScaleEoSfrag}, we see actually that for values $M<10^{-1}\,m_p$, inflaton fragmentation will take place before backreaction of $\chi$, invalidating our analysis based on the assumption that the inflaton condensate remains homogeneous. This rules out the corresponding parameter space, as indicated by the blue shaded regions in Fig.~\ref{fig:BRandPRtime}. While the parameter space for the blue region of $N_{\rm RH}$ is simply ruled out to support our mechanism for proper reheating into $\chi$, this does not change the fact that the universe will transit into RD, once inflaton fragmentation becomes effective. In other words, for very small scales, say $M \ll 0.1m_p$, the universe enters into RD almost immediately after inflation (actually in much less than $\sim 1$ efold after the end of inflation, see bottom panel in Fig.~\ref{fig:BRandPRtime} for $M < 0.01m_p$). Proper reheating, however, would require that the inflaton decays into other relativistic species through direct couplings. 

In summary, due to inflaton fragmentation, only relatively large energy models with $M \gtrsim 0.1 m_p$, can lead to proper reheating in our scenario. This is reflected in Table~\ref{tab:NRHforp6}, where we show the time scale and temperature of reheating for the allowed cases, with $T_{\rm RH}$ computed from $\rho_{NMC}(N_{\rm RH}) =\frac{\pi^2}{30} \; g_*(T) T_{\rm RH}^4$. In order to obtain concrete numbers, we have taken $g_*(T) = 106.75$, as in the Standard Model.
\renewcommand{\arraystretch}{2}
\begin{table}[t]
  \hspace*{-3.5mm}
  \centering
  \begin{tabular}{|c||c|c|c|}
    \hline

& \makecell{$M = 10^{-1}\, m_p$  \\ \hline {\footnotesize $N_{\rm RH}$~~~~~~~$T_{\rm RH}$ [GeV]}}  
& \makecell{$M =  m_p$  \\ \hline{\footnotesize $N_{\rm RH}$~~~~~~~$T_{\rm RH}$ [GeV]}} 
& \makecell{$M = 5\, m_p$  \\ \hline{\footnotesize $N_{\rm RH}$~~~~~~~$T_{\rm RH}$ [GeV]}} 
\\
\hline
    \, & \, & \, & \,  \vspace*{-9.2mm}\\

\hline
$\xi = 50$ 
    & \makecell{$4.514$} ~~ \makecell{$5.465 \times 10^{12}$}
    & \makecell{$2.992$} ~~ \makecell{$6.836 \times 10^{13}$} 
    & \makecell{$2.258$} ~~ \makecell{$1.7561 \times 10^{14}$} 
    \\
    \hline
$\xi = 100$ 
    & \makecell{$3.5$}~~ \makecell{$1.706 \times 10^{13}$} 
    & \makecell{$1.2223$}~~ \makecell{$4.692 \times 10^{14}$} 
    & \makecell{$1.5585$}~~ \makecell{$3.759 \times 10^{14}$} 
    \\
    \hline
$\xi = 200$ 
    & \makecell{$2.75$}~~ \makecell{$3.947 \times 10^{13}$}
    & \makecell{$1.905$}~~ \makecell{$2.253 \times 10^{14}$} 
    & \makecell{$1.391$}~~ \makecell{$4.835 \times 10^{14}$}
    \\
    \hline
$\xi = 300$ 
    & \makecell{$2.49$} \makecell{$5.287 \times 10^{13}$}
    & \makecell{$1.6143$} \makecell{$3.077 \times 10^{14}$}
    & \makecell{$1.3285$} \makecell{$5.319\times 10^{14}$} 
    \\
    \hline
$\xi = 500$ 
    & \makecell{$2.37$} \makecell{$6.043 \times 10^{13}$}
    & \makecell{$1.243$} \makecell{$4.659 \times 10^{14}$}
    & \makecell{$1.276$} \makecell{$5.753 \times 10^{14}$}
    \\
    \hline
  \end{tabular}
  \caption{Time scale  and temperature of (p)reheating for the allowed cases in the parameter space of bottom Fig.~\ref{fig:BRandPRtime}
  }
  \label{tab:NRHforp6}
\end{table}

\subsubsection{Including NMC field self-interactions} 
\label{subsec:selfinteractions}

In sections~\ref{subsec:p2}-\ref{subsec:p6} we have considered that interactions of the NMC are negligible, simply setting $V_{\rm NMC}(\chi) = 0$. In this section we modify this hypothesis and consider self-interactions of the NMC field as $V_{\rm NMC}(\chi) = {\lambda\over 4} \chi^4$, with $\lambda$ a dimensionless coupling. If the NMC field sustains sufficiently strong 
self-interactions, the system dynamics may also develop non-linearities, regularizing the tachyonic instability. This is due to the backreaction of the field onto itself, as during the tachyonic growth it develops a positive (squared) mass $m_{\rm int}^2 = 3\lambda \langle \chi^2 \rangle > 0$. If $\lambda$ is sufficiently large, $m_{\rm int}^2$ may grow enough to screen the tachyonic mass $m_{\rm grav}^2 = \xi \overline{R} < 0$, so that the total effective mass of the NMC field, $m_{\rm tot}^2 = m_{\rm int}^2 + \xi \overline{R}$, flips from negative to positive, ending the tachyonic instability. The condition to reach to  self-interaction backreaction is simple: it must hold that $m_{\rm tot}^2 = 0$. This leads to determine the NMC field amplitude at that moment as 
\begin{eqnarray}
\chi_{\rm br} \equiv \sqrt{\langle \chi^2 \rangle}\Big|_{m_{\rm tot}^2 = 0} \simeq \sqrt{\xi\over 3\lambda}  \sqrt{\overline R}\,.
\end{eqnarray}
While backreaction due to self-interactions can take place for any exponent $p$, from the point of reheating this phenomenon is interesting only for $p > 4$, i.e.~for $p = 6$ in our case. The reason is simple to understand: once the tachyonic mass is screened, the NMC field dynamics becomes dictated by the self-interactions, which for a quartic potential implies an  energy density scaling as radiation\footnote{This is only technically correct once the gravitational mass $|m_{\rm grav}^2| = \xi |\overline{R}|$ becomes negligible as compared to the self-interaction mass $m_{\rm int}^2 = 3\lambda \langle \chi^2 \rangle$, i.e.~when $\xi \overline{R} \ll 3\lambda \langle \chi^2 \rangle$. This is always expected to happen, as due to the expansion of the universe, eventually the inflaton only oscillates sampling the positive curvature part of its potential, and form that moment onward $\overline R / \langle \chi^2 \rangle$ becomes a decreasing function of time.}~\cite{Figueroa:2015rqa}. This means that once the tachyonic growth is regularized, the energy density of the NMC starts decreasing (typically after a short period of time) as $\rho_{\chi} \propto a^{-4}$, while the energy density of the inflaton continues scaling as $\rho_{\phi}\propto a^{-{6p/(p+2)}}$. After backreaction, the NMC-to-inflaton energy ratio behaves therefore as $\rho_{\chi}/\rho_{\phi} \propto a^{2(p-4)/(p+2)}$, growing, remaining constant, or decreasing in time, depending on whether $p > 4$, $p = 4$, or $p < 4$, respectively. At the onset of backreaction due to self-interactions, the energy of the NMC field is however initially subdominant, $\rho_{\chi}/\rho_{\phi} \ll 1$, so this small ratio can only increase in time, allowing for proper reheating ($\rho_\chi \geq \rho_\phi$), if $p > 4$. 

\begin{figure*}[tbp]
    \begin{center}
    \includegraphics[width=0.45\textwidth,height=5cm]{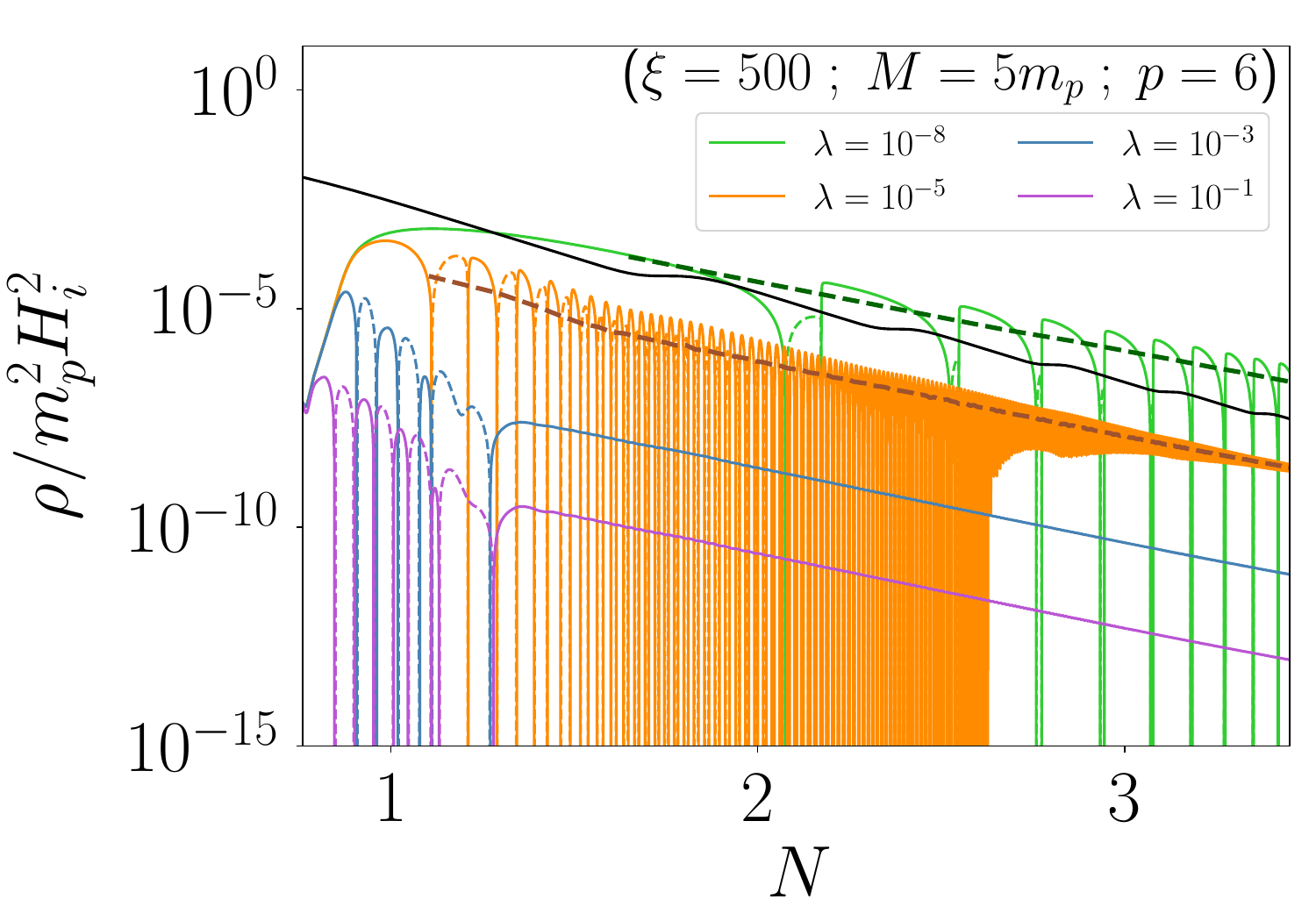} 
\includegraphics[width=0.45\textwidth,height=5cm]{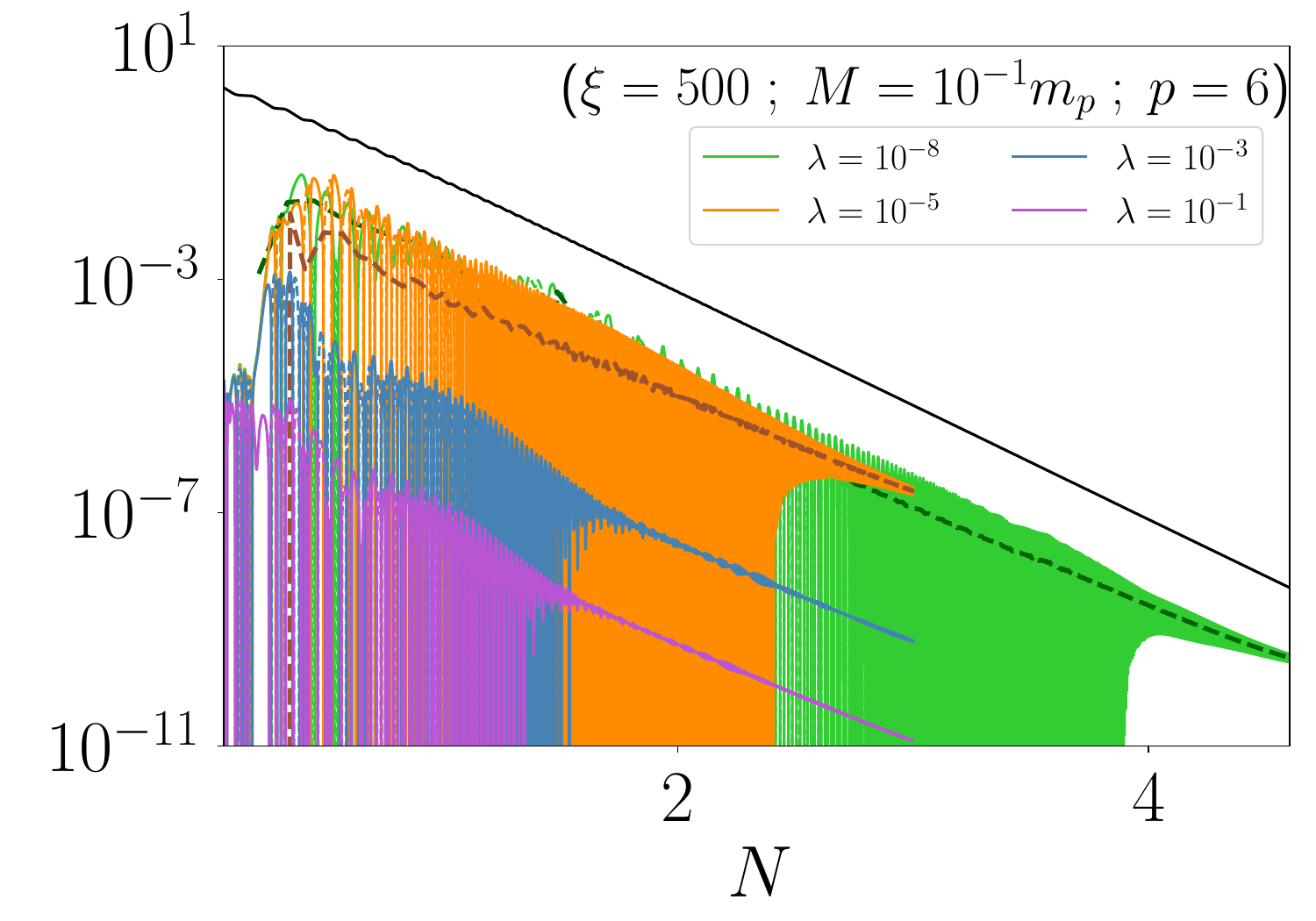} 

    \end{center}
    \vspace*{-0.5cm}
    \caption{Energy density time evolution for different choices of $\lambda$ and $M$ with fixed $\xi=500$ value. Dashed lines depict average per oscillation values.} \label{fig:EnergiesLambdaxi500p6} \vspace*{-0.3cm}
\end{figure*}
\begin{figure*}[tbp]
\begin{center}
\includegraphics[width=0.45\textwidth,height=5cm]{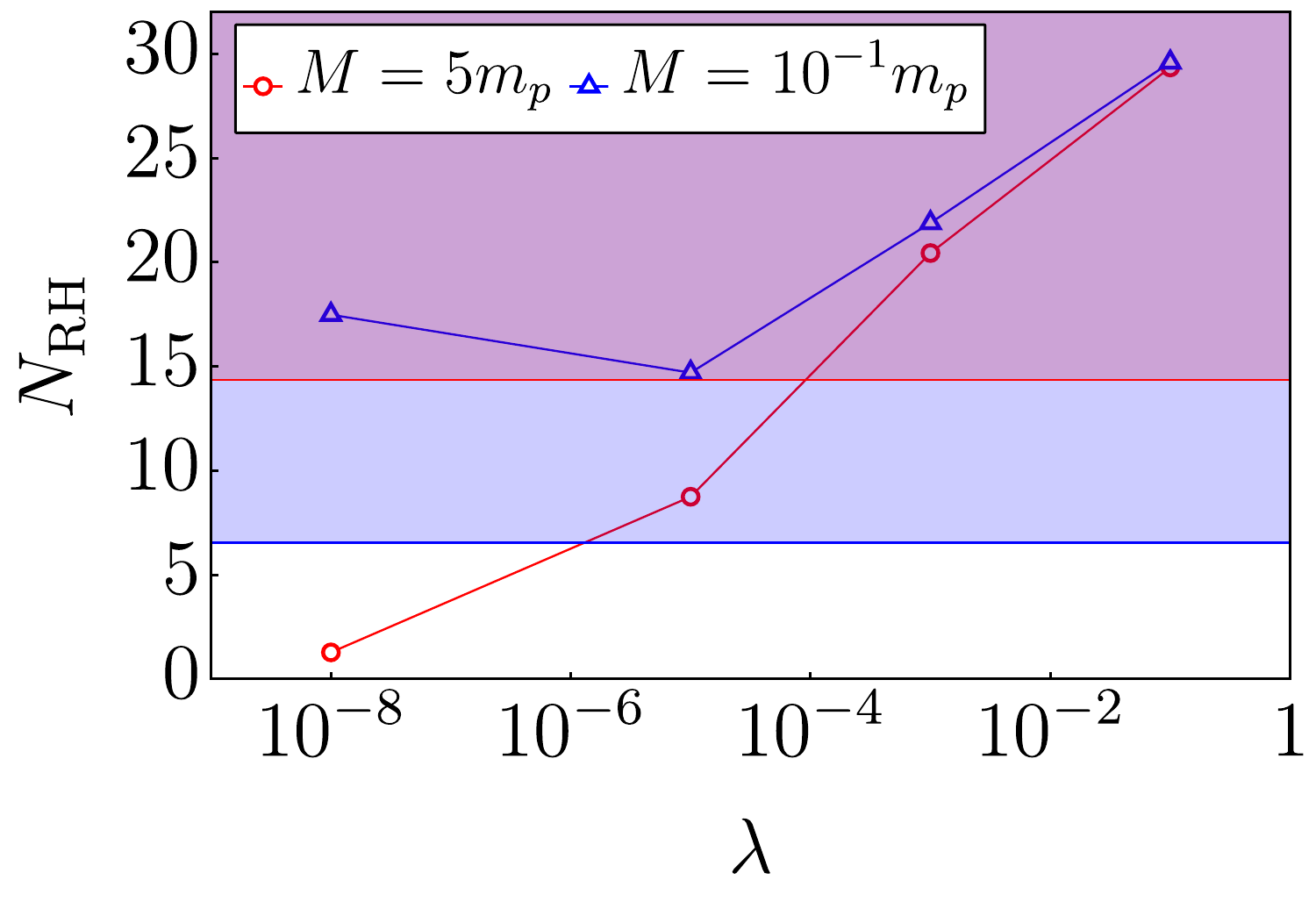} 
    \end{center}
\caption{Time scale for reheating as a function of $\lambda$ and $M$ for fixed $\xi=500$ value. Colored regions depict excluded region for reheating due to inflaton fragmentation, each for the corresponding $M$ case.} \label{fig:Nrhlambdap6}
\end{figure*}

To characterize the impact of $\lambda$, together with $\xi$ and $M$, on the reheating time scale for $p = 6$, we have simulated the case of non-vanishing self-interaction of the NMC $\lambda \neq 0$. We have measured the number of efoldings $N_{\rm RH}$ it takes for the energy of the NMC field to reach the energy of the inflaton. If needed, we have obtained $N_{\rm RH}$ at times longer than our simulation time, by extrapolating to later times the scaling behavior of both the inflaton and NMC energy densities, as measured from our lattice simulations. We find that for low energy scenarios, even the smallest chosen self-interaction delays so much reheating, that inflaton fragmentation takes place before. For larger scales, proper reheating is only possible for the largest mass scales (e.g.~$M = 5m_p$ in the figure), and very small coupling constants, $\lambda \lesssim 10^{-6}$. In conclusion, adding self-interactions for the NMC field, basically spoils the possibility of proper reheating, unless $\lambda$ is extremely small. Our numerical findings are summarized by Fig.~\ref{fig:Nrhlambdap6} and by the allowed cases shown in Table~\ref{tab:NRHforlambdap6} 

\renewcommand{\arraystretch}{2}
\begin{table}[t]
  \hspace*{-3.5mm}
  \centering
  \begin{tabular}{|c||c|}
    \hline
& \makecell{$M = 5\, m_p$  \\ \hline {\footnotesize $N_{\rm RH}$~~~~~~~$T_{\rm RH}$ [GeV]}} \\
\hline
$\lambda = 10^{-8}$ 
    & \makecell{$1.281$}~~\makecell{$6.511 \times 10^{14}$} 
    \\
    \hline
$\lambda = 10^{-5}$ 
    & \makecell{$8.747$}~~\makecell{$3.382 \times 10^{11}$}
    \\
    \hline
  \end{tabular}
  \caption{Time scale and temperature of (p)reheating for the allowed cases in the parameter space Fig.~\ref{fig:Nrhlambdap6}
  }
  \label{tab:NRHforlambdap6}
\end{table}
\section{Summary and Conclusions}
\label{sec:discussion}

In this paper, we have analyzed the feasibility of geometric reheating into a NMC scalar field $\chi$, in $\alpha$-attractor models with tunable mass scale $M$. The latter controls monotonically both the inflationary energy scale $\Lambda$, and the width of the monomial shape $V_{\rm inf} \propto \phi^p$ of the inflaton potential around $\phi = 0$. Choosing small values of ${M}$, say well below the value $M_{\rm max}$ that saturates the current tensor-to-scalar CMB bound, may affect the inflaton dynamics after inflation in various different ways. We find that, $1)$ the inflaton amplitude at the end of inflation is larger than the inflection point of the potential, as soon as $M$ is a factor $\sim 1/8$ smaller than $M_{\rm max}$. We also find that $2)$ as long as the inflaton remains homogeneous, its oscillations after inflation become fast compared to the expansion rate, the faster the smaller the value of $M$. These two effects work against the efficiency of preheating, as they prevent the Ricci scalar $R$ to be sufficiently negative, hence making inefficient the energy transfer into $\chi$. If we choose sufficiently small values of $M$, we also find that $3)$ the inflaton condensate fragments soon enough after the onset of oscillations. This breaks the spatial coherence of the inflaton, deactivating the geometric preheating mechanism. The time scale for inflaton fragmentation drastically decreases when reducing $M$. 

Given that inflaton fragmentation and geometric (p)reheating are independent mechanisms from each other, we have studied their effects and time scales separately, determining what fraction of parameter space support viable reheating, and which part must be discarded when inflaton fragmentation occurs before gravitational backreaction. We have studied the cases $p=2$, $p=4$ and $p=6$, which behave very differently. For the case $p=2$, preheating is inefficient independently of the choice of the NMC coupling $\xi$. This happens because inflaton oscillations are mainly positive in average. For the case $p=4$, preheating ceases to be efficient when ${M}$ is already of order $10^{-1}m_p$, as for that mass scale (and smaller), both effects $1)$ and $2)$  mentioned above, make inefficient the production of $\chi$-particles. For the case $p=6$, the energy transfer during preheating can be achieved for low scale scenarios, thanks to the stiff equation of state that leads to a Ricci scalar negative in average, $\bar R < 0$. This is a sufficient condition for $\chi$-particles to be excited efficiently. However, for scenarios with $M \lesssim 10^{-2}m_p$, inflaton fragmentation happens so fast, that it is not possible to transfer enough energy into $\chi$, and proper reheating is therefore frustrated. For the case $p = 6$, we have also considered the effect of the NMC field self-interactions, finding that these block very effectively energy transfer into $\chi$. For certain cases, proper reheating is still feasible before inflaton fragmentation, but only for large scale scenarios and very small NMC self-couplings. The parameter space that lead to proper preheating can be found summarised in Tables~\ref{tab:NRHforp6} and~\ref{tab:NRHforlambdap6}. 

\acknowledgments 

We thank Toby Opferkuch and Ben Stefanek for collaboration on related projects. We are also very thankful for the use of computational resources provided by Finis-Terrae III cluster of CESGA (Centro de Supercomputaci\'{o}n de Galicia), Vives and Tirant clusters of Universitat de Val\`{e}ncia, and {\it Graviton} cluster of IFIC. Our work is supported by the Generalitat Valenciana grant PROMETEO/2021/083, and and by Spanish Ministerio de Ciencia e Innovacion grant
PID2020-113644GB-I00.

\newpage

\appendix
\section{Initial Conditions from Linear Analysis to \CL~}\label{app:IC}
In order to fix the initial conditions of the Non-minimal coupled field, we solve Eq.\eqref{eq:modeseqnconfchi} in cosmic time. We consider a set of modes that start being very sub-horizon during inflation such that the curvature is negligible and the modes can be taken as the \textit{Bunch-Davis} vacuum 
\begin{equation}
    \widetilde{\chi}_k(k \tau \rightarrow \infty) = \frac{1}{\sqrt{2k}}e^{-ik\tau}
\end{equation}
where $\tau$ is the conformal time. \CL~needs as initial condition the power spectrum of the field and its time derivative as a function of the comoving momenta. The comoving momenta value is dependent on the normalization of the scale factor, this matter because in the linear analysis we set $a=1$ for $t=0$, and \CL~sets $a=1$ for the beginning of the simulation so we need to match both normalizations. We can match both the momenta and the field value by equating quantities that are scale-factor normalization invariant, like the physical momenta $p=k/a$ and the power spectrum ${\Delta}_{\chi}$
\begin{equation}
    \langle \widetilde{\chi}^2 \rangle = \langle 0|\hat{\widetilde{\chi}}(\tau,x)\hat{\widetilde{\chi}}(\tau,x)|0\rangle = \int \frac{dk}{k} \Delta_{\widetilde{\chi}}(k,\tau)
\end{equation}
We can relate the power spectrum of the original field by
\begin{equation}
    \Delta_{\chi}(k,\tau) = \frac{1}{a^2}\Delta_{\widetilde{\chi}} = \frac{k^3}{2\pi^2 a^2} |\widetilde{\chi}_k(\tau)|^2 \equiv \frac{k^3}{2\pi^2} |\chi_k(\tau)|^2
\end{equation}
by equating the physical momenta and the power spectrum we can convert the solution of the linear analysis to the initial conditions for the lattice simulation
\begin{eqnarray}
    p_{\rm CL} &=& p_{\rm LA} \implies \frac{k_{\rm CL}}{a_{\rm CL}} = \frac{k_{\rm LA}} {a_{\rm LA}} \rightarrow k_{\rm CL}=\frac{k_{\rm LA}} {a_{\rm LA}}\;, \\
     \Delta_{\chi}(k,\tau) &=& \frac{1}{a^2}\Delta_{\widetilde{\chi}}(k,\tau) \rightarrow \frac{k_{\rm CL}^3}{2\pi^2} |\chi_{k,{\rm CL}}|^2 = \frac{k_{\rm LA}^3}{2\pi^2 a_{\rm LA}^2} |\widetilde{\chi}_{k,{\rm LA}}(\tau)|^2 \nonumber \;,\\
     &\implies& |\chi_{k,{\rm CL}}|^2 = a_{\rm LA}|\widetilde{\chi}_{k,{\rm LA}}(\tau)|^2\;.
\end{eqnarray}
Subscript ${\rm LA}$ stands for linear analysis and ${\rm CL}$ for \CL. Meanwhile for the time derivative we have 
\begin{equation}
     |\chi'_{k,{\rm CL}}|^2 = a_{\rm LA}|\widetilde{\chi'}_{k,{\rm LA}}(\tau)-\mathcal{H}\widetilde{\chi}_{k,{\rm LA}}(\tau)|^2\;,
\end{equation}
where $\mathcal{H}=a'/a$. Finally, \CL~admits the initial conditions in units of GeV for the the momenta, (GeV)$^{-1}$ for the amplitude square of the modes functions, and (GeV)$^{-1}$ for time. Initially to write equation Eq.~\eqref{eq:modeseqnconfchi} we choose to measure time and momenta in units of $H_i$ (Hubble rate at the beginning of linear analysis), therefore the input values to \CL~are given as
\begin{eqnarray}
    k_{\rm CL}&=&\frac{k_{\rm LA}} {a_{\rm LA}} H_i \;,\\
    |\chi_{k,{\rm CL}}|^2 &=& \frac{a_{\rm LA}}{H_i}|\widetilde{\chi}_{k,{\rm LA}}(\tau)|^2 \;,\\
     |\chi'_{k,{\rm CL}}|^2 &=& a_{\rm LA}H_i|\widetilde{\chi'}_{k,{\rm LA}}(\tau)-\mathcal{H}\widetilde{\chi}_{k,{\rm LA}}(\tau)|^2\;.
\end{eqnarray}
\footnotesize{
\bibliographystyle{JHEP}
\bibliography{auto,manual}}

\providecommand{\href}[2]{#2}\begingroup\raggedright\begin{thebibliography}{100}

\bibitem{Martin:2013tda}
J.~Martin, C.~Ringeval and V.~Vennin, \emph{{Encyclop\ae{}dia Inflationaris}},
  \href{https://doi.org/10.1016/j.dark.2014.01.003}{\emph{Phys. Dark Univ.}
  {\bfseries 5-6} (2014) 75} [\href{https://arxiv.org/abs/1303.3787}{{\ttfamily
  1303.3787}}].

\bibitem{Planck:2018jri}
{\scshape Planck} collaboration, \emph{{Planck 2018 results. X. Constraints on
  inflation}}, \href{https://doi.org/10.1051/0004-6361/201833887}{\emph{Astron.
  Astrophys.} {\bfseries 641} (2020) A10}
  [\href{https://arxiv.org/abs/1807.06211}{{\ttfamily 1807.06211}}].

\bibitem{Planck:2018vyg}
{\scshape Planck} collaboration, \emph{{Planck 2018 results. VI. Cosmological
  parameters}},
  \href{https://doi.org/10.1051/0004-6361/201833910}{\emph{Astron. Astrophys.}
  {\bfseries 641} (2020) A6}
  [\href{https://arxiv.org/abs/1807.06209}{{\ttfamily 1807.06209}}].

\bibitem{BICEP:2021xfz}
{\scshape BICEP, Keck} collaboration, \emph{{Improved Constraints on Primordial
  Gravitational Waves using Planck, WMAP, and BICEP/Keck Observations through
  the 2018 Observing Season}},
  \href{https://doi.org/10.1103/PhysRevLett.127.151301}{\emph{Phys. Rev. Lett.}
  {\bfseries 127} (2021) 151301}
  [\href{https://arxiv.org/abs/2110.00483}{{\ttfamily 2110.00483}}].

\bibitem{Tristram:2021tvh}
M.~Tristram et~al., \emph{{Improved limits on the tensor-to-scalar ratio using
  BICEP and Planck data}},
  \href{https://doi.org/10.1103/PhysRevD.105.083524}{\emph{Phys. Rev. D}
  {\bfseries 105} (2022) 083524}
  [\href{https://arxiv.org/abs/2112.07961}{{\ttfamily 2112.07961}}].

\bibitem{Kawasaki:1999na}
M.~Kawasaki, K.~Kohri and N.~Sugiyama, \emph{{Cosmological constraints on late
  time entropy production}},
  \href{https://doi.org/10.1103/PhysRevLett.82.4168}{\emph{Phys. Rev. Lett.}
  {\bfseries 82} (1999) 4168}
  [\href{https://arxiv.org/abs/astro-ph/9811437}{{\ttfamily
  astro-ph/9811437}}].

\bibitem{Kawasaki:2000en}
M.~Kawasaki, K.~Kohri and N.~Sugiyama, \emph{{MeV scale reheating temperature
  and thermalization of neutrino background}},
  \href{https://doi.org/10.1103/PhysRevD.62.023506}{\emph{Phys. Rev. D}
  {\bfseries 62} (2000) 023506}
  [\href{https://arxiv.org/abs/astro-ph/0002127}{{\ttfamily
  astro-ph/0002127}}].

\bibitem{Hannestad:2004px}
S.~Hannestad, \emph{{What is the lowest possible reheating temperature?}},
  \href{https://doi.org/10.1103/PhysRevD.70.043506}{\emph{Phys. Rev. D}
  {\bfseries 70} (2004) 043506}
  [\href{https://arxiv.org/abs/astro-ph/0403291}{{\ttfamily
  astro-ph/0403291}}].

\bibitem{Hasegawa:2019jsa}
T.~Hasegawa, N.~Hiroshima, K.~Kohri, R.S.L.~Hansen, T.~Tram and S.~Hannestad,
  \emph{{MeV-scale reheating temperature and thermalization of oscillating
  neutrinos by radiative and hadronic decays of massive particles}},
  \href{https://doi.org/10.1088/1475-7516/2019/12/012}{\emph{JCAP} {\bfseries
  12} (2019) 012} [\href{https://arxiv.org/abs/1908.10189}{{\ttfamily
  1908.10189}}].

\bibitem{Linde:1981mu}
A.D.~Linde, \emph{{A New Inflationary Universe Scenario: A Possible Solution of
  the Horizon, Flatness, Homogeneity, Isotropy and Primordial Monopole
  Problems}}, \href{https://doi.org/10.1016/0370-2693(82)91219-9}{\emph{Phys.
  Lett. B} {\bfseries 108} (1982) 389}.

\bibitem{Albrecht:1982mp}
A.~Albrecht, P.J.~Steinhardt, M.S.~Turner and F.~Wilczek, \emph{{Reheating an
  Inflationary Universe}},
  \href{https://doi.org/10.1103/PhysRevLett.48.1437}{\emph{Phys. Rev. Lett.}
  {\bfseries 48} (1982) 1437}.

\bibitem{Dolgov:1982th}
A.D.~Dolgov and A.D.~Linde, \emph{{Baryon Asymmetry in Inflationary Universe}},
  \href{https://doi.org/10.1016/0370-2693(82)90292-1}{\emph{Phys. Lett. B}
  {\bfseries 116} (1982) 329}.

\bibitem{Abbott:1982hn}
L.F.~Abbott, E.~Farhi and M.B.~Wise, \emph{{Particle Production in the New
  Inflationary Cosmology}},
  \href{https://doi.org/10.1016/0370-2693(82)90867-X}{\emph{Phys. Lett. B}
  {\bfseries 117} (1982) 29}.

\bibitem{Traschen:1990sw}
J.H.~Traschen and R.H.~Brandenberger, \emph{{Particle Production During
  Out-of-equilibrium Phase Transitions}},
  \href{https://doi.org/10.1103/PhysRevD.42.2491}{\emph{Phys. Rev. D}
  {\bfseries 42} (1990) 2491}.

\bibitem{Kofman:1994rk}
L.~Kofman, A.D.~Linde and A.A.~Starobinsky, \emph{{Reheating after inflation}},
  \href{https://doi.org/10.1103/PhysRevLett.73.3195}{\emph{Phys. Rev. Lett.}
  {\bfseries 73} (1994) 3195}
  [\href{https://arxiv.org/abs/hep-th/9405187}{{\ttfamily hep-th/9405187}}].

\bibitem{Shtanov:1994ce}
Y.~Shtanov, J.H.~Traschen and R.H.~Brandenberger, \emph{{Universe reheating
  after inflation}},
  \href{https://doi.org/10.1103/PhysRevD.51.5438}{\emph{Phys. Rev. D}
  {\bfseries 51} (1995) 5438}
  [\href{https://arxiv.org/abs/hep-ph/9407247}{{\ttfamily hep-ph/9407247}}].

\bibitem{Kaiser:1995fb}
D.I.~Kaiser, \emph{{Post inflation reheating in an expanding universe}},
  \href{https://doi.org/10.1103/PhysRevD.53.1776}{\emph{Phys. Rev. D}
  {\bfseries 53} (1996) 1776}
  [\href{https://arxiv.org/abs/astro-ph/9507108}{{\ttfamily
  astro-ph/9507108}}].

\bibitem{Kofman:1997yn}
L.~Kofman, A.D.~Linde and A.A.~Starobinsky, \emph{{Towards the theory of
  reheating after inflation}},
  \href{https://doi.org/10.1103/PhysRevD.56.3258}{\emph{Phys. Rev. D}
  {\bfseries 56} (1997) 3258}
  [\href{https://arxiv.org/abs/hep-ph/9704452}{{\ttfamily hep-ph/9704452}}].

\bibitem{Greene:1997fu}
P.B.~Greene, L.~Kofman, A.D.~Linde and A.A.~Starobinsky, \emph{{Structure of
  resonance in preheating after inflation}},
  \href{https://doi.org/10.1103/PhysRevD.56.6175}{\emph{Phys. Rev. D}
  {\bfseries 56} (1997) 6175}
  [\href{https://arxiv.org/abs/hep-ph/9705347}{{\ttfamily hep-ph/9705347}}].

\bibitem{Kaiser:1997mp}
D.I.~Kaiser, \emph{{Preheating in an expanding universe: Analytic results for
  the massless case}},
  \href{https://doi.org/10.1103/PhysRevD.56.706}{\emph{Phys. Rev. D} {\bfseries
  56} (1997) 706} [\href{https://arxiv.org/abs/hep-ph/9702244}{{\ttfamily
  hep-ph/9702244}}].

\bibitem{Kaiser:1997hg}
D.I.~Kaiser, \emph{{Resonance structure for preheating with massless fields}},
  \href{https://doi.org/10.1103/PhysRevD.57.702}{\emph{Phys. Rev. D} {\bfseries
  57} (1998) 702} [\href{https://arxiv.org/abs/hep-ph/9707516}{{\ttfamily
  hep-ph/9707516}}].

\bibitem{Greene:1998nh}
P.B.~Greene and L.~Kofman, \emph{{Preheating of fermions}},
  \href{https://doi.org/10.1016/S0370-2693(99)00020-9}{\emph{Phys. Lett. B}
  {\bfseries 448} (1999) 6}
  [\href{https://arxiv.org/abs/hep-ph/9807339}{{\ttfamily hep-ph/9807339}}].

\bibitem{Greene:2000ew}
P.B.~Greene and L.~Kofman, \emph{{On the theory of fermionic preheating}},
  \href{https://doi.org/10.1103/PhysRevD.62.123516}{\emph{Phys. Rev. D}
  {\bfseries 62} (2000) 123516}
  [\href{https://arxiv.org/abs/hep-ph/0003018}{{\ttfamily hep-ph/0003018}}].

\bibitem{Peloso:2000hy}
M.~Peloso and L.~Sorbo, \emph{{Preheating of massive fermions after inflation:
  Analytical results}},
  \href{https://doi.org/10.1088/1126-6708/2000/05/016}{\emph{JHEP} {\bfseries
  05} (2000) 016} [\href{https://arxiv.org/abs/hep-ph/0003045}{{\ttfamily
  hep-ph/0003045}}].

\bibitem{Felder:2000hj}
G.N.~Felder, J.~Garcia-Bellido, P.B.~Greene, L.~Kofman, A.D.~Linde and
  I.~Tkachev, \emph{{Dynamics of symmetry breaking and tachyonic preheating}},
  \href{https://doi.org/10.1103/PhysRevLett.87.011601}{\emph{Phys. Rev. Lett.}
  {\bfseries 87} (2001) 011601}
  [\href{https://arxiv.org/abs/hep-ph/0012142}{{\ttfamily hep-ph/0012142}}].

\bibitem{Felder:2001kt}
G.N.~Felder, L.~Kofman and A.D.~Linde, \emph{{Tachyonic instability and
  dynamics of spontaneous symmetry breaking}},
  \href{https://doi.org/10.1103/PhysRevD.64.123517}{\emph{Phys. Rev. D}
  {\bfseries 64} (2001) 123517}
  [\href{https://arxiv.org/abs/hep-th/0106179}{{\ttfamily hep-th/0106179}}].

\bibitem{Copeland:2002ku}
E.J.~Copeland, S.~Pascoli and A.~Rajantie, \emph{{Dynamics of tachyonic
  preheating after hybrid inflation}},
  \href{https://doi.org/10.1103/PhysRevD.65.103517}{\emph{Phys. Rev. D}
  {\bfseries 65} (2002) 103517}
  [\href{https://arxiv.org/abs/hep-ph/0202031}{{\ttfamily hep-ph/0202031}}].

\bibitem{GarciaBellido:2002aj}
J.~Garcia-Bellido, M.~Garcia~Perez and A.~Gonzalez-Arroyo, \emph{{Symmetry
  breaking and false vacuum decay after hybrid inflation}},
  \href{https://doi.org/10.1103/PhysRevD.67.103501}{\emph{Phys. Rev. D}
  {\bfseries 67} (2003) 103501}
  [\href{https://arxiv.org/abs/hep-ph/0208228}{{\ttfamily hep-ph/0208228}}].

\bibitem{Rajantie:2000nj}
A.~Rajantie, P.M.~Saffin and E.J.~Copeland, \emph{{Electroweak preheating on a
  lattice}}, \href{https://doi.org/10.1103/PhysRevD.63.123512}{\emph{Phys. Rev.
  D} {\bfseries 63} (2001) 123512}
  [\href{https://arxiv.org/abs/hep-ph/0012097}{{\ttfamily hep-ph/0012097}}].

\bibitem{Copeland:2001qw}
E.J.~Copeland, D.~Lyth, A.~Rajantie and M.~Trodden, \emph{{Hybrid inflation and
  baryogenesis at the TeV scale}},
  \href{https://doi.org/10.1103/PhysRevD.64.043506}{\emph{Phys. Rev. D}
  {\bfseries 64} (2001) 043506}
  [\href{https://arxiv.org/abs/hep-ph/0103231}{{\ttfamily hep-ph/0103231}}].

\bibitem{Smit:2002yg}
J.~Smit and A.~Tranberg, \emph{{Chern-Simons number asymmetry from CP violation
  at electroweak tachyonic preheating}},
  \href{https://doi.org/10.1088/1126-6708/2002/12/020}{\emph{JHEP} {\bfseries
  12} (2002) 020} [\href{https://arxiv.org/abs/hep-ph/0211243}{{\ttfamily
  hep-ph/0211243}}].

\bibitem{GarciaBellido:2003wd}
J.~Garcia-Bellido, M.~Garcia-Perez and A.~Gonzalez-Arroyo, \emph{{Chern-Simons
  production during preheating in hybrid inflation models}},
  \href{https://doi.org/10.1103/PhysRevD.69.023504}{\emph{Phys. Rev. D}
  {\bfseries 69} (2004) 023504}
  [\href{https://arxiv.org/abs/hep-ph/0304285}{{\ttfamily hep-ph/0304285}}].

\bibitem{Tranberg:2003gi}
A.~Tranberg and J.~Smit, \emph{{Baryon asymmetry from electroweak tachyonic
  preheating}},
  \href{https://doi.org/10.1088/1126-6708/2003/11/016}{\emph{JHEP} {\bfseries
  11} (2003) 016} [\href{https://arxiv.org/abs/hep-ph/0310342}{{\ttfamily
  hep-ph/0310342}}].

\bibitem{Skullerud:2003ki}
J.-I.~Skullerud, J.~Smit and A.~Tranberg, \emph{{W and Higgs particle
  distributions during electroweak tachyonic preheating}},
  \href{https://doi.org/10.1088/1126-6708/2003/08/045}{\emph{JHEP} {\bfseries
  08} (2003) 045} [\href{https://arxiv.org/abs/hep-ph/0307094}{{\ttfamily
  hep-ph/0307094}}].

\bibitem{vanderMeulen:2005sp}
M.~van~der Meulen, D.~Sexty, J.~Smit and A.~Tranberg, \emph{{Chern-Simons and
  winding number in a tachyonic electroweak transition}},
  \href{https://doi.org/10.1088/1126-6708/2006/02/029}{\emph{JHEP} {\bfseries
  02} (2006) 029} [\href{https://arxiv.org/abs/hep-ph/0511080}{{\ttfamily
  hep-ph/0511080}}].

\bibitem{DiazGil:2007dy}
A.~Diaz-Gil, J.~Garcia-Bellido, M.~Garcia~Perez and A.~Gonzalez-Arroyo,
  \emph{{Magnetic field production during preheating at the electroweak
  scale}}, \href{https://doi.org/10.1103/PhysRevLett.100.241301}{\emph{Phys.
  Rev. Lett.} {\bfseries 100} (2008) 241301}
  [\href{https://arxiv.org/abs/0712.4263}{{\ttfamily 0712.4263}}].

\bibitem{DiazGil:2008tf}
A.~Diaz-Gil, J.~Garcia-Bellido, M.~Garcia~Perez and A.~Gonzalez-Arroyo,
  \emph{{Primordial magnetic fields from preheating at the electroweak scale}},
  \href{https://doi.org/10.1088/1126-6708/2008/07/043}{\emph{JHEP} {\bfseries
  07} (2008) 043} [\href{https://arxiv.org/abs/0805.4159}{{\ttfamily
  0805.4159}}].

\bibitem{Dufaux:2010cf}
J.-F.~Dufaux, D.G.~Figueroa and J.~Garcia-Bellido, \emph{{Gravitational Waves
  from Abelian Gauge Fields and Cosmic Strings at Preheating}},
  \href{https://doi.org/10.1103/PhysRevD.82.083518}{\emph{Phys. Rev. D}
  {\bfseries 82} (2010) 083518}
  [\href{https://arxiv.org/abs/1006.0217}{{\ttfamily 1006.0217}}].

\bibitem{Berges:2010zv}
J.~Berges, D.~Gelfand and J.~Pruschke, \emph{{Quantum theory of fermion
  production after inflation}},
  \href{https://doi.org/10.1103/PhysRevLett.107.061301}{\emph{Phys. Rev. Lett.}
  {\bfseries 107} (2011) 061301}
  [\href{https://arxiv.org/abs/1012.4632}{{\ttfamily 1012.4632}}].

\bibitem{Tranberg:2017lrx}
A.~Tranberg, S.~T\"ahtinen and D.J.~Weir, \emph{{Gravitational waves from
  non-Abelian gauge fields at a tachyonic transition}},
  \href{https://doi.org/10.1088/1475-7516/2018/04/012}{\emph{JCAP} {\bfseries
  04} (2018) 012} [\href{https://arxiv.org/abs/1706.02365}{{\ttfamily
  1706.02365}}].

\bibitem{Deskins:2013lfx}
J.T.~Deskins, J.T.~Giblin and R.R.~Caldwell, \emph{{Gauge Field Preheating at
  the End of Inflation}},
  \href{https://doi.org/10.1103/PhysRevD.88.063530}{\emph{Phys. Rev. D}
  {\bfseries 88} (2013) 063530}
  [\href{https://arxiv.org/abs/1305.7226}{{\ttfamily 1305.7226}}].

\bibitem{Adshead:2015pva}
P.~Adshead, J.T.~Giblin, T.R.~Scully and E.I.~Sfakianakis,
  \emph{{Gauge-preheating and the end of axion inflation}},
  \href{https://doi.org/10.1088/1475-7516/2015/12/034}{\emph{JCAP} {\bfseries
  12} (2015) 034} [\href{https://arxiv.org/abs/1502.06506}{{\ttfamily
  1502.06506}}].

\bibitem{Adshead:2016iae}
P.~Adshead, J.T.~Giblin, T.R.~Scully and E.I.~Sfakianakis,
  \emph{{Magnetogenesis from axion inflation}},
  \href{https://doi.org/10.1088/1475-7516/2016/10/039}{\emph{JCAP} {\bfseries
  10} (2016) 039} [\href{https://arxiv.org/abs/1606.08474}{{\ttfamily
  1606.08474}}].

\bibitem{Lozanov:2016hid}
K.D.~Lozanov and M.A.~Amin, \emph{{Equation of State and Duration to Radiation
  Domination after Inflation}},
  \href{https://doi.org/10.1103/PhysRevLett.119.061301}{\emph{Phys. Rev. Lett.}
  {\bfseries 119} (2017) 061301}
  [\href{https://arxiv.org/abs/1608.01213}{{\ttfamily 1608.01213}}].

\bibitem{Lozanov:2016pac}
K.D.~Lozanov and M.A.~Amin, \emph{{The charged inflaton and its gauge fields:
  preheating and initial conditions for reheating}},
  \href{https://doi.org/10.1088/1475-7516/2016/06/032}{\emph{JCAP} {\bfseries
  06} (2016) 032} [\href{https://arxiv.org/abs/1603.05663}{{\ttfamily
  1603.05663}}].

\bibitem{Lozanov:2017hjm}
K.D.~Lozanov and M.A.~Amin, \emph{{Self-resonance after inflation: oscillons,
  transients and radiation domination}},
  \href{https://doi.org/10.1103/PhysRevD.97.023533}{\emph{Phys. Rev. D}
  {\bfseries 97} (2018) 023533}
  [\href{https://arxiv.org/abs/1710.06851}{{\ttfamily 1710.06851}}].

\bibitem{Figueroa:2017qmv}
D.G.~Figueroa and M.~Shaposhnikov, \emph{{Lattice implementation of Abelian
  gauge theories with Chern\textendash{}Simons number and an axion field}},
  \href{https://doi.org/10.1016/j.nuclphysb.2017.12.001}{\emph{Nucl. Phys. B}
  {\bfseries 926} (2018) 544}
  [\href{https://arxiv.org/abs/1705.09629}{{\ttfamily 1705.09629}}].

\bibitem{Adshead:2017xll}
P.~Adshead, J.T.~Giblin and Z.J.~Weiner, \emph{{Non-Abelian gauge preheating}},
  \href{https://doi.org/10.1103/PhysRevD.96.123512}{\emph{Phys. Rev. D}
  {\bfseries 96} (2017) 123512}
  [\href{https://arxiv.org/abs/1708.02944}{{\ttfamily 1708.02944}}].

\bibitem{Adshead:2018doq}
P.~Adshead, J.T.~Giblin and Z.J.~Weiner, \emph{{Gravitational waves from gauge
  preheating}}, \href{https://doi.org/10.1103/PhysRevD.98.043525}{\emph{Phys.
  Rev. D} {\bfseries 98} (2018) 043525}
  [\href{https://arxiv.org/abs/1805.04550}{{\ttfamily 1805.04550}}].

\bibitem{Cuissa:2018oiw}
J.R.C.~Cuissa and D.G.~Figueroa, \emph{{Lattice formulation of axion inflation.
  Application to preheating}},
  \href{https://doi.org/10.1088/1475-7516/2019/06/002}{\emph{JCAP} {\bfseries
  06} (2019) 002} [\href{https://arxiv.org/abs/1812.03132}{{\ttfamily
  1812.03132}}].

\bibitem{Adshead:2019lbr}
P.~Adshead, J.T.~Giblin, M.~Pieroni and Z.J.~Weiner, \emph{{Constraining axion
  inflation with gravitational waves from preheating}},
  \href{https://doi.org/10.1103/PhysRevD.101.083534}{\emph{Phys. Rev. D}
  {\bfseries 101} (2020) 083534}
  [\href{https://arxiv.org/abs/1909.12842}{{\ttfamily 1909.12842}}].

\bibitem{Adshead:2019igv}
P.~Adshead, J.T.~Giblin, M.~Pieroni and Z.J.~Weiner, \emph{{Constraining Axion
  Inflation with Gravitational Waves across 29 Decades in Frequency}},
  \href{https://doi.org/10.1103/PhysRevLett.124.171301}{\emph{Phys. Rev. Lett.}
  {\bfseries 124} (2020) 171301}
  [\href{https://arxiv.org/abs/1909.12843}{{\ttfamily 1909.12843}}].

\bibitem{Figueroa:2019jsi}
D.G.~Figueroa, A.~Florio and M.~Shaposhnikov, \emph{{Chiral charge dynamics in
  Abelian gauge theories at finite temperature}},
  \href{https://doi.org/10.1007/JHEP10(2019)142}{\emph{JHEP} {\bfseries 10}
  (2019) 142} [\href{https://arxiv.org/abs/1904.11892}{{\ttfamily
  1904.11892}}].

\bibitem{Antusch:2020iyq}
S.~Antusch, D.G.~Figueroa, K.~Marschall and F.~Torrenti, \emph{{Energy
  distribution and equation of state of the early Universe: matching the end of
  inflation and the onset of radiation domination}},
  \href{https://doi.org/10.1016/j.physletb.2020.135888}{\emph{Phys. Lett. B}
  {\bfseries 811} (2020) 135888}
  [\href{https://arxiv.org/abs/2005.07563}{{\ttfamily 2005.07563}}].

\bibitem{Antusch:2021aiw}
S.~Antusch, D.G.~Figueroa, K.~Marschall and F.~Torrenti, \emph{{Characterizing
  the postinflationary reheating history: Single daughter field with
  quadratic-quadratic interaction}},
  \href{https://doi.org/10.1103/PhysRevD.105.043532}{\emph{Phys. Rev. D}
  {\bfseries 105} (2022) 043532}
  [\href{https://arxiv.org/abs/2112.11280}{{\ttfamily 2112.11280}}].

\bibitem{Ford:1986sy}
L.H.~Ford, \emph{{Gravitational Particle Creation and Inflation}},
  \href{https://doi.org/10.1103/PhysRevD.35.2955}{\emph{Phys. Rev. D}
  {\bfseries 35} (1987) 2955}.

\bibitem{Spokoiny:1993kt}
B.~Spokoiny, \emph{{Deflationary universe scenario}},
  \href{https://doi.org/10.1016/0370-2693(93)90155-B}{\emph{Phys. Lett. B}
  {\bfseries 315} (1993) 40}
  [\href{https://arxiv.org/abs/gr-qc/9306008}{{\ttfamily gr-qc/9306008}}].

\bibitem{Bassett:1997az}
B.A.~Bassett and S.~Liberati, \emph{{Geometric reheating after inflation}},
  \href{https://doi.org/10.1103/PhysRevD.60.049902}{\emph{Phys. Rev. D}
  {\bfseries 58} (1998) 021302}
  [\href{https://arxiv.org/abs/hep-ph/9709417}{{\ttfamily hep-ph/9709417}}].

\bibitem{Tsujikawa:1999jh}
S.~Tsujikawa, K.-i.~Maeda and T.~Torii, \emph{{Resonant particle production
  with nonminimally coupled scalar fields in preheating after inflation}},
  \href{https://doi.org/10.1103/PhysRevD.60.063515}{\emph{Phys. Rev. D}
  {\bfseries 60} (1999) 063515}
  [\href{https://arxiv.org/abs/hep-ph/9901306}{{\ttfamily hep-ph/9901306}}].

\bibitem{Tsujikawa:1999iv}
S.~Tsujikawa, K.-i.~Maeda and T.~Torii, \emph{{Preheating with nonminimally
  coupled scalar fields in higher curvature inflation models}},
  \href{https://doi.org/10.1103/PhysRevD.60.123505}{\emph{Phys. Rev. D}
  {\bfseries 60} (1999) 123505}
  [\href{https://arxiv.org/abs/hep-ph/9906501}{{\ttfamily hep-ph/9906501}}].

\bibitem{Tsujikawa:1999me}
S.~Tsujikawa, K.-i.~Maeda and T.~Torii, \emph{{Preheating of the nonminimally
  coupled inflaton field}},
  \href{https://doi.org/10.1103/PhysRevD.61.103501}{\emph{Phys. Rev. D}
  {\bfseries 61} (2000) 103501}
  [\href{https://arxiv.org/abs/hep-ph/9910214}{{\ttfamily hep-ph/9910214}}].

\bibitem{DeCross:2015uza}
M.P.~DeCross, D.I.~Kaiser, A.~Prabhu, C.~Prescod-Weinstein and
  E.I.~Sfakianakis, \emph{{Preheating after Multifield Inflation with
  Nonminimal Couplings, I: Covariant Formalism and Attractor Behavior}},
  \href{https://doi.org/10.1103/PhysRevD.97.023526}{\emph{Phys. Rev. D}
  {\bfseries 97} (2018) 023526}
  [\href{https://arxiv.org/abs/1510.08553}{{\ttfamily 1510.08553}}].

\bibitem{DeCross:2016fdz}
M.P.~DeCross, D.I.~Kaiser, A.~Prabhu, C.~Prescod-Weinstein and
  E.I.~Sfakianakis, \emph{{Preheating after multifield inflation with
  nonminimal couplings, II: Resonance Structure}},
  \href{https://doi.org/10.1103/PhysRevD.97.023527}{\emph{Phys. Rev. D}
  {\bfseries 97} (2018) 023527}
  [\href{https://arxiv.org/abs/1610.08868}{{\ttfamily 1610.08868}}].

\bibitem{DeCross:2016cbs}
M.P.~DeCross, D.I.~Kaiser, A.~Prabhu, C.~Prescod-Weinstein and
  E.I.~Sfakianakis, \emph{{Preheating after multifield inflation with
  nonminimal couplings, III: Dynamical spacetime results}},
  \href{https://doi.org/10.1103/PhysRevD.97.023528}{\emph{Phys. Rev. D}
  {\bfseries 97} (2018) 023528}
  [\href{https://arxiv.org/abs/1610.08916}{{\ttfamily 1610.08916}}].

\bibitem{Figueroa:2016dsc}
D.G.~Figueroa and C.T.~Byrnes, \emph{{The Standard Model Higgs as the origin of
  the hot Big Bang}},
  \href{https://doi.org/10.1016/j.physletb.2017.01.059}{\emph{Phys. Lett. B}
  {\bfseries 767} (2017) 272}
  [\href{https://arxiv.org/abs/1604.03905}{{\ttfamily 1604.03905}}].

\bibitem{Opferkuch:2019zbd}
T.~Opferkuch, P.~Schwaller and B.A.~Stefanek, \emph{{Ricci Reheating}},
  \href{https://doi.org/10.1088/1475-7516/2019/07/016}{\emph{JCAP} {\bfseries
  07} (2019) 016} [\href{https://arxiv.org/abs/1905.06823}{{\ttfamily
  1905.06823}}].

\bibitem{Dimopoulos:2018wfg}
K.~Dimopoulos and T.~Markkanen, \emph{{Non-minimal gravitational reheating
  during kination}},
  \href{https://doi.org/10.1088/1475-7516/2018/06/021}{\emph{JCAP} {\bfseries
  06} (2018) 021} [\href{https://arxiv.org/abs/1803.07399}{{\ttfamily
  1803.07399}}].

\bibitem{Bettoni:2021zhq}
D.~Bettoni, A.~Lopez-Eiguren and J.~Rubio, \emph{{Hubble-induced phase
  transitions on the lattice with applications to Ricci reheating}},
  \href{https://doi.org/10.1088/1475-7516/2022/01/002}{\emph{JCAP} {\bfseries
  01} (2022) 002} [\href{https://arxiv.org/abs/2107.09671}{{\ttfamily
  2107.09671}}].

\bibitem{Figueroa:2021iwm}
D.G.~Figueroa, A.~Florio, T.~Opferkuch and B.A.~Stefanek, \emph{{Lattice
  simulations of non-minimally coupled scalar fields in the Jordan frame}},
  \href{https://doi.org/10.21468/SciPostPhys.15.3.077}{\emph{SciPost Phys.}
  {\bfseries 15} (2023) 077}
  [\href{https://arxiv.org/abs/2112.08388}{{\ttfamily 2112.08388}}].

\bibitem{Figueroa:2024asq}
D.G.~Figueroa, T.~Opferkuch and B.A.~Stefanek, \emph{{Ricci Reheating on the
  Lattice}},  \href{https://arxiv.org/abs/2404.17654}{{\ttfamily 2404.17654}}.

\bibitem{Ema:2016dny}
Y.~Ema, R.~Jinno, K.~Mukaida and K.~Nakayama, \emph{{Violent Preheating in
  Inflation with Nonminimal Coupling}},
  \href{https://doi.org/10.1088/1475-7516/2017/02/045}{\emph{JCAP} {\bfseries
  02} (2017) 045} [\href{https://arxiv.org/abs/1609.05209}{{\ttfamily
  1609.05209}}].

\bibitem{Nguyen:2019kbm}
R.~Nguyen, J.~van~de Vis, E.I.~Sfakianakis, J.T.~Giblin and D.I.~Kaiser,
  \emph{{Nonlinear Dynamics of Preheating after Multifield Inflation with
  Nonminimal Couplings}},
  \href{https://doi.org/10.1103/PhysRevLett.123.171301}{\emph{Phys. Rev. Lett.}
  {\bfseries 123} (2019) 171301}
  [\href{https://arxiv.org/abs/1905.12562}{{\ttfamily 1905.12562}}].

\bibitem{vandeVis:2020qcp}
J.~van~de Vis, R.~Nguyen, E.I.~Sfakianakis, J.T.~Giblin and D.I.~Kaiser,
  \emph{{Time scales for nonlinear processes in preheating after multifield
  inflation with nonminimal couplings}},
  \href{https://doi.org/10.1103/PhysRevD.102.043528}{\emph{Phys. Rev. D}
  {\bfseries 102} (2020) 043528}
  [\href{https://arxiv.org/abs/2005.00433}{{\ttfamily 2005.00433}}].

\bibitem{Allahverdi:2010xz}
R.~Allahverdi, R.~Brandenberger, F.-Y.~Cyr-Racine and A.~Mazumdar,
  \emph{{Reheating in Inflationary Cosmology: Theory and Applications}},
  \href{https://doi.org/10.1146/annurev.nucl.012809.104511}{\emph{Ann. Rev.
  Nucl. Part. Sci.} {\bfseries 60} (2010) 27}
  [\href{https://arxiv.org/abs/1001.2600}{{\ttfamily 1001.2600}}].

\bibitem{Amin:2014eta}
M.A.~Amin, M.P.~Hertzberg, D.I.~Kaiser and J.~Karouby, \emph{{Nonperturbative
  Dynamics Of Reheating After Inflation: A Review}},
  \href{https://doi.org/10.1142/S0218271815300037}{\emph{Int. J. Mod. Phys. D}
  {\bfseries 24} (2014) 1530003}
  [\href{https://arxiv.org/abs/1410.3808}{{\ttfamily 1410.3808}}].

\bibitem{Lozanov:2019jxc}
K.D.~Lozanov, \emph{{Lectures on Reheating after Inflation}},
  \href{https://arxiv.org/abs/1907.04402}{{\ttfamily 1907.04402}}.

\bibitem{Allahverdi:2020bys}
R.~Allahverdi et~al., \emph{{The First Three Seconds: a Review of Possible
  Expansion Histories of the Early Universe}},
  \href{https://arxiv.org/abs/2006.16182}{{\ttfamily 2006.16182}}.

\bibitem{Parker2009}
L.~Parker and D.~Toms, \emph{{Quantum Field Theory in Curved Spacetime:
  Quantized Fields and Gravity}}, Cambridge Monographs on Mathematical Physics
  (2009).

\bibitem{Birrell1984}
N.~Birrell and P.~Davies, \emph{{Quantum Fields in Curved Space}}, Cambridge
  Monographs on Mathematical Physics (1984).

\bibitem{Peebles:1998qn}
P.J.E.~Peebles and A.~Vilenkin, \emph{{Quintessential inflation}},
  \href{https://doi.org/10.1103/PhysRevD.59.063505}{\emph{Phys. Rev. D}
  {\bfseries 59} (1999) 063505}
  [\href{https://arxiv.org/abs/astro-ph/9810509}{{\ttfamily
  astro-ph/9810509}}].

\bibitem{Damour:1995pd}
T.~Damour and A.~Vilenkin, \emph{{String theory and inflation}},
  \href{https://doi.org/10.1103/PhysRevD.53.2981}{\emph{Phys. Rev. D}
  {\bfseries 53} (1996) 2981}
  [\href{https://arxiv.org/abs/hep-th/9503149}{{\ttfamily hep-th/9503149}}].

\bibitem{Peloso:1999dm}
M.~Peloso and F.~Rosati, \emph{{On the construction of quintessential inflation
  models}}, \href{https://doi.org/10.1088/1126-6708/1999/12/026}{\emph{JHEP}
  {\bfseries 12} (1999) 026}
  [\href{https://arxiv.org/abs/hep-ph/9908271}{{\ttfamily hep-ph/9908271}}].

\bibitem{Huey:2001ae}
G.~Huey and J.E.~Lidsey, \emph{{Inflation, brane worlds and quintessence}},
  \href{https://doi.org/10.1016/S0370-2693(01)00808-5}{\emph{Phys. Lett. B}
  {\bfseries 514} (2001) 217}
  [\href{https://arxiv.org/abs/astro-ph/0104006}{{\ttfamily
  astro-ph/0104006}}].

\bibitem{Majumdar:2001mm}
A.S.~Majumdar, \emph{{From brane assisted inflation to quintessence through a
  single scalar field}},
  \href{https://doi.org/10.1103/PhysRevD.64.083503}{\emph{Phys. Rev. D}
  {\bfseries 64} (2001) 083503}
  [\href{https://arxiv.org/abs/astro-ph/0105518}{{\ttfamily
  astro-ph/0105518}}].

\bibitem{Dimopoulos:2001ix}
K.~Dimopoulos and J.W.F.~Valle, \emph{{Modeling quintessential inflation}},
  \href{https://doi.org/10.1016/S0927-6505(02)00115-9}{\emph{Astropart. Phys.}
  {\bfseries 18} (2002) 287}
  [\href{https://arxiv.org/abs/astro-ph/0111417}{{\ttfamily
  astro-ph/0111417}}].

\bibitem{Wetterich:2013jsa}
C.~Wetterich, \emph{{Variable gravity Universe}},
  \href{https://doi.org/10.1103/PhysRevD.89.024005}{\emph{Phys. Rev. D}
  {\bfseries 89} (2014) 024005}
  [\href{https://arxiv.org/abs/1308.1019}{{\ttfamily 1308.1019}}].

\bibitem{Wetterich:2014gaa}
C.~Wetterich, \emph{{Inflation, quintessence, and the origin of mass}},
  \href{https://doi.org/10.1016/j.nuclphysb.2015.05.019}{\emph{Nucl. Phys. B}
  {\bfseries 897} (2015) 111}
  [\href{https://arxiv.org/abs/1408.0156}{{\ttfamily 1408.0156}}].

\bibitem{Hossain:2014xha}
M.W.~Hossain, R.~Myrzakulov, M.~Sami and E.N.~Saridakis, \emph{{Variable
  gravity: A suitable framework for quintessential inflation}},
  \href{https://doi.org/10.1103/PhysRevD.90.023512}{\emph{Phys. Rev. D}
  {\bfseries 90} (2014) 023512}
  [\href{https://arxiv.org/abs/1402.6661}{{\ttfamily 1402.6661}}].

\bibitem{Rubio:2017gty}
J.~Rubio and C.~Wetterich, \emph{{Emergent scale symmetry: Connecting inflation
  and dark energy}},
  \href{https://doi.org/10.1103/PhysRevD.96.063509}{\emph{Phys. Rev. D}
  {\bfseries 96} (2017) 063509}
  [\href{https://arxiv.org/abs/1705.00552}{{\ttfamily 1705.00552}}].

\bibitem{Figueroa:2018twl}
D.G.~Figueroa and E.H.~Tanin, \emph{{Inconsistency of an inflationary sector
  coupled only to Einstein gravity}},
  \href{https://doi.org/10.1088/1475-7516/2019/10/050}{\emph{JCAP} {\bfseries
  10} (2019) 050} [\href{https://arxiv.org/abs/1811.04093}{{\ttfamily
  1811.04093}}].

\bibitem{Caprini:2018mtu}
C.~Caprini and D.G.~Figueroa, \emph{{Cosmological Backgrounds of Gravitational
  Waves}}, \href{https://doi.org/10.1088/1361-6382/aac608}{\emph{Class. Quant.
  Grav.} {\bfseries 35} (2018) 163001}
  [\href{https://arxiv.org/abs/1801.04268}{{\ttfamily 1801.04268}}].

\bibitem{Clarke:2020bil}
T.J.~Clarke, E.J.~Copeland and A.~Moss, \emph{{Constraints on primordial
  gravitational waves from the Cosmic Microwave Background}},
  \href{https://doi.org/10.1088/1475-7516/2020/10/002}{\emph{JCAP} {\bfseries
  10} (2020) 002} [\href{https://arxiv.org/abs/2004.11396}{{\ttfamily
  2004.11396}}].

\bibitem{Giovannini:2009kg}
M.~Giovannini, \emph{{Stochastic backgrounds of relic gravitons: a theoretical
  appraisal}}, \href{https://doi.org/10.1186/1754-0410-4-1}{\emph{PMC Phys. A}
  {\bfseries 4} (2010) 1} [\href{https://arxiv.org/abs/0901.3026}{{\ttfamily
  0901.3026}}].

\bibitem{Boyle:2007zx}
L.A.~Boyle and A.~Buonanno, \emph{{Relating gravitational wave constraints from
  primordial nucleosynthesis, pulsar timing, laser interferometers, and the
  CMB: Implications for the early Universe}},
  \href{https://doi.org/10.1103/PhysRevD.78.043531}{\emph{Phys. Rev. D}
  {\bfseries 78} (2008) 043531}
  [\href{https://arxiv.org/abs/0708.2279}{{\ttfamily 0708.2279}}].

\bibitem{Figueroa:2019paj}
D.G.~Figueroa and E.H.~Tanin, \emph{{Ability of LIGO and LISA to probe the
  equation of state of the early Universe}},
  \href{https://doi.org/10.1088/1475-7516/2019/08/011}{\emph{JCAP} {\bfseries
  08} (2019) 011} [\href{https://arxiv.org/abs/1905.11960}{{\ttfamily
  1905.11960}}].

\bibitem{Bernal:2020ywq}
N.~Bernal, A.~Ghoshal, F.~Hajkarim and G.~Lambiase, \emph{{Primordial
  Gravitational Wave Signals in Modified Cosmologies}},
  \href{https://doi.org/10.1088/1475-7516/2020/11/051}{\emph{JCAP} {\bfseries
  11} (2020) 051} [\href{https://arxiv.org/abs/2008.04959}{{\ttfamily
  2008.04959}}].

\bibitem{Fu:2019qqe}
C.~Fu, P.~Wu and H.~Yu, \emph{{Nonlinear preheating with nonminimally coupled
  scalar fields in the Starobinsky model}},
  \href{https://doi.org/10.1103/PhysRevD.99.123526}{\emph{Phys. Rev. D}
  {\bfseries 99} (2019) 123526}
  [\href{https://arxiv.org/abs/1906.00557}{{\ttfamily 1906.00557}}].

\bibitem{Kallosh:2013hoa}
R.~Kallosh and A.~Linde, \emph{{Universality Class in Conformal Inflation}},
  \href{https://doi.org/10.1088/1475-7516/2013/07/002}{\emph{JCAP} {\bfseries
  07} (2013) 002} [\href{https://arxiv.org/abs/1306.5220}{{\ttfamily
  1306.5220}}].

\bibitem{Baumann:2009ds}
D.~Baumann, \emph{{Inflation}},  in \emph{{Theoretical Advanced Study Institute
  in Elementary Particle Physics}: {Physics of the Large and the Small}},
  pp.~523--686, 2011, \href{https://doi.org/10.1142/9789814327183_0010}{DOI}
  [\href{https://arxiv.org/abs/0907.5424}{{\ttfamily 0907.5424}}].

\bibitem{Turner:1983he}
M.S.~Turner, \emph{{Coherent Scalar Field Oscillations in an Expanding
  Universe}}, \href{https://doi.org/10.1103/PhysRevD.28.1243}{\emph{Phys. Rev.
  D} {\bfseries 28} (1983) 1243}.

\bibitem{Figueroa:2021yhd}
D.G.~Figueroa, A.~Florio, F.~Torrenti and W.~Valkenburg, \emph{{CosmoLattice: A
  modern code for lattice simulations of scalar and gauge field dynamics in an
  expanding universe}},
  \href{https://doi.org/10.1016/j.cpc.2022.108586}{\emph{Comput. Phys. Commun.}
  {\bfseries 283} (2023) 108586}
  [\href{https://arxiv.org/abs/2102.01031}{{\ttfamily 2102.01031}}].

\bibitem{Figueroa:2020rrl}
D.G.~Figueroa, A.~Florio, F.~Torrenti and W.~Valkenburg, \emph{{The art of
  simulating the early Universe -- Part I}},
  \href{https://doi.org/10.1088/1475-7516/2021/04/035}{\emph{JCAP} {\bfseries
  04} (2021) 035} [\href{https://arxiv.org/abs/2006.15122}{{\ttfamily
  2006.15122}}].

\bibitem{Figueroa:2023xmq}
D.G.~Figueroa, A.~Florio and F.~Torrenti, \emph{{Present and future of
  CosmoLattice}},  \href{https://arxiv.org/abs/2312.15056}{{\ttfamily
  2312.15056}}.

\bibitem{Figueroa:2015rqa}
D.G.~Figueroa, J.~Garcia-Bellido and F.~Torrenti, \emph{{Decay of the standard
  model Higgs field after inflation}},
  \href{https://doi.org/10.1103/PhysRevD.92.083511}{\emph{Phys. Rev. D}
  {\bfseries 92} (2015) 083511}
  [\href{https://arxiv.org/abs/1504.04600}{{\ttfamily 1504.04600}}].

\end{thebibliography}\endgroup

\end{document}